\documentclass[letterpaper,11pt,fleqn]{article}
\pdfoutput=1
\usepackage{jheppub}

\usepackage{graphicx}
\usepackage[figuresright]{rotating}
\usepackage{bm,amsmath,amssymb}
\usepackage[mathscr]{eucal}
\usepackage{color}
\usepackage{caption}

\long\def\comment#1{ }
\newcommand{\eqn}[1]{Eq.~\eqref{#1}}
\newcommand{\beq}{\begin{eqnarray}}
  \newcommand{\eeq}{\end{eqnarray}}

\newcommand{\nn}{\nonumber\\}
\newcommand{\dif}{{\rm d}}
\newcommand{\rmd}{{\rm d}}
\newcommand{\rme}{{\rm e}}
\newcommand{\rmi}{{\rm i}}

\newcommand{\rmI}{{\rm I}}
\newcommand{\rmJ}{{\rm J}}
\newcommand{\del}{\partial}

\newcommand{\order}[1]{\mcal{O}{(#1)}}
\newcommand{\mcal}{\mathcal}

\newcommand{\bk}{\bm{k}}
\newcommand{\be}{\bm{e}}

\newcommand{\bp}{\bm{p}}
\newcommand{\bx}{\bm{x}}
\newcommand{\by}{\bm{y}}

\newcommand{\bv}{\bm{v}}
\newcommand{\bz}{\bm{z}}

\newcommand{\abar}{\bar{\alpha}_s}

\newcommand{\sdla}{{\rm \scriptscriptstyle DLA}}

\newcommand{\ssA}{{\rm \scriptscriptstyle A}}
\newcommand{\ssB}{{\rm \scriptscriptstyle B}}
\newcommand{\ssC}{{\rm \scriptscriptstyle C}}
\newcommand{\ssE}{{\rm \scriptscriptstyle E}}

\newcommand{\CF}{C_{\rm F}}

\newcommand{\calC}{\mathcal{C}}

\newcommand{\calA}{\mathcal{A}}

\title{\Large  Resumming double non-global logarithms in the evolution of a jet}

\author[a]{Y. Hatta,}

\author[b]{E. Iancu,}

\author[c]{A.H. Mueller,}

\author[d]{and D.N.~Triantafyllopoulos\,}

\affiliation[a]{Yukawa Institute for Theoretical Physics, Kyoto University, Kyoto 606-8502, Japan}

\affiliation[b]{Institut de physique th\'{e}orique, Universit\'{e} Paris Saclay, CNRS, CEA, F-91191 Gif-sur-Yvette, France}

\affiliation[c]{Department of Physics, Columbia University, New York, NY 10027, USA}

\affiliation[d]{European Centre for Theoretical Studies in Nuclear Physics and Related Areas (ECT*)\\and Fondazione Bruno Kessler, Strada delle Tabarelle 286, I-38123 Villazzano (TN), Italy}

\emailAdd{hatta@yukawa.kyoto-u.ac.jp}
\emailAdd{edmond.iancu@ipht.fr}
\emailAdd{ahm4@columbia.edu}
\emailAdd{trianta@ectstar.eu}

\abstract{We consider the Banfi-Marchesini-Smye (BMS) equation which resums `non-global' energy logarithms in the QCD evolution of the energy lost by a pair of jets via soft radiation at large angles.  We identify a new physical regime where, besides the energy logarithms, one has to also resum (anti)collinear logarithms. Such a regime occurs when the jets are highly collimated (boosted) and the relative angles between successive soft gluon emissions are strongly increasing. These anti-collinear emissions can violate the correct time-ordering for time-like cascades and result in large radiative corrections enhanced by double collinear logs, making the BMS evolution unstable beyond leading order. We isolate the first such a correction in a recent calculation of the BMS equation to next-to-leading order by Caron-Huot. To overcome this difficulty, we construct a `collinearly-improved' version of the leading-order BMS equation which resums the double collinear logarithms to all orders. Our construction is inspired by a recent treatment of the Balitsky-Kovchegov (BK) equation for the high-energy evolution of a space-like wavefunction, where similar time-ordering issues occur.  We show that the conformal mapping relating the leading-order BMS and BK equations correctly predicts the physical time-ordering, but it fails to predict the detailed structure of the collinear improvement.}


\begin{document}
\maketitle

\section{Introduction}
\label{sec:intro}

The problem of the non-global logarithms \cite{Dasgupta:2001sh,Dasgupta:2002bw,Banfi:2002hw,Dokshitzer:2003uw,Marchesini:2003nh,Weigert:2003mm,Avsar:2009yb,Hatta:2013iba,Caron-Huot:2015bja,Larkoski:2015zka,Neill:2015nya,Neill:2016stq}
refers to the radiation by a jet at large angles w.r.t.~the jet axis, where the standard collinear radiation --- which controls the hadron multiplicity produced by the jet and is enhanced by collinear logs of the type $\ln(1/\theta^2_{\rm jet})$, with $\theta_{\rm jet}\ll 1$  the jet opening angle --- is suppressed. 

\begin{figure}[t] \centerline{
\includegraphics[width=.6\textwidth]{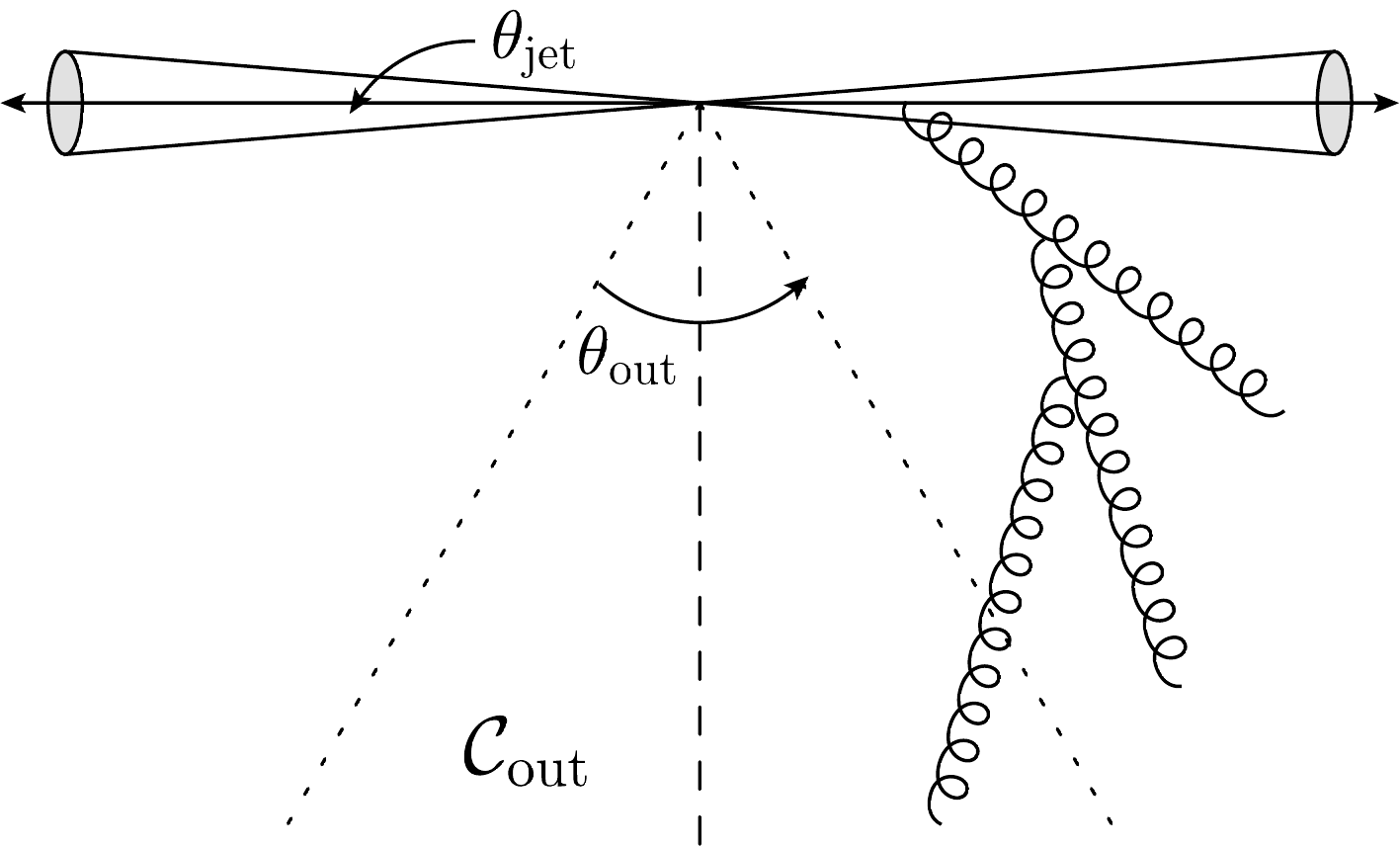}}
 \caption{\small A typical set-up for the emergence of non-global logarithms, as resummed by the BMS equation.}
 \label{fig:general}
\end{figure}

For definiteness, consider a pair of jets produced by the decay of a heavy particle, such as a $Z$ boson, or the virtual photon in the case of $e^+ e^-$ annihilation. In the center-of-mass (COM) frame, where the two initial partons --- say, a quark-antiquark ($q\bar q$) pair --- are propagating back-to-back and with equal energies ($E_a=E_b\equiv E$) ---, we define an `exclusion region' $\calC_{\rm out}$ which is separated from the jet axis by large angles and ask for the probability $P(E,E_0)$ that the total energy radiated by the jet within that region be smaller than a given value $E_0$, necessarily smaller than $E$ (see Fig.~\ref{fig:general}). When $E_0\ll E$, which is indeed the typical situation since radiation at large angles is strongly suppressed, the calculation of this probability in perturbative QCD receives large radiative corrections, of order $\big(\alpha_s\ln(E/E_0)\big)^n$ with $n\ge 1$, associated with successive emissions of soft gluons which are strongly ordered in energy and which propagate at larger and larger angles w.r.t.~the jet axis, within the `allowed' region between the jet and $\calC_{\rm out}$ (see Fig.~\ref{fig:general}). The last (softest) among these gluons can radiate a gluon with $\omega > E_0$ which propagates into $\calC_{\rm out}$, thus reducing the probability $P(E,E_0)$. The ensemble of this evolution with increasing `rapidity' $Y\equiv \ln(E/E_0)$ is described by the BMS equation (from Banfi, Marchesini, and Smye) \cite{Banfi:2002hw}. This equation, which is non-linear (as required by probability conservation), has recently been extended to next-to-leading order (NLO) accuracy \cite{Caron-Huot:2015bja}. The energy logarithms resummed by this equation are generally referred to as {\em non-global, single, logarithms}, to emphasise that (a) they refer to radiation in a restricted region of the angular phase-space and (b) the energy logarithms are not accompanied by collinear logs (unlike for the usual intra-jet evolution, where the successive emissions are strongly ordered in both energy and angles).

\begin{figure}[t] \centerline{
\includegraphics[width=.42\textwidth]{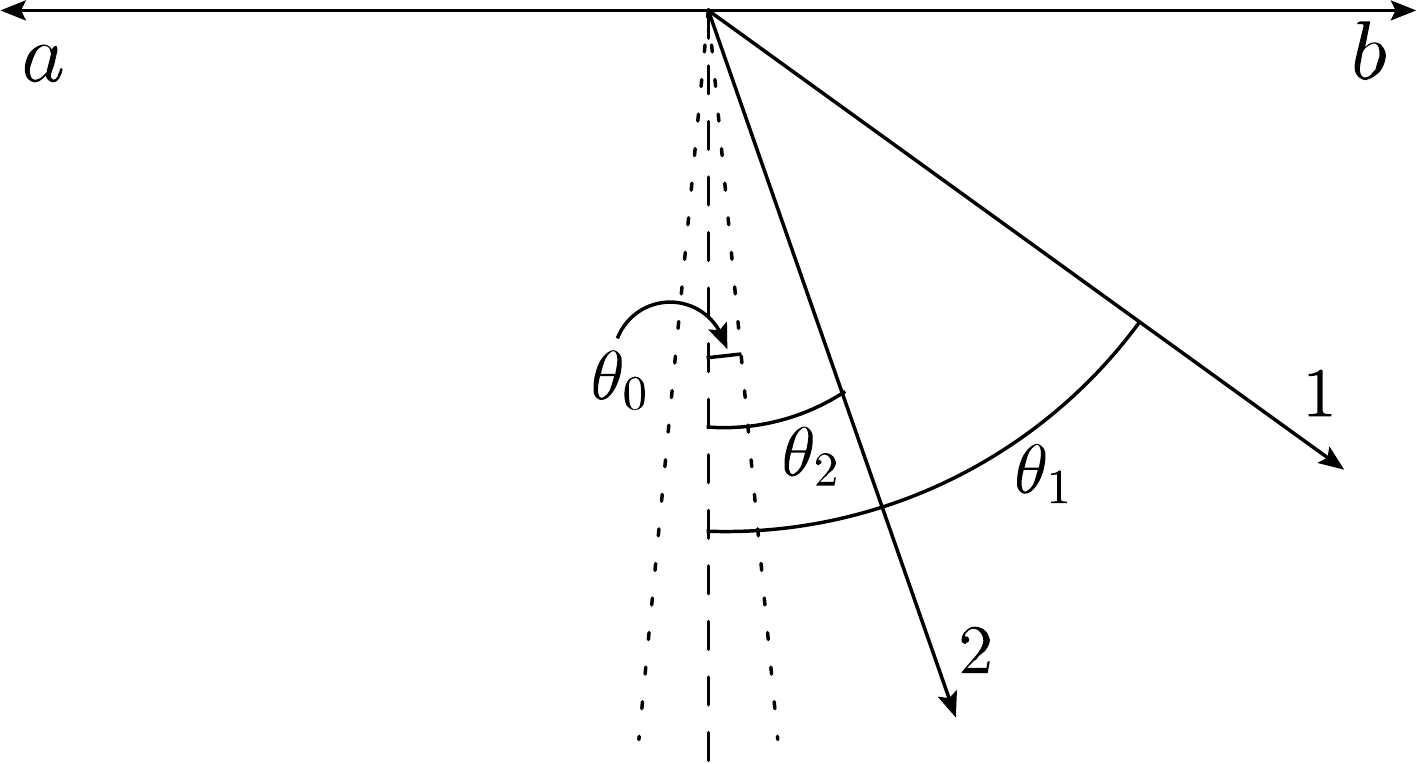}\qquad
\includegraphics[width=.42\textwidth]{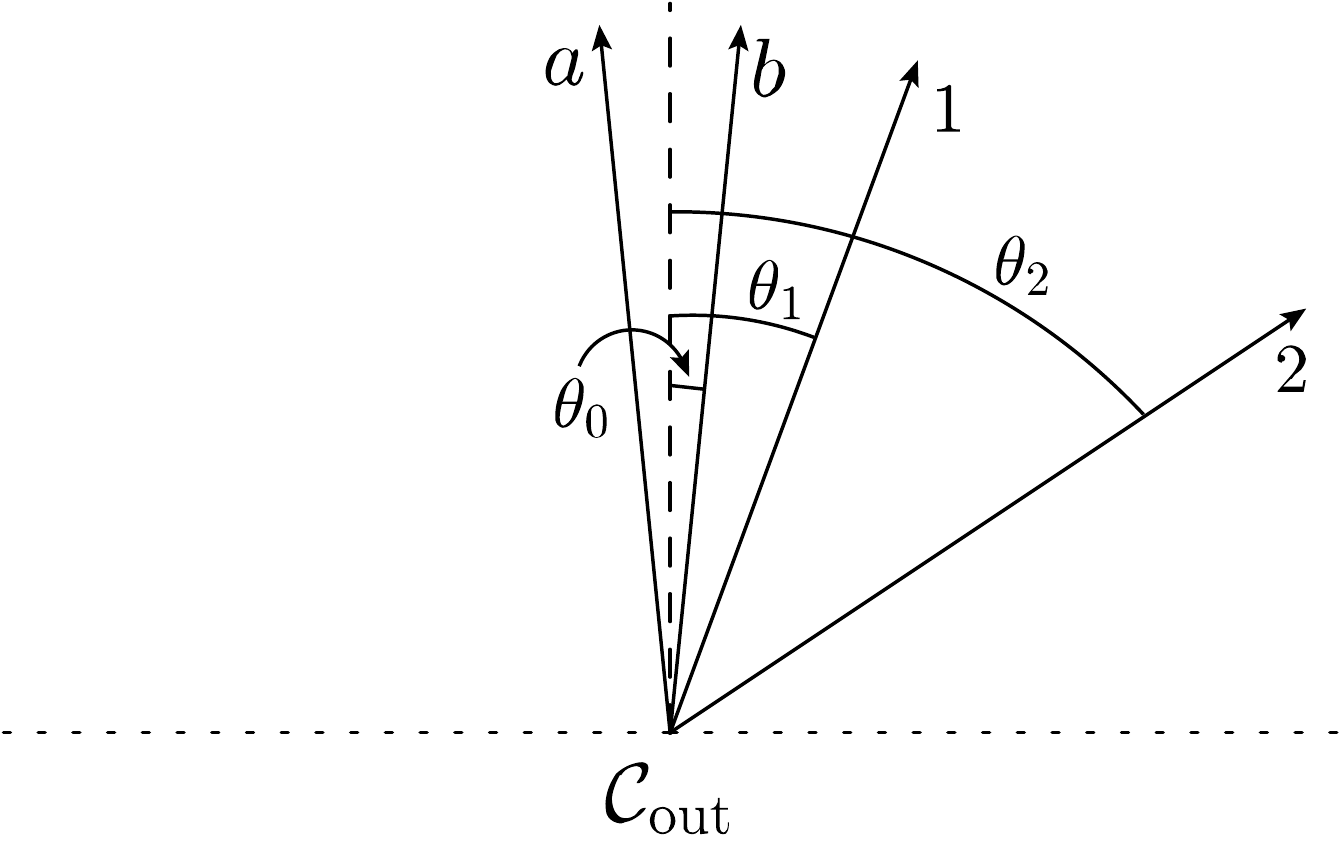}}
 \caption{\small Di-jet set-ups giving rise to double non-global logarithms. The original partons initiating the two jets are denoted with labels $a$ and $b$ respectively.  Successive gluon emissions contributing to the double-logarithmic BMS evolution are indicated by numbers 1, 2, \dots. (gluon 2 being emitted after gluon 1 etc).  Left: a pair of back-to-back jets in the COM frame; the exclusion region $\calC_{\rm out}$ has a small opening angle $\theta_0\ll 1$ around an axis perpendicular to the jet axis. Right: a pair of boosted jets making a small relative angle $\theta_0\ll 1$; the exclusion region $\calC_{\rm out}$ occupies the whole backward hemisphere. }
 \label{fig:Cout}
\end{figure}

Yet, as we shall argue in what follows, {\em double} non-global logarithms --- energy and collinear --- can exist as well: within the COM set-up that we have so far considered, they emerge when the opening angle $\theta_{\rm out}\equiv 2\theta_0$ characterising the excluded region $\calC_{\rm out}$ is small enough,  $\theta_0\ll 1$. This situation is illustrated in the left panel of Fig.~\ref{fig:Cout}. The probability for radiation inside the excluded region seems {\em a priori} small, since proportional to  $\theta_0^2$, yet this can be strongly amplified (and thus become of order one) by the multiple emission of soft and collinear gluons. Specifically, we shall argue that, when $\theta_0\ll 1$, radiative corrections enhanced by the double logarithm $\ln(E/E_0)\ln(1/\theta_0^2)$ are generated by successive gluon emissions which accumulate towards $\calC_{\rm out}$: the angles made by these gluons with the central axis of $\calC_{\rm out}$ are strongly decreasing from one emission to the next one. This is turn implies that the relative angles between 2 successive gluon emissions are strongly decreasing. This situation is illustrated in the left hand side of Fig.~\ref{fig:Cout}. As we shall further argue, these double logarithms are properly encoded in the (leading-order) BMS equation.

For our present purposes, it is however more interesting to visualise and compute this evolution in a boosted frame where the two jets (more precisely, the two initial quarks) make a small angle equal to $2\theta_0$, whereas the excluded region occupies the whole backward hemisphere at $-1<\cos\theta<0$ (see the figure in the right panel of Fig.~\ref{fig:Cout}). This frame is obtained by boosting the COM frame with a boost factor $\gamma=1/\theta_0$ along the positive $z$ axis. In this frame, the double-logarithmic contributions are generated by successive gluon emissions in the {\em anti}--collinear regime, namely such that the emission angles are small, but strongly increasing from one emission to the next one: $\theta_0\ll \theta_1\ll \theta_2\ll \dots\ll 1$.  (Strictly speaking, the emission angles are differences like $\Delta  \theta_{i,i-1}\equiv \theta_{i}-\theta_{i-1}$, but for the anti-collinear regime under consideration one has $\Delta  \theta_{i,i-1}\simeq \theta_{i}$.)
Since by assumption $\gamma=1/\theta_0\gg 1$, the original quarks have a large longitudinal momentum $p_z\simeq \gamma E$. Hence the natural variables for energy ordering in this frame are the gluons longitudinal momenta $k_{iz}$, which are strongly decreasing from one emission to the next one: $p_z\gg k_{1z}\gg k_{2z}\gg ...$


Besides being conceptually intriguing, for reasons to be shortly explained, this boosted picture is also physically interesting: it corresponds to the actual physical situation for {\em boosted jets}, say as created by the decay of a particle which is very energetic in the laboratory frame (e.g.~a $Z$ boson with energy much larger than its mass).

What is rather intriguing about the evolution in this boosted frame is the fact that the simultaneous ordering with decreasing energy ($k_z$) and increasing angle ($\theta$) may lead to violations of the physical condition that the gluon {\em formation time} $\tau_{\rm f}\sim 1/(k_z\theta^2)$ must increase along a time-like cascade (since the time-like evolution proceeds towards decreasing virtuality).  When computing this evolution from Feynman graphs within light-cone (time-ordered) perturbation theory, the proper time-ordering is introduced by the energy denominators, as we shall later check. However, the respective corrections are of higher order in $\alpha_s$ --- they start at next-to-leading order (NLO) ---, hence do not matter for the leading-order (LO) version of the BMS equation. And indeed, the latter includes contributions which violate the proper time-ordering in the anti-collinear regime. Albeit they are formally of higher orders, the radiative corrections associated with time-ordering can be numerically large, since enhanced by double (anti)collinear logarithms. For the problem at hand, they bring corrections to the kernel of the BMS equation in the form of a series in powers of $\abar\ln^2 (1/\theta_0^2)$, with $\abar\equiv\alpha_s N_c/\pi$. The first such a correction is indeed present in the NLO version of the BMS equation \cite{Caron-Huot:2015bja}, albeit this is perhaps not manifest in the original expressions in Ref.~\cite{Caron-Huot:2015bja}. (We shall isolate this contribution from the full NLO kernel in Appendix \ref{sect:nlo}.) From the experience with the respective {\em space-like} evolution --- the BFKL equation  \cite{Lipatov:1976zz,Kuraev:1977fs,Balitsky:1978ic} and its non-linear generalisations, the Balitsky-Kovchegov (BK) equation  \cite{Balitsky:1995ub,Kovchegov:1999yj} and the Balitsky-JIMWLK hierarchy \cite{Balitsky:1995ub,JalilianMarian:1997jx,JalilianMarian:1997gr,Kovner:2000pt,Iancu:2000hn,Iancu:2001ad,Ferreiro:2001qy} ---, where a similar problem arises, we expect such double collinear logs to lead to instabilities in the NLO evolution and, in any case, to jeopardise the  convergence of the perturbative expansion.

In order to restore the predictive power of perturbation theory, it becomes necessary to resum these large radiative corrections to all orders in $\abar$. Methods in that sense have been  developed in the context of the BFKL/BK evolution \cite{Beuf:2014uia,Iancu:2015vea,Iancu:2015joa} (see also \cite{Kwiecinski:1997ee,Salam:1998tj,Ciafaloni:1999yw,Ciafaloni:2003rd,Vera:2005jt} for earlier resummations proposed in the context of the linear BFKL evolution)
and in what follows we shall extend them to the case of the time-like evolution described by the BMS equation. More precisely, we shall construct a {\em collinearly-improved} version of the BMS equation, applicable in situations where the proper time-ordering is not automatically satisfied (like the boosted jets in the right panel of Fig.~\ref{fig:Cout}), by following exactly the same steps as for the respective improvement of the BK equation \cite{Iancu:2015vea}. This improvement works in the same way in both cases: it amounts to modifying the leading-order (BMS  or BK) kernel by a multiplicative factor and the respective initial condition at low energy by an additional term. Both types of corrections resum double collinear logarithms to all orders. The corrective functions turn out to be the same for the BFKL and BMS equations, only the corresponding arguments are different. From the experience with the BK equation, we expect this collinear improvement to considerably slow down the evolution \cite{Iancu:2015vea,Iancu:2015joa}  and thus dramatically modify the physical results in the regime where $E\gg E_0$ and $\theta_0\ll 1$.

Returning to the problem of the back-to-back jets, cf.~the left figure in Fig.~\ref{fig:Cout}, it is easy to see that, in that context, the LO BMS evolution {\em does} respect the proper time-ordering: the typical emission angles are (strongly) decreasing  from one emission to the next one, simultaneously with the gluon energies, so the associated formation times are increasing, as they should. Hence, in that COM set-up one does not expect higher order corrections enhanced by double collinear logs and there is no need for resummation. This will be explicitly checked at NLO in
Appendix \ref{sect:nlo}. It is therefore important to understand why two physical situations which are {\em a priori} equivalent, since related by a boost, can admit such different mathematical descriptions: the usual (unresummed) BMS equation in the COM frame and, respectively, the collinearly-improved version of this equation in the boosted frame. 

As we shall later explain in more detail, the answer to the above question is related to the difference between the energy phase-spaces available in the two frames: in the COM frame, this is simply $\ln(E/E_0)$, as already discussed, but in the boosted frame it is considerably increased by the boost, to a value $\ln(E/E_0)+ \ln(1/\theta_0^2)$. Roughly speaking, the collinearly-improved evolution over the larger energy phase-space available in the boosted frame should produce the same results as the LO BMS evolution over the smaller phase-space corresponding to the COM frame. This equivalence is however {\em not} exact: it holds only in the double logarithmic approximation (DLA), in which one resums {\em just} the perturbative corrections enhanced by double logarithms (energy-collinear, or collinear-collinear). In general however, the solutions to the two equations --- `bare' and `resummed' --- are expected to be different from each other, even after properly matching the respective phase-spaces, because of the different ways in which they treat the `BFKL diffusion' (the non-locality in angles). It would be interesting to study these differences via numerical solutions, but this goes beyond the scope of the present paper.

Another aspect that we shall address is the interplay between the collinear resummation and the conformal transformation relating the space-like and time-like evolutions \cite{Hatta:2008st,Avsar:2009yb}.  Let us first recall that, to leading order at least, the BK and BMS  evolutions are precisely related to each other by a stereographic projection mapping angles on the 2-dimensional sphere (on the BMS side) onto coordinates in the 2-dimensional transverse space (on the BK side) \cite{Hatta:2008st}. This projection is in fact a subset of a more general conformal transformation in 4-dimensions, which at least in a conformal theory like the $\mcal{N}=4$ supersymmetric Yang--Mills theory, has been conjectured \cite{Hofman:2008ar,Hatta:2008st} to relate space-like and time-like evolutions to all orders in the coupling. Within the context of $\mcal{N}=4$ SYM, this correspondence has already been checked to NLO in perturbation theory \cite{Caron-Huot:2015bja} and also in the strong-coupling limit, via the AdS/CFT correspondence  \cite{Hofman:2008ar}.

The interplay between the conformal transformation and the collinear resummation turns out to be quite subtle. For the problems at hand, the conformal symmetry is explicitly broken by the physical set-up, i.e.~by the large separation of scales between the relative angle between the two jets on one hand and the angular opening of the excluded region $\calC_{\rm out}$ on the other hand. This in turn implies an asymmetry in the evolution: the dominant evolution --- the one which generates double logarithms ---proceeds either towards increasing emission angles (in the boosted frame), or towards decreasing angles (in the COM frame).  To leading logarithmic accuracy, both evolutions are described by the leading-order BMS equation, which has conformal symmetry. Yet, they are physically different --- one can violate the proper time-ordering, while the other cannot ---, so they receive different higher-order corrections. This difference is visible in the radiative corrections enhanced by double collinear logs, which first appear at NLO and break the conformal symmetry. This is why these corrections can be large in one regime (when the emissions angles are increasing, as in the right panel of Fig.~\ref{fig:Cout}), but small in the other (when the emissions angles are decreasing, as in the left panel of Fig.~\ref{fig:Cout}).  

These conclusions are indeed supported by the NLO corrections to the BMS kernel, as computed in \cite{Caron-Huot:2015bja}, but  in order to render them manifest it is important {\em not} to use the `conformal scheme', in which the kernel is by construction conformally symmetric. In that scheme, double-logs appear at NLO in both the `collinear' and the `anti-collinear' regimes, that is, for both decreasing and increasing angles. This conformal scheme is reminiscent of the symmetric choice of the energy scale \cite{Salam:1998tj,Ciafaloni:1999yw,Ciafaloni:2003rd}
in the framework of the NLO BFKL equation  \cite{Fadin:1995xg,Fadin:1997zv,Camici:1996st,Camici:1997ij,Fadin:1998py,Ciafaloni:1998gs}. It is likely that the change from the `non-conformal' to the `conformal' scheme can be viewed too as a redefinition of the variable used for the energy evolution, albeit this is not obvious in the manipulations in \cite{Caron-Huot:2015bja}.  The standard evolution variable (the logarithm of the energy fraction carried by an emitted gluon, a.k.a. `rapidity') corresponds in fact to the `non-conformal' scheme. This is the scheme where the physical picture is most transparent and where the collinear resummations are naturally associated with the time-ordering of the successive emissions. From the above discussion, it should however be clear that collinear resummations are also needed in the conformal scheme (in both the collinear and the anti-collinear regime). 


The lack of conformal invariance for the double collinear logarithms also implies that
the collinear improvements of the BK and BMS equations are not simply related to each other via a conformal transformation, except in the special limit where all the angles are small. (This last condition refers both to the emission angles and to the angles involved in the stereographic projection; see Sect.~\ref{sec:BK} for details.) This being said, the conformal transformation is powerful enough to predict the {\em need} for time ordering and hence for collinear improvement. This is so since the transformation law for {\em energy} scales which is inherent this correspondence also involves the scale of the {\em collinear} logarithms --- the dipole transverse sizes in the case of the BK equation and the emission angles for the BMS equation. When this collinear logarithm is relatively large, the emerging evolution variable is not just the energy anymore, but the formation time in the time-like case (respectively, the lifetime of the fluctuations in the space-like evolution). 
This will be explained in Sect.~\ref{sec:BK}, where we will see that {\em both} time-like evolutions that we have discussed so far --- that in the COM frame and that in the boosted frame ---,  are actually mapped onto the {\em same} space-like evolution --- that where the dipole sizes are strongly increasing from one step to the next one. This particular dipole evolution requires explicit time-ordering\footnote{In the space-like case, the gluon lifetimes must decrease in the course of the evolution \cite{Beuf:2014uia,Iancu:2015vea}.} \cite{Beuf:2014uia,Iancu:2015vea,Iancu:2015joa} and this is indeed predicted by the conformal mapping, as we shall see.

This paper is organised as follows. In Sect.~\ref{sec:BMS} we introduce the (leading-order) BMS equation and demonstrate the emergence of (anti-)collinear logarithms in the 2 regimes illustrated in  Fig.~\ref{fig:Cout}. We also explain the difference between the associated energy phase-spaces. In Sect.~\ref{sec:LCPT} we focus on the anti-collinear evolution for boosted jets and present two arguments for the time-ordering of the soft gluons. The first argument uses a Lorentz transformations from the COM frame (energy ordering in the COM frame together with the boost implies time-ordering in the boosted frame); the second one is based on an explicit diagrammatic calculation of up to 2 gluon emissions in light-cone perturbation theory. Incidentally, this calculation also provides a pedagogical derivation of the kernel of the LO BMS equation (a.k.a. the {\em antenna pattern}). In Sect.~\ref{sec:coll} we shall present our main result: the collinearly-improved version of the BMS equation (see \eqn{collBMS}). Finally, in Sect.~\ref{sec:BK} we discuss the conformal mapping relating time-like and space-like evolutions, i.e.~the BMS and BK equations, in connection with time-ordering. In Appendix \ref{sect:nlo} we shall revisit the result for the NLO correction to the BMS kernel \cite{Caron-Huot:2015bja}, with the purpose of extracting the piece enhanced by a double collinear logarithm.

\section{Collinear logarithms in the BMS evolution}
\label{sec:BMS}

In this section, we shall introduce the leading-order BMS equation and demonstrate that under special circumstances --- namely, for the configurations  illustrated in  Fig.~\ref{fig:Cout} --- this equation also resums (anti)-collinear logarithms, on top of the energy logarithms that it was originally meant for. We shall moreover argue that, when these (anti)-collinear logs are sufficiently large, the BMS equation is not boost-invariant anymore: it can still be used as it stands in the di-jet COM frame, but not also in a boosted frame where the energy phase-space available to the evolution is much larger.

\subsection{The BMS equation}

In order to write down the BMS equation, we shall consider the final state of $e^+e^-$ annihilation, as viewed in an arbitrary Lorentz frame. (The complications with the choice of a frame formally go beyond the leading-logarithmic approximation at high energy and will be discussed later, starting with the next subsection.) We shall use $P_{ab}(E)\equiv P_{ab}(E,E_0,\calC_{\rm out})$ to denote the probability to deposit a total energy lower than $E_0$ inside the `away-from-jet region' $\calC_{\rm out}$ via radiation from the di-jets initiated by two primary partons, the quark $a$ and the antiquark $b$, whose total energy $p_a^0+p_b^0$ is equal to $2E$. 

The two energies aforementioned, $E_0$ and $E$, refer both to the COM frame of the original $q\bar q$ pair (the frame where $p_a^0=p_b^0=E$ while $\vec p_a+\vec p_b=0$), but the probability $P_{ab}(E,E_0,\calC_{\rm out})$ can in principle be computed in any frame (of course, the geometry of the excluded region $\calC_{\rm out}$ can change as well when changing the frame). In general, the function $P_{ab}$ also depends upon the directions of motions of the primary partons in the Lorentz frame at hand, that is, upon the two null 4-vectors $v_a$ and $v_b$, with $v_a^\mu\equiv p^\mu_a/p^0_a=(1,\vec v_a)$, etc. We shall assume that $E\gg E_0$, so that the radiative corrections enhanced by powers of $\abar\ln(E/E_0)$ must be resummed to all orders.


This resummation is the scope of the BMS equation. This equation
has been originally formulated \cite{Banfi:2002hw} in the limit of a large number of colors\footnote{See also Refs.~\cite{Weigert:2003mm,Hatta:2013iba} for generalisations to an arbitrary value for $N_c$, that we shall however not consider in this paper.} $N_c$, in which the emission of a soft gluon by the original quark-antiquark pair can be viewed as the splitting of the color dipole (or ``color antenna'') $(ab)$  into two new dipoles $(ac)$ and $(cb)$; the index $c$ refers to the direction of motion (the null-vector $v_c$) of the emitted gluon. By iterating this argument, the whole high-energy evolution can be described as a change in the distribution of dipoles. Then the leading-order BMS equation reads \cite{Banfi:2002hw}
\begin{align}
\label{BMSLO}
E\del_E P_{ab}(E) =& -\abar\int_{\calC_{\rm out}} 
\frac{\rmd \Omega_c}{4\pi}\,w_{abc}\,P_{ab}(E) +\nn
&+\abar\int_{\calC_{\rm in}}  \frac{\rmd \Omega_c}{4\pi}\,w_{abc}\,\Big[P_{ac}(E)P_{bc}(E)-P_{ab}(E)\Big]\,,
 \end{align}
where the kernel $w_{abc}$ describes the angular distribution of the radiation (the `antenna pattern'):
\beq\label{wabc}
w_{abc}\,\equiv\,\frac{v_a\cdot v_b}{(v_a\cdot v_c)(v_c\cdot v_b)}\,=\,
\frac{1- \cos\theta_{ab}}{(1- \cos\theta_{ac})(1- \cos\theta_{cb})}\,.
\eeq
Here, $\theta_{ab}$ is the relative angle between the momenta of the quarks $a$ and $b$,
$\vec v_a\cdot \vec v_b= \cos\theta_{ab}$, etc. The angular integrals in the two terms in the r.h.s. of  \eqn{BMSLO} run over the directions of the unit vector $\vec v_c$, but they have different --- actually, complementary --- supports: that in the first term (the `source', or `Sudakov', term) runs over the excluded region $\calC_{\rm out}$, whereas that in the second (`evolution') term runs over the complementary region of space, $\calC_{\rm in}=S^2\setminus \calC_{\rm out}$, which in particular includes the 2 jets (see e.g.~Fig.~\ref{fig:general}).  As anticipated in the Introduction, the most important region for the evolution is the intermediate region between the jets and the excluded region.

\eqn{BMSLO} must be solved with the initial condition that, when $E=E_0$, $P_{ab}=1$ for any dipole $(ab)$. By itself, the `evolution' term in the r.h.s. of \eqn{BMSLO} vanishes for this particular initial condition, hence the evolution is initiated by the `source' term, which describes a direct emission from the dipole $(ab)$ to the excluded region. More generally, during the later stages of the evolution, this `source' term will describe the reduction in the probability $P_{ab}$ due to emissions from any of the dipoles produced by the evolution towards $\calC_{\rm out}$.  For this reason, it is also known as the `Sudakov term'. 

The `evolution' term in the r.h.s. of \eqn{BMSLO} is itself built with two pieces. The first piece, which is positive and quadratic in the probability, describes a real gluon emission. At large $N_c$, this emission effectively replaces the original dipole  $(ab)$ by the two dipoles $(ac)$ and $(cb)$, which subsequently develop their own evolutions (leading to the probabilities $P_{ac}$ and $P_{bc}$, respectively). The second piece, which is negative and linear in $P_{ab}$, comes from Feynman graphs describing a virtual emission and represents the reduction in the survival probability for the original dipole. Note that the collinear singularities of the kernel \eqref{wabc} at $\theta_{ac}\to 0$ or $\theta_{cb}\to 0$ cancel between real and virtual corrections: when e.g.~$\theta_{ac}\to 0$, one has $P_{ac}\to 1$, since there is no emission from a colorless antenna with zero opening angle. 


As manifest from \eqn{BMSLO}, the high-energy evolution  is {\em logarithmic} --- the probability $P_{ab}$ depends upon the energy variables $E$ and $E_0$ via the `rapidity' variable $Y\equiv \ln(E/E_0)$ ---, hence it can be equivalently formulated as an evolution with increasing $E$ at fixed $E_0$, or with decreasing $E_0$ at fixed $E$. In what follows, we shall adopt the second point of view, that is, we shall choose the `running' value of the rapidity as $Y=\ln(E/k)$ where $k=|\vec k|$ is the energy of the last (softest) emitted gluon and obeys $E\gg k \gg E_0$. Correspondingly, we shall replace $E\del_E \to \del_Y$ in the l.h.s. of \eqn{BMSLO}.

So far, we made no special assumption about the geometry of the exclusion region. From now on, we shall focus on the situation where the associated opening angle $\theta_{\rm out}$ as measured in the COM frame is small: $\theta_{\rm out}=2\theta_0$, with $\theta_0\ll 1$ (cf.~the left  panel of Fig.~\ref{fig:Cout}). In this case, we shall see that the evolution generated by \eqn{BMSLO} also generates `collinear logarithms', i.e.~radiative corrections proportional to the double logarithm $\ln(E/E_0)\ln(1/\theta_0^2)$. The physical picture underlying these corrections and also their magnitude turns out to be strongly frame-dependent, so in what follows we shall separately discuss the evolution in the boosted frame and that in the COM frame.

\subsection{Collinear logarithms in the boosted frame}
\label{sec:collBOOST}

The emergence of the collinear logarithms is conceptually more transparent when the problem is analysed in the boosted frame, so we shall start by discussing this case. Starting in the COM frame, we perform a boost along the positive $z$ axis with boost factor $\gamma = 1/\sin\theta_0\simeq 1/\theta_0$ (see Sect.~\ref{sec:boost} for more details on this Lorentz transformation). In the boosted frame, the quark and the antiquark propagate nearly along the $z$ axis, with a small relative angle $\theta_{ab}=2\theta_0\ll 1$, whereas the excluded region occupies the whole backward hemisphere at $z<0$ (see the right panel of Fig.~\ref{fig:Cout}). 

As anticipated in the Introduction, the collinear logs are generated by soft emissions whose emission angles are strongly increasing, yet remain small: $\theta_0\ll \theta_1\ll \theta_2\ll \dots\ll 1$. At first sight, this might look intriguing, as it is well known that large-angle emissions by a colorless antenna are strongly suppressed (see also the discussion of \eqn{wabcboost} below). However, we shall see that in the problem at hand the large-angle suppression in the emission probability is exactly compensated by the rapid rise in the observable that we measure: the deviation $R_{ab}\equiv 1- P_{ab}$  of the probability $P_{ab}$ from unity  (that we shall refer to as the `radiance'). This mechanism is in fact similar to that responsible for the emergence of anti-collinear logs in the context of the BK evolution (see the discussion in \cite{Beuf:2014uia,Iancu:2015vea}), but to our knowledge it was not previously noticed in the context of the time-like, BMS, evolution.

To see this, it is convenient to solve \eqn{BMSLO} via iterations. As already mentioned, the `evolution' term in the r.h.s. vanishes when evaluated with the initial condition  $P_{ab}(Y=0)=1$, hence the first iteration involves solely the `source' term. For the kinematics in the boosted frame, the latter can be estimated as (with $P_{ab}\to 1$)
\begin{align}\label{sourceboost}
 -\abar\int_{\calC_{\rm out}} 
\frac{\rmd \Omega_c}{4\pi}\,w_{abc}\, \simeq -\abar \frac{\theta_{ab}^2}{2} \int_0^{2\pi}\frac{\rmd \phi_c}{4\pi}
\int_{\pi/2}^\pi \frac{\sin\theta_c\,\rmd\theta_c}{(1-\cos\theta_c)^2} \simeq -\abar\frac{\theta_{ab}^2}{8}\,,
\end{align}
where we have used $\theta_{ab}\ll 1$ together with $\theta_{c}\in(\pi/2,\,\pi)$ to approximate $1- \cos\theta_{ab}\simeq \theta_{ab}^2/2$ and $\theta_{ac}\simeq \theta_{bc} \simeq \theta_{c}$.
This implies the following approximation for $R_{ab}=1- P_{ab}$ to linear order in $\abar Y$:
\beq\label{Rablin}
R_{ab}(Y)\,\simeq\,\frac{1}{8}\,\abar\theta_{ab}^2Y\,.\eeq
This is a legitimate approximation so long as $R_{ab}\ll 1$.
Similar estimates can be used for the other radiances which enter the `evolution' term, that is, $R_{ac}$ and $R_{bc}$: indeed, the polar angle $\theta_c$ made by the first evolution gluon is small as well (albeit large compared to $\theta_0$). We thus deduce
\beq\label{Pablin}
P_{ac}(Y)P_{bc}(Y)-P_{ab}(Y)\,\simeq\,-R_{ac} - R_{bc} + R_{ab}\,\simeq\,-2R_{ac}
\,\simeq\,-\frac{1}{4}\,\abar\theta_{c}^2Y\,\eeq
for the combination of `real' and `virtual' terms which enter  \eqn{BMSLO}.
We have successively used the fact that all the $R$'s are small enough to neglect the quadratic term $R_{ac}R_{bc}$ and the fact that $R_{ac} \simeq R_{bc} \gg R_{ab}$ (since $\theta_{ac}\simeq \theta_{bc} \simeq \theta_{c}\gg \theta_{ab}$) to neglect the `virtual' contribution. Notice that it was essential for the previous argument that $R_{ab}$ is proportional to $\theta_{ab}^2$ at small angles. As we shall shortly see, this property remains true after resumming the energy logarithms $\abar Y$ to all orders. This is the time-like analog of the `color-transparency' property of the solution to the BK equation (the fact that the dipole scattering amplitude vanishes like $r^2$ in the limit $r\to 0$). The net result in \eqn{Pablin} comes from the `real' terms alone and it rapidly grows with the emission angle, like $\theta_{c}^2$.

The `evolution' term in \eqn{BMSLO} also involves the emission probability \eqref{wabc}, which for the kinematics at hand simplifies to
\beq\label{wabcboost}
w_{abc}\,\simeq\,\frac{2\theta_{ab}^2}{\theta_{ac}^2 \theta_{bc}^2}
\,\simeq \,\frac{2\theta_{ab}^2}{\theta_{c}^4}
\,.
\eeq
This rapid decrease $\propto 1/\theta_{c}^4$ of the emission rate with increasing $\theta_c$ reflects the aforementioned fact that the radiation by a colorless antenna is suppressed at large angles. However, within \eqn{BMSLO} this decrease is partially compensated by the increase of the radiance of the daughter gluons, $R_{ac} \simeq R_{bc}\propto \theta_{c}^2$. The resulting integral over $\theta_c^2$ is {\em logarithmic}, as anticipated.

Specifically, by inserting  Eqs.~\eqref{Pablin}--\eqref{wabcboost} into \eqn{BMSLO}, one deduces the following second-order (in $\abar Y$) approximation for  $R_{ab}(Y)$, valid in the {\em double-logarithmic approximation} (i.e.~by neglecting second-order terms which are not enhanced by a collinear log), or `DLA' :
\beq\label{Rabquad}
R_{ab}(Y)\,\simeq\,\frac{\theta_{ab}^2}{8}\,\abar Y\bigg[ 1+\frac{1}{2}\abar Y\ln\frac{1}{\theta_{ab}^2}\bigg]\,,
\eeq
where the collinear log $\ln({1}/{\theta_{ab}^2})$ has been generated by integrating over $\theta_c^2$ between $\theta_{ab}^2$ and 1 (the precise upper limit is irrelevant to logarithmic accuracy). 

\eqn{Rabquad} illustrates the effects of the high-energy evolution (here, to DLA): the original contribution \eqref{Rablin} of one Sudakov emission, which is small since proportional to $\theta_{ab}^2$, receives radiative corrections enhanced by powers of $\abar Y \ln(1/\theta_{ab}^2)$ and thus can become quite large --- meaning that the probability $P_{ab}(Y)$ can become significantly smaller than unity --- for sufficiently high energy and/or small angle $\theta_{ab}$. This enhancement will be apparent in a moment.

The above discussion also shows that higher-order iterations of the Sudakov term are unimportant in the kinematics of interest, since they are {\em power-suppressed} --- i.e.~multiplied by higher powers of $\theta_{ab}^2$ --- compared to its first iteration. In fact, it is easy to resum multiple Sudakov emissions by the primary dipole to all orders: this amounts to solving a simplified version of \eqn{BMSLO} in which one keeps the Sudakov term alone, evaluated as in  \eqn{sourceboost}. One thus finds  $P_{ab}^{\rm\scriptscriptstyle  Sudakov}=\exp{(-\theta_{ab}^2 \abar Y/8)}$. In the exponent, $\theta_{ab}^2$ appears to be multiplied by $\abar Y$, but this product is still small in the regime of interest for us here. So, in what follows we shall keep only `leading-twist' terms which are linear in $\theta_{ab}^2$, but which as a result of the evolution can involve arbitrary powers of $\abar Y \ln(1/\theta_{ab}^2)$. 

It should be quite clear that the above arguments extend to the subsequent emissions of soft gluons, so long as the respective angles are strongly increasing, yet small in absolute value. One can easily write down an approximate version of the (LO) BMS equation, valid at DLA --- that is, an equation which resums solely the terms enhanced by the double logarithm $Y\ln({1}/{\theta_{0}^2})$. To that aim, we first note first that, at DLA, the probability $P_{ab}(Y)$ depends upon the unit vectors $\vec v_a$ and $\vec v_b$ only via their relative angle $\theta_{ab}$. It is convenient to isolate the dominant dependence upon $\theta_{ab}$ by writing $R(\theta_{ab}, Y)\equiv\theta_{ab}^2\calA(\theta_{ab}, Y)$. Then, the DLA version of the BMS equation reads
\begin{align}\label{DLAboost}
\frac{\del\calA(\theta_{ab}, Y)}{\del Y} = \frac{\abar}{8} +\abar
\int_{\theta_{ab}^2}^1  \frac{\rmd\theta_c^2}{\theta_c^2}\,\calA(\theta_c, Y)\,,
 \end{align}
where to the accuracy of interest one could as well replace $\theta_{ab}$ by $\theta_0$. 
(We recall that $\theta_{ab}=2\theta_0$, but the relative factor of 2 is irrelevant inside the logarithms.) 
This equation can be easily solved via iterations, with the initial condition
$\calA(\theta, 0)=0$, to yield 
\begin{align}\label{Aboost}
\calA(\rho, Y)\,=\,\frac{\abar Y}{8}\sum_{n=0}^\infty\, \frac{(\abar Y \rho)^n}{n! (n+1)!}\,=\,\frac{1}{8}
\sqrt{\frac{\abar Y}{\rho}}\,\rmI_1(2\sqrt{\abar Y\rho})
 \end{align}
where we have introduced the compact notation $\rho\equiv \ln(1/\theta_{ab}^2)$ for the collinear logarithm and $\rmI_1(x)$ is the modified Bessel function of first rank. This function grows rapidly\footnote{We recall the asymptotic behavior of the modified Bessel function: $\rmI_1(x)\simeq \rme^x/\sqrt{2\pi x}$.} with the product $Y\rho$, but the present approximation is of course valid only so long as  $R(\theta_{ab}, Y)=\theta_{ab}^2\calA(\theta_{ab}, Y)\ll 1$. Eqs.~\eqref{DLAboost} and \eqref{Aboost} describe the dominant evolution leading to the increase in the radiance $R$ in the regime where the latter is still small. Note that, in this DLA regime, the emission towards the excluded region $\calC_{\rm out}$ is most likely sourced by the last emitted `evolution' gluon, since this makes the largest polar angle $\theta$ and since $R(\theta, Y)\propto \theta^2$.

In general, i.e.~if one needs to go beyond the double-logarithmic approximation and to also cover the non-linear regime where $R_{ab}(Y)\sim \order{1}$, one must use the full BMS equation \eqref{BMSLO}. This being said, none of these equations --- the original BMS equation \eqref{BMSLO} or its DLA version \eqref{DLAboost} --- is fully right for the case of boosted jets: indeed, as we shall argue at length in what follows, these equations do not properly cover the phase-space for soft emissions at large angles. As a first step in that sense, let us clarify here the energy phase-space available to the evolution in the boosted frame.

One may be tempted to identify the (maximal value of the) rapidity $Y$ in the solution  \eqref{Aboost} with $Y=\ln(E/E_0)$, but in the boosted frame this would {\em not} be right. Note first that the energy of one of the primary partons in this frame is $\gamma E$, hence the running value of the rapidity variable is $Y=\ln(\gamma E/k)
\simeq \ln(\gamma E/k_z)$. We have used the fact that $k\simeq k_z$ for the gluons which matter in the double-logarithmic regime: their emission angles obey $\theta_0\ll \theta\ll 1$, which together with $\theta\simeq k_\perp/k_z$  implies indeed $\theta_0k_z \ll k_\perp\ll k_z$. To deduce the upper limit on $Y$, one must understand the lower limit on $k_z$. To that aim, it is easier to argue in the COM frame and then make a boost. 

The softest emissions which matter to DLA in the COM frame have an energy $k\sim E_0$ and make a very small angle w.r.t.~the negative $z$ axis (see the left panel of Fig.~\ref{fig:Cout}); for them $k_z$ is negative and the energy is mostly longitudinal: $k\simeq |k_z| \gg k_\perp$. When boosting in the positive $z$ direction, both $|k_z|$ and $k$ will be reduced by the boost factor $\gamma$ (see Sect. \ref{sec:boost}  for more details on the boost). Accordingly, in the boosted frame, the softest emissions have an energy $k\simeq k_z\sim E_0/\gamma$. This discussion implies that the rapidity range available to evolution in the boosted frame is $Y=\ln(\gamma^2 E/E_0)=Y_0+\rho$, where $Y_0\equiv \ln(E/E_0)$ is the corresponding range in the COM frame and $\rho= \ln\gamma^2 =\ln(1/\theta_{0}^2)$ is the collinear logarithm. When $\theta_{0}\ll 1$, there is therefore an excess in the phase-space for the energy evolution in the boosted frame as compared to the COM frame. As we shall later argue, this excess corresponds to spurious emissions, which are included in the LLA but do not respect the proper time-ordering condition. Such emissions can be removed by hand, by enforcing time ordering.  Incidentally, the above discussion also shows that the upper limit on the energy that can be emitted within the backward hemisphere by the boosted jets is not $E_0$, but the much smaller value $\theta_0E_0$.


\subsection{Collinear logarithms in the COM frame}
\label{sec:collCOM}

It is straightforward to `boost back' the gluon kinematics from the boosted frame to the COM frame and thus establish that the emissions responsible for double logarithms are those which accumulates towards the negative $z$ axis (the central axis of the excluded region $\calC_{\rm out}$), as illustrated in the left panel of Fig.~\ref{fig:Cout}. The corresponding Lorentz transformations will be presented in Sect. \ref{sec:boost}. Here however we would like to develop the argument for double logs directly in the COM frame. To that aim, we shall consider the two successive emissions exhibited in the left panel of Fig.~\ref{fig:Cout}, whose propagation angles $\theta_i$ as measured w.r.t.~the {\em negative} $z$ axis obey $\theta_0\ll\theta_2\ll\theta_1\ll 1$.

Consider first the emission of gluon 1 from the original color antenna $(ab)$. In this frame, $\theta_{ab}=\pi$ and $\theta_{a1}\simeq \theta_{1b}\simeq \pi/2$, hence $w_{abc}\simeq 2$ : the emission probability shows no sign of collinear enhancement, as expected for a large angle emission. However, in  \eqn{BMSLO} the emission kernel $w_{abc}$ is multiplied by
\beq\label{Pablin2}
P_{a1} P_{1b} -P_{ab}\,\simeq\,-R_{a1} - R_{1b} + R_{ab}\,\eeq
where we have again ignored the quadratic term $R_{a1} R_{1b}$ since we are in the regime where all the probabilities are close to 1. To leading order in $\abar Y$, the radiances are determined by the Sudakov term in \eqn{BMSLO} as, e.g.
\beq\label{Ra1}
R_{a1}\,\simeq\,\abar Y\int_{\calC_{\rm out}} 
\frac{\rmd \Omega_c}{4\pi}\,w_{a1c} \,\simeq\,\abar Y\int_0^{2\pi}\frac{\rmd \phi_c}{4\pi}
\int_0^{\theta_0} \frac{\theta_c\rmd\theta_c}{\theta_{1c}^2/2}\,=\,\frac{\abar Y}{2}\, \frac{\theta_0^2}{\theta_{1}^2}\,,
\eeq
where we have used $\theta_{a1}\simeq\theta_{ac}\simeq \pi/2$ and $\theta_{1c}\simeq \theta_{1}-\theta_c\simeq  \theta_{1}\ll 1$. This result provides the appropriate factor $1/\theta_{1}^2$ to render the ensuing integral over $\theta_{1}^2$ logarithmic. Clearly, this factor expresses the collinear enhancement for the small-angle emission of the gluon $c$ from the evolution gluon 1. There is a similar enhancement for $R_{1b}$, but not also for $R_{ab}$; hence, $R_{a1}\simeq R_{1b}\gg R_{ab}$. Using the above estimates within  \eqn{BMSLO}, one finds the second-order  (in $\abar Y$) estimate for $R_{ab}$ as follows:
\beq\label{Rabquad2}
R_{ab}(Y)\,\simeq\,\frac{\theta_{0}^2}{2}\,\abar Y\bigg[ 1+\frac{\abar Y}{2}\ln\frac{1}{\theta_{0}^2}\bigg]\,.
\eeq
This agrees indeed with \eqn{Rabquad}, in view of the fact that $\theta_{ab}=2\theta_0$, with $\theta_{ab}$
the di-jet angle in the boosted frame.

It is furthermore instructive to consider the emission of the second soft gluon --- the one denoted as `2' in the left panel of Fig.~\ref{fig:Cout} --- since the corresponding geometry is quite different compared to the first emission. Gluon 2 can be emitted from either the dipole $(a1)$, or from the dipole $(1b)$, and we shall consider the first case for definiteness. The relevant emission kernel is $w_{a12}\simeq 2/\theta_{12}^2\simeq 2/\theta_{1}^2$, where we have used the fact that $\theta_2\ll\theta_1\ll 1$. This factor $1/\theta_{1}^2$ provides the logarithmic enhancement for the emission of the parent gluon 1. That is, the daughter gluon 2 plays here the same role as the emission inside $\calC_{\rm out}$ discussed in relation with \eqn{Ra1}. As in that case, the enhancement is associated with a small-angle emission --- here, of the gluon 2 --- by the gluon 1.  

Similarly, a factor $1/\theta_{2}^2$ will be generated by a small-angle emission from the gluon 2 --- either the emission of a third `evolution' gluon at an even smaller angle $\theta_3$, with $\theta_0\ll\theta_3\ll\theta_2$, or an emission inside $\calC_{\rm out}$.  Consider the second case: an emission from gluon 2 towards the excluded region $\calC_{\rm out}$. By studying this case, one can compute the second order correction to $R_{a1}(Y)$ and thus exhibit the first collinear logarithm within $R_{a1}$. To that aim, one also needs (cf.~\eqn{Ra1})
\beq\label{Ra2}
R_{a2}\,\simeq\,R_{21}\,\simeq\,\,\frac{\abar Y}{2}\, \frac{\theta_0^2}{\theta_{2}^2}\,\gg\,
R_{a1}\,\simeq\,\frac{\abar Y}{2}\, \frac{\theta_0^2}{\theta_{1}^2}\,.
\eeq
By combining the above results, one finds 
\beq\label{Ra1quad}
R_{a1}(Y)\,\simeq\,\frac{\abar Y}{2}\frac{\theta_{0}^2}{\theta_{1}^2}\bigg[ 1+\frac{\abar Y}{2}\ln\frac{\theta_{1}^2}{\theta_{0}^2}\bigg]\,,
\eeq
where the collinear log has been generated by integrating over $\theta_2^2$ between $\theta_{0}^2$ and $\theta_{1}^2$.

It should be clear by now what is the general pattern of the evolution: when emitting softer and softer gluons which make smaller and smaller angles $\theta_i$ w.r.t.~the negative $z$ axis (with $\theta_i\gg\theta_0$ though), the associated radiances are larger and larger, since proportional to ${\theta_{0}^2}/{\theta_{i}^2}$, and the associated emission kernels, which scale like $1/{\theta_{i-1}^2}$, provide the collinear enhancement for their parent gluon. It is straightforward to write an approximate (`DLA') version of the BMS equation which resums the double logarithms (energy times collinear) alone.  To that aim,  we shall rewrite $R(\theta, Y)\equiv (\theta_{0}^2/\theta^2)\calA(\theta/\theta_{0}, Y)$ where it is understood that $\theta=1$ for the original antenna $(ab)$ and $\theta=\theta_i$ for an antenna which includes the evolution gluon $i$, with $i\ge 1$ (in particular, $\theta_0\ll \theta_i\ll 1$). The equation obeyed by the new function $\calA(\theta/\theta_{0}, Y)$ to the accuracy of interest reads
\begin{align}\label{DLACOM}
\frac{\del\calA(\theta/\theta_{0},  Y)}{\del Y} = \frac{\abar}{2} +\abar
\int_{\theta_{0}^2}^{\theta^2}  \frac{\rmd\theta_c^2}{\theta_c^2}\,\calA(\theta_{c}/\theta_0, Y)\,,
 \end{align}
to be solved with the initial condition  $\calA(\theta/\theta_{0}, 0)=0$. Clearly, the solution is the same as shown in \eqn{Aboost}, except for the change $1/8\to 1/2$ in the overall normalization and for the meaning of the logarithmic variable $\rho$, now defined as $\rho\equiv \ln(\theta^2/\theta_{0}^2)$. Given the general solution $\calA(\theta/\theta_{0}, Y)$, the radiance $R_{ab}(Y)$ of the original dipole $(ab)$ is obtained by letting $\theta\to 1$, that is, $R_{ab}(Y)=\theta_{0}^2 \calA(1/\theta_{0}, Y)$. 

From the above discussion, it should be clear that the present calculation of $R_{ab}(Y)$ in the COM frame gives the same result as its previous calculation in the boosted frame, based on \eqn{DLAboost}. (To check this, one should also recall the relation $\theta_{ab}=2\theta_0$ for the di-jet angle in the boosted frame.) In the present context too, the emission towards $\calC_{\rm out}$ is predominantly sourced by the last `evolution' gluon --- the one which makes the smallest angle $\theta$ w.r.t.~the negative $z$ axis and thus gives the largest value for the radiance $R(\theta)\propto 1/\theta^2$.

This formal equivalence between the calculations of $R_{ab}(Y)$ in the COM frame and respectively the boosted frame seems to comfort the boost-invariance of the leading-order BMS equation. However the situation is more subtle. As explained at the end of the previous section, the phase-space for the energy evolution, that is, the maximal value $Y_{\rm max}$ of the rapidity variable $Y$, is different in the two frames: this is equal to  $Y_0= \ln(E/E_0)$ in the COM frame, but it is larger, namely $Y_{\rm max}=Y_0+\rho$ with  $\rho=\ln(1/\theta_{0}^2)$,  in the boosted frame. The boost invariance is in fact broken by our choice of the `energy' variable: in both frames, it is natural to measure the energy of the emitted gluons by their longitudinal momentum, but in the boosted frame this momentum $k_z$ is oriented along the {\em positive} $z$ axis, whereas in the COM frame it is rather oriented along the {\em negative} $z$ axis. In other terms, the natural energy variable is the {\em modulus} $|k_z|$, or more precisely its ratio w.r.t.~the energy of one of the primary partons; this ratio is not boost invariant.

To render this discussion more transparent, it is useful to introduce the light-cone variables $k^+\equiv(k_0+k_z)/\sqrt{2}$ and $k^-\equiv(k_0-k_z)/\sqrt{2}$. Then, the `rapidity' variables are $Y^+\equiv \ln(p^+_a/k^+)$, with $p^+_a= \sqrt{2}\gamma E$, in the boosted frame and, respectively,  $Y^-\equiv \ln(p^-_a/k^-)$, with $p^-_a= E/ \sqrt{2}$, in the COM frame. These variables $Y^+$ and $Y^-$ are individually boost-invariant, but their interchange $Y^+\leftrightarrow Y^-$ is {\em not}.  Indeed, $Y^+-Y^-=\rho$, as it can be easily checked (see below). The use of different variables in different frames is not just a matter of convenience, rather it is imposed by the corresponding kinematics.

It might look surprising that the LO BMS equation can lead to violations of the Lorentz symmetry. But one should recall that this equation has been derived for the case where the energy logarithm $Y_0= \ln(E/E_0)$ is the only large logarithm in the problem. That is, in the original derivation one has implicitly assumed that $Y_0\gg \rho$ and hence $Y^+\simeq Y^-$ to the accuracy of interest. Here however we are interested in the very asymmetric situation where $\theta_0\ll 1$, so the collinear logarithm  $\rho=\ln(1/\theta_{0}^2)$ can be large and comparable to the rapidity $Y_0$. In this situation, the BMS dynamics is genuinely different in the two frames at hand: the evolution in the COM frame (which consists in simultaneously decreasing $k^-$ and the emission angle $\theta$) automatically preserves the proper time-ordering, whereas that in the boosted frame (with decreasing $k^+$ but increasing $\theta$) may lead to violations of the time-ordering condition, as we shall later explain.

Since the difference between $Y^+$ and $Y^-$ is important for our present purposes, let us provide here another argument for its value, which corroborates the one presented in Sect.~\ref{sec:collBOOST}. We start in the COM frame,  cf.~the left panel of Fig.~\ref{fig:Cout}: the typical evolution gluons make a small angle $\theta$ w.r.t.~the negative $z$ axis, hence their transverse and longitudinal momenta are related by $k_\perp \simeq \theta |k_z|$. Since $|k_z|\ge E_0$ and $\theta\ge \theta_0$, we conclude that the smallest allowed value for $k_\perp$ is $k_{\perp{\rm min}}=\theta_0 E_0$. Since the transverse momentum is boost-invariant, this kinematical limit also applies in the boosted frame. But in that frame, the typical evolution gluons make a small angle $\theta\ll 1$ w.r.t.~the positive $z$ axis, so for them $k_z\ge k_\perp$. We thus deduce a lower limit on $k_z$ in this boosted frame, namely $k_z\ge \theta_0 E_0$, which agrees with the one found at the end of Sect.~\ref{sec:collBOOST} via a different argument. It is now easy to check that the maximal values for $Y^+$ and $Y^-$ are indeed equal to $Y_{\rm max}^+=Y_0+\rho$ and $Y_{\rm max}^-=Y_0$, respectively.

\section{Time ordering from light-cone perturbation theory}
\label{sec:LCPT}

In this section, we shall clarify the origin of the condition of time ordering in the time-like evolution. We shall first present a simple kinematical argument in that sense: we will show that, via the Lorentz transformation relating the two frames illustrated in Fig.~\ref{fig:Cout}, the condition of energy ordering in the COM frame gets mapped onto the condition of time ordering in the boosted frame. Then we shall discuss the origin of the latter within perturbative QCD. To that aim we shall employ light-cone perturbation theory (LCPT), which renders the temporal picture of the high-energy evolution manifest. To see the need for time-ordering, we will eventually need to consider two successive emissions. But before doing that, we will consider the case of a single emission, as a warm up, and thus derive the `antenna pattern' \eqref{wabc} from LCPT.

\subsection{Time ordering from Lorentz transformations}
\label{sec:boost}

In this section, we shall study some consequences of the Lorentz transformation relating the kinematics of the high-energy evolution in the COM frame and the boosted frame, respectively.  We start in the COM frame, where the LO BMS evolution with decreasing energies automatically respects the proper time ordering of the successive emissions --- the formation time increases from one emission to the next one --- because the emission angles are not (strongly) increasing. We then
perform a boost to the frame where the primary quarks make a small angle $2\theta_0$, as shown in the right panel of Fig.~\ref{fig:Cout}.

The COM and boosted frames are related by a boost factor $\gamma=1/\sqrt{1-v^2}$ with velocity $v=\cos \theta_0$; we deduce that $\gamma={1}/{\sin \theta_0}\simeq 1/\theta_0$ is large when $\theta_0\ll 1$.  
 Let us denote the four-momentum of the $i$-th emitted gluon in the COM and boosted frames by $\bar{k}_i^\mu=\bar{\omega}_i(1,\sin\bar{\theta}_i,0, -\cos\bar{\theta}_i)$ and $k_i^\mu=\omega_i(1,\sin \theta_i,0,\cos \theta_i)$, respectively. (Note that in the former case the angles are measured with respect to the negative $z$-axis.) As explained in the previous sections, the angles are strongly ordered in the two frames, but in the opposite directions:  $1\gg \bar{\theta}_1\gg \bar{\theta}_{2}\gg \cdots \gg \theta_0$  and  $\theta_0\ll \theta_1 \ll \theta_{2} \ll \cdots \ll 1$. 
Using the Lorentz transformation law for the energy together with the boost-invariance of the transverse momentum, one finds
\begin{align}
\omega_i = \bar{\omega}_i\,\frac{1-\cos\bar{\theta}_i \cos \theta_0}{\sin \theta_0} \simeq \bar{\omega}_i \,\frac{\theta_0^2 + \bar{\theta}_i^2}{2\theta_0} \simeq  \frac{ \bar{\omega}_i\bar{\theta}_i^2}{2\theta_0} , \label{first}
\qquad
 \omega_i  =\bar{\omega}_i\, \frac{\sin \bar{\theta}_i}{\sin \theta_i } \simeq \frac{\bar{\omega}_i \bar{\theta}_i}{\theta_i},
\end{align} 
from where one deduces $\theta_i \bar{\theta}_i \simeq 2\theta_0$. From (\ref{first}) we immediately see that the ordering in energies $\bar{\omega}_i\gg \bar{\omega}_{i+1}$  in the COM frame corresponds to an ordering in formation times in the boosted frame: 
\begin{align}
\tau_i \simeq \frac{1}{\omega_i \theta_i^2} = \frac{1}{2\bar{\omega}_i\theta_0}  \ll \tau_{i+1}.
\end{align}

This correspondence has an important implication in how the double logarithms (energy times collinear) arise in the two frames. In the COM frame, they correspond to a collinear regime, where the energies and the emission angles are simultaneously decreasing:  $\bar{\omega}_i \gg \bar{\omega}_{i+1}$ and $\bar{\theta}_i \gg \bar{\theta}_{i+1}$ (which automatically imply the proper time ordering: $\bar{\tau}_i \ll \bar{\tau}_{i+1}$).  The double logarithm then simply arises from the unconstrained, double, integral 
\begin{align}\label{2logCOM}
I=2\bar{\alpha}_s\int_{\theta_0}^{1}\frac{\rmd\bar{\theta}_1}{\bar{\theta}_1} \int_{E_0}^E \frac{\rmd\bar{\omega}_1}{\bar{\omega}_1} = \bar{\alpha}_s \ln \frac{1}{\theta_0^2} \ln \frac{E}{E_0}.
\end{align}

Going to the boosted frame, it is clear that the {\em same} double logarithm will be generated by integrating over a {\em different} domain in phase-space, where the respective variables $\theta$ and $\omega$ are now constrained by time-ordering. It is furthermore clear that one can return to unconstrained integrations by changing variables from $\theta$ and $\omega$ to $\theta$ and $\tau$. In order to deduce the respective integration limits, we shall perform a change of variables in \eqn{2logCOM} in the form of the Lorentz transformation from the COM frame to the boosted frame. It is easy to check that $\frac{\rmd\bar{\theta}_1\rmd\bar{\omega}_1}{\bar{\theta}_1\bar{\omega}_1}=\frac{\rmd\theta_1 \rmd\omega_1}{\theta_1\omega_1}$, meaning that the integrand preserves the same form but the integration region in $\omega_1$ depends on $\theta_1$ in such a way that effectively one has a logarithmic integral over the formation time $\tau_1=1/(\omega_1\theta_1^2)$:
\begin{align}\label{2logBOOST}
I=2\bar{\alpha}_s\int_{2\theta_0}^{2}\frac{\rmd\theta_1}{\theta_1} \int_{\frac{2\theta_0E_0}{\theta_1^2}}^{\frac{2\theta_0E}{\theta_1^2}} \frac{\rmd\omega_1}{\omega_1} = 2\bar{\alpha}_s\int_{2\theta_0}^{2}\frac{\rmd\theta_1}{\theta_1}\int_{\tau_0}^{\tau_{\rm max}} \frac{\rmd\tau_1}{\tau_1}=\bar{\alpha}_s \ln \frac{1}{\theta_0^2} \ln \frac{E}{E_0}.
\end{align}
where $\tau_{\rm max}\equiv {1}/({2E_0\theta_0})$ and $\tau_0\equiv {1}/({2E\theta_0 })$. It is interesting to study the integration limits for $\omega_1$ and also for $\tau_1$ in more detail. The upper limit $2\theta_0E/{\theta_1^2}\sim
(\theta_0/\theta_1)^2\gamma E$ on $\omega_1$  is much smaller, by a factor $(\theta_0/\theta_1)^2\ll 1$, than the would-be absolute upper limit on the energy of an emitted gluon, as set by the energy $p_a^0=\gamma E$ of the primary quark. Similarly, the lower limit ${2\theta_0E_0}/{\theta_1^2}$ is much larger, by the factor $1/{\theta_1^2}\gg 1$, than the lowest value $\omega_{\rm min}=\theta_0E_0$ for the gluon energy that was argued in our previous discussions, in Sects.~\ref{sec:collBOOST} and \ref{sec:collCOM}. Accordingly, the logarithmic phase-space for the integration over $\omega_1$   is effectively reduced, by the condition of time ordering, from its `naive' value $Y^+\simeq \ln(p_a^0/\omega_{\rm min})$ down to $Y^-=\ln(E/E_0)=Y^+ -\rho$, with $\rho=\ln(1/\theta_0)^2$. This reduction was anticipated in Sect.~\ref{sec:collBOOST}.

Consider finally the integration limits on $\tau_1$. The upper limit  $\tau_{\rm max}\sim  1/\omega_{\rm min}$ is recognized as the formation time for a gluon with energy $\omega\sim \omega_{\rm min}$ and which makes a polar angle of order one. This is softest gluon which matter to the DLA  evolution, as discussed in Sects.~\ref{sec:collBOOST} and \ref{sec:collCOM}. The lower limit, that can be rewritten as $\tau_0={1}/({2\gamma E \theta^2_0})$, is the coherence time associated with the original (boosted) dipole, with energy $\gamma E$ and opening angle $2\theta_0$. This is also the shortest possible formation time, as it corresponds to a very hard gluon emission with energy $\omega \sim \gamma E$ and which is nearly collinear with its parent quark.

\subsection{One gluon emission from a boosted antenna}

In this section, we shall use the rules of light-cone perturbation theory (LCPT), in which emission vertices are explicitly ordered in time and gluons are described by physical polarization vectors, to compute the emission of a soft gluon by a boosted antenna. The result for the emission probability that we shall obtain is of course standard ---  the `antenna pattern' \eqref{wabc} ---, but its present calculation is perhaps less familiar. Indeed, such calculations are generally performed within the covariant formalism (notably, the Feynman gauge), which is more economical. Yet, the formalism to be used has the virtue to make the physical picture more transparent.

Consider a boosted dipole with small opening angle $\theta_{ab}\ll 1$, as illustrated in the right panel of Fig.~\ref{fig:Cout}. The primary quarks, $a$ and $b$, have large longitudinal momenta $p_a^z=p_a^0\cos\theta_a\simeq p_a^0$, but comparatively small transverse momenta $p_{a\perp}=p_a^0\sin\theta_a\simeq p_a^0\theta_a$ (and similarly\footnote{In the context of the previous section, we have considered the symmetric situation where $\theta_a=\theta_b=\theta_0$ and $p_a^0=p_b^0$. For the present purposes, such a strict symmetry is not needed. We shall merely assume that the angles $\theta_a$ and $\theta_b$ are both small, such that $\theta_{ab}=\theta_a+\theta_b\ll 1$, and that the initial energies  $p_a^0$ and $p_b^0$ are comparable with each other.}  for the antiquark $b$). The radiated gluon is much softer, $k_z\ll p_a^0$, but in general it still makes a small angle w.r.t.~the $z$ axis, $\theta_k\simeq k_\perp/k_z\ll 1$. It is then convenient to use light-cone (LC) variables, e.g.~$k^\pm=(k_0\pm k_z)/\sqrt{2}$. The `large component' $k^+\simeq\sqrt{2} k_z$ is the LC longitudinal momentum, whereas the `small component' $k^-$ is the LC energy and is equal to $k^-=k_\perp^2/2k^+$ for an on-shell gluon.

\begin{figure*}
\begin{center}
\begin{minipage}[b]{0.49\textwidth}
\begin{center}
\includegraphics[width=0.8\textwidth,angle=0]{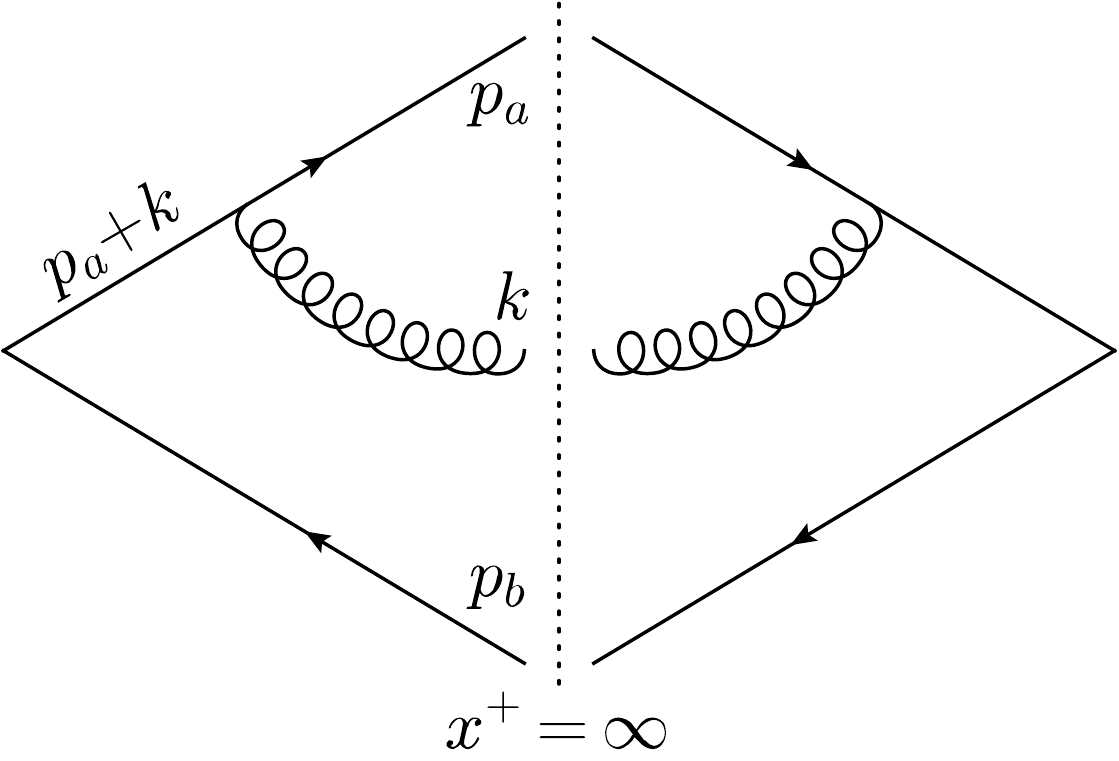}\\(A)\vspace{0.5cm}
\end{center}
\end{minipage}
\begin{minipage}[b]{0.49\textwidth}
\begin{center}
\includegraphics[width=0.8\textwidth,angle=0]{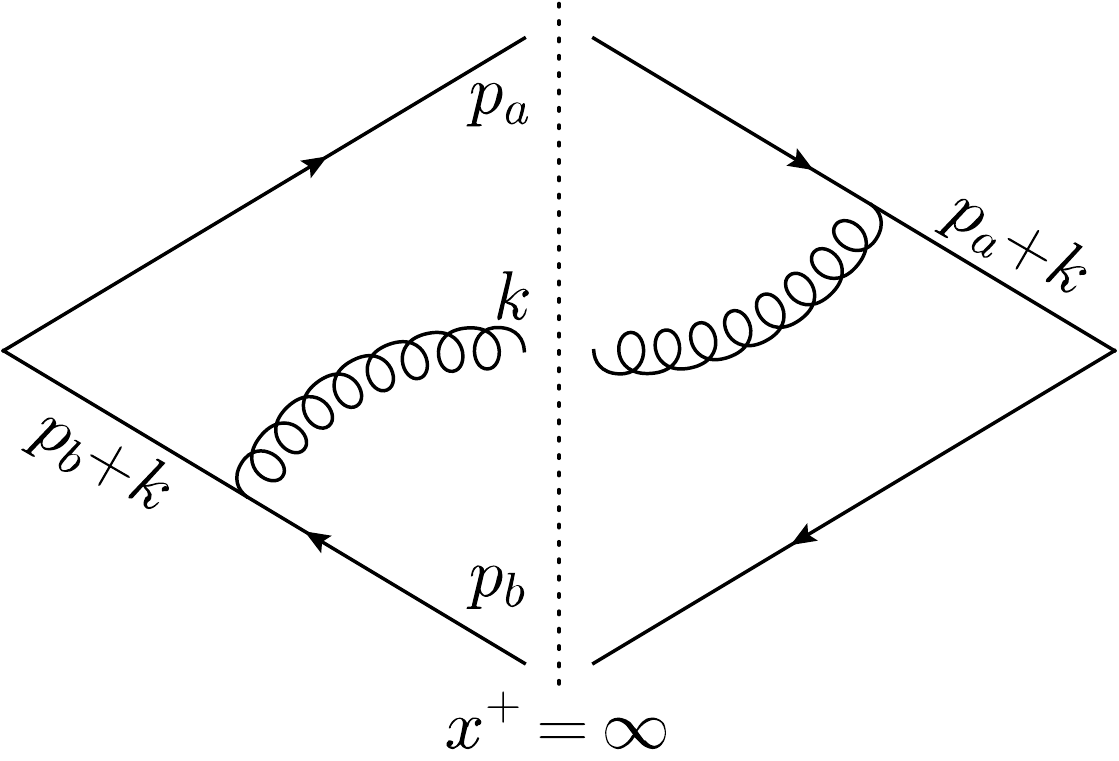}\\(B)\vspace{0.5cm}
\end{center}
\end{minipage}
\begin{minipage}[b]{0.49\textwidth}
\begin{center}
\includegraphics[width=0.8\textwidth,angle=0]{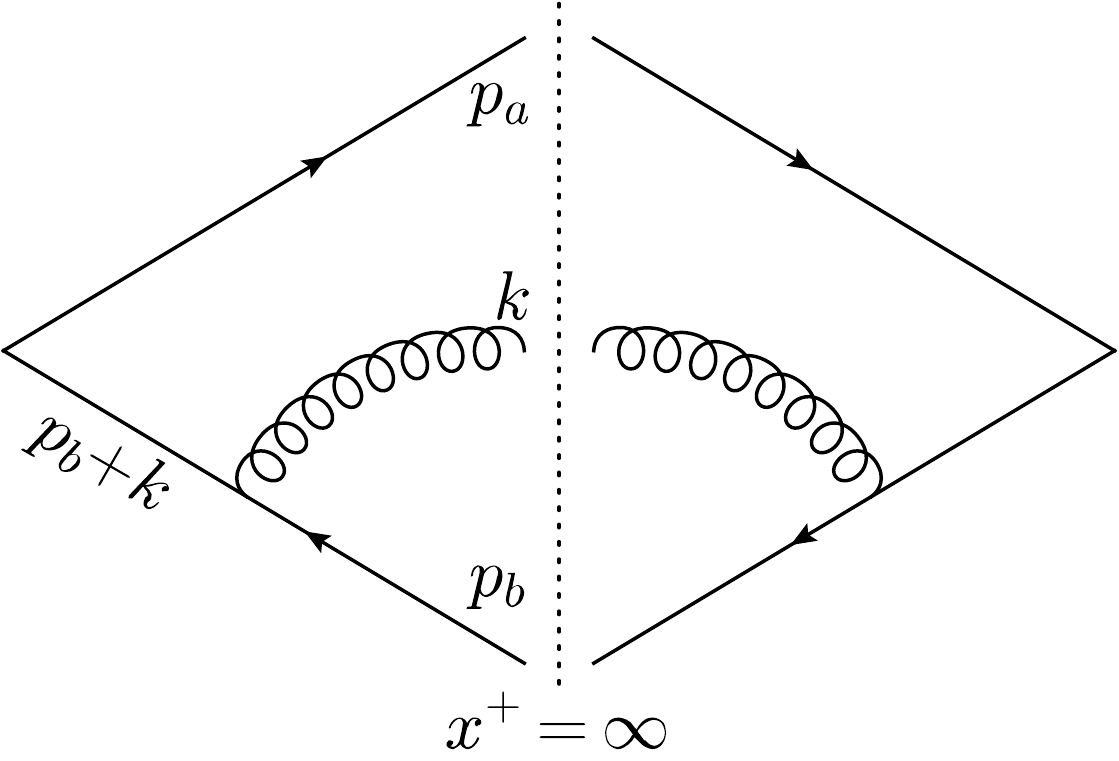}\\(C)\vspace{0cm}
\end{center}
\end{minipage}
\begin{minipage}[b]{0.49\textwidth}
\begin{center}
\includegraphics[width=0.8\textwidth,angle=0]{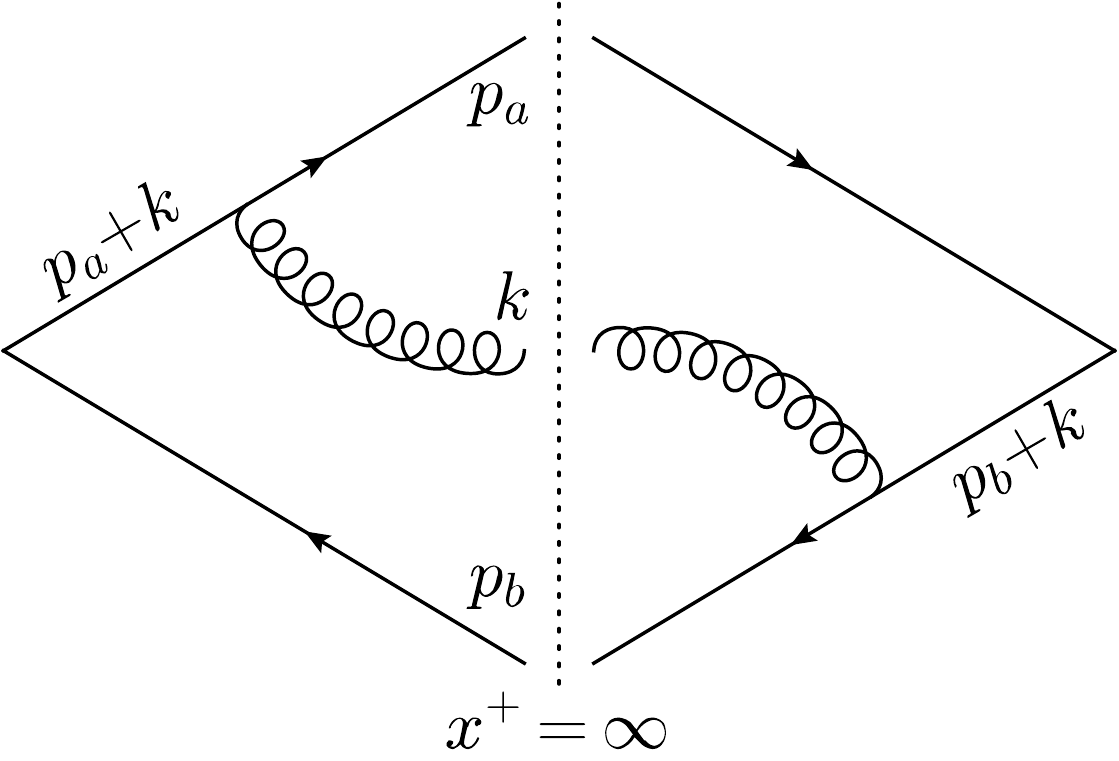}\\(D)\vspace{0cm}
\end{center}
\end{minipage}
\end{center}
\caption{\label{fig:1g} \small The 4 graphs which contribute to the emission probability of a soft gluon by an antenna in LCPT.}
\end{figure*}

We would like to compute the differential probability $\rmd P/\rmd k^+\rmd^2\bk_\perp$ for the emission of a soft gluon. To that aim, one needs to evaluate the 4 Feynman graphs displayed in Fig.~\ref{fig:1g}.
Graphs $A$ and $C$ describe direct emissions, by either the quark $a$ or the antiquark $b$, while graphs $B$ and $D$ describe interference effects between the emissions by the two fermions. It is perhaps interesting to notice that in the Feynman gauge with gluon propagator $G^{\mu\nu}\propto g^{\mu\nu}$, the contributions of the direct emissions, graphs $A$ and $C$, are both equal to zero\footnote{Indeed, in the eikonal approximation appropriate for soft gluons, the emission vertices are simply proportional to the 4-momenta of the on-shell quarks; hence, the contribution  of a  `direct' graph, say graph $A$, is proportional to $p^\mu_aG_{\mu\nu}(k) p^\nu_a\propto p_a^2=0$.}, so the whole result comes from the interference terms alone ($B$ and $D$). This is a peculiarity of the Feynman gauge, without any deep physical meaning. The physical picture becomes manifest only in the LC gauge $A^+=0$, that we shall employ here. 

We present more details for the interference graph $B$. After applying the Feynman rules of LCPT
in the LC gauge $A^+=0$  (see e.g.~\cite{Kovchegov:2012mbw}), one finds
\begin{align}\label{Bgraph}
G_B&=-g^2 \CF\int\frac{\rmd^3 k}{(2\pi)^32k^+}\frac{1}{2(p_a^++k^+)}\frac{1}{2(p_b^++k^+)}\,
\big[\bar u(p_b) \gamma\cdot \epsilon_{(\lambda)}  u(p_b+k)\big]
\big[\bar u(p_a) \gamma\cdot \epsilon_{(\lambda)}^*  u(p_a+k)\big]
\nonumber\\
&\qquad \times \frac{1}{\frac{p_{a\perp}^2}{2p_a^+}+ \frac{k_{\perp}^2}{2k^+} -
\frac{(\bp_{a\perp}+\bk_\perp)^2}{2(p_a^++k^+)}}\,
 \frac{1}{\frac{p_{b\perp}^2}{2p_b^+}+ \frac{k_{\perp}^2}{2k^+} -
\frac{(\bp_{b\perp}+\bk_\perp)^2}{2(p_b^++k^+)}}\,,
 \end{align}
where the kinematics of the emitted gluon has been integrated over (with the notation $\rmd^3 k
=\rmd k^+\rmd^2 \bk_\perp$), the proper limits being implicitly understood (in particular, $k^+$ is positive and soft). The overall minus sign occurs because the quark and the antiquark have opposite color charges. The sum over the gluon polarization states ($\lambda=1,\,2$) is understood.

The two energy denominators in the second line of \eqn{Bgraph} have been obtained after integrating over the gluons emission times --- $x_b^+$ for the emission by the antiquark $b$ in the direct amplitude (DA) and, respectively, $x_a^+$ for that by the quark $a$ in the complex conjugate amplitude (CCA); e.g.,
\beq\label{DaDa}
\frac{1}{D_a}\,\equiv\,i\int_0^\infty\rmd x_a^+  \,\rme^{-i\Delta E_a x_a^+ } 
\,=\,\frac{1}{\Delta E_a}
\,,\eeq
where $\Delta E_a\equiv p_a^-  + k^--(p_a+k)^-$ is the difference between the LC energies at the emission vertex. [$(p_a+k)^-$ denotes the LC energy of the parton with 3-momentum $(p_a^++k^+,\, \bp_{a\perp}+\bk_\perp)$.]  Accordingly,
\beq
D_a=\Delta E_a= p_a^-  + k^- - (p_a+k)^-=\frac{p_{a\perp}^2}{2p_a^+}+ \frac{k_{\perp}^2}{2k^+} -
\frac{(\bp_{a\perp}+\bk_\perp)^2}{2(p_a^++k^+)}\,.
\eeq
A similar expression holds for the other energy denominator $D_b$. 

Let $z\equiv {k^+}/{(p_a^++k^+)}$
denote the longitudinal momentum fraction taken by the gluon; 
this is small, $z\ll 1$, for a soft emission. Then one can successively write 
\begin{align}\label{Da-eval}
D_a&=\frac{1}{2(p_a^++k^+)}\left(\frac{p_{a\perp}^2}{1-z}+\frac{k_{\perp}^2}{z}
-\big(\bp_{a\perp}+\bk_\perp\big)^2\right)\nonumber\\
&=\frac{z(1-z)}{2(p_a^++k^+)}\left(\frac{\bp_{a\perp}}{1-z}-\frac{\bk_{\perp}}{z}\right)^2
\nonumber\\
&=\frac{k^+(1-z)}{4}\left(\bv_{a}-\bv_{k}\right)^2\,.
 \end{align}
In the last line we introduced the transverse velocities of the quark and the gluon,
\beq\label{veldef}
\bv_{a}\equiv\sqrt{2}\,\frac{\bp_{a\perp}}{p^+_a}\,,\qquad
\bv_{k}\equiv\sqrt{2}\,\frac{\bk_{\perp}}{k^+}\,,\eeq
which are convenient since directly related to the respective polar angles. For instance, the gluon angle reads $\theta_k\simeq k_{\perp}/k_z \simeq v_k$, with $v_k\equiv |\bv_k|$. 

The standard BMS regime corresponds to a situation where the various polar angle are comparable to each other, $\theta_k\sim \theta_a$, so the two terms inside the last parenthesis in \eqn{Da-eval} are equally important.  
Together, the two conditions  $z\ll 1$ and $\theta_k\sim \theta_a$ imply $k_\perp/p_{a\perp}\sim k^+/p_a^+ \simeq z \ll 1$. That is, the emitted gluon is (relatively) `soft' not only for its longitudinal momentum, but also for its transverse one. In spite of that, in evaluating  \eqn{Da-eval} it was important not to perform kinematical approximations {\em too early}. [For instance, within the braces in the first line, the would-be dominant term  $p_{a\perp}^2$ cancels out in the difference ${p_{a\perp}^2}/{(1-z)}-\big(\bp_{a\perp}+\bk_\perp\big)^2$.] But of course it is possible to replace $1-z\simeq 1$ in the prefactor occurring in the final result. We thus conclude that
\begin{align}\label{DaDb}
D_a\simeq \frac{k^+}{4}\left(\bv_{a}-\bv_{k}\right)^2\,,\qquad
D_b\simeq \frac{k^+}{4}\left(\bv_{b}-\bv_{k}\right)^2\,, \end{align}
where $D_b$ refers to the energy denominator for the DA.

Consider now the numerators in \eqn{Da-eval}, which are built with spinors, Dirac matrices, and the gluon polarization vector $ \epsilon_{(\lambda)}^\mu$. In the LC gauge $A^+=0$, one has 
\beq
 \epsilon_{(\lambda)}^\mu(k)\equiv \left( \epsilon_{(\lambda)}^+, \epsilon_{(\lambda)}^-, \be_{(\lambda)}\right)=\left(0, \frac{\bk_\perp\cdot \be_{(\lambda)}}{k^+},\, \be_{(\lambda)}\right),\qquad
 \sum_{\lambda=1,2} e_{(\lambda)}^i e_{(\lambda)}^j=\delta^{ij}\,.
 \eeq
Using the fact that the emitted gluon is soft, in the sense that $k^+\ll p^+_a$ and $k_{\perp}\ll p_{a\perp}$, together with $\bar u(p_a) \gamma^\mu u(p_a)=2p^\mu$, one finds e.g.
\beq
\bar u(p_a) \gamma\cdot \epsilon_{(\lambda)}^*  u(p_a+k)\simeq
2 p_a\cdot  \epsilon_{(\lambda)}^{*}\,=\,\frac{2p^+_a}{k^+}\,\bk_\perp\cdot \be_{(\lambda)}
- 2\bp_{a\perp}\cdot \be_{(\lambda)}=\sqrt{2}p^+_a(\bv_k-\bv_a)\cdot \be_{(\lambda)}\,.
\eeq
As already discussed in relation with \eqn{Da-eval}, both terms in the above result --- the one proportional to $\bv_k$ and that proportional to $\bv_a$ --- are equally important. 
 At a first sight, this seems to go beyond the standard eikonal approximation, which instructs us to keep only the coupling between the `large' component $p^+_a$ and the `minus' component $\epsilon_{(\lambda)}^-$ of the polarization vector, which is enhanced at small $k^+$. But a moment of thinking reveals that the scope of the eikonal approximation must be enlarged in this case, in order to keep trace of the polar angle made by the parent parton: indeed, although small, this angle is essential for computing dipole radiation\footnote{The opening angle $\theta_{ab}$ plays the same role within the time-like evolution of the antenna as the dipole transverse size in the context of the space-like evolution.}. As a matter of fact, we {\em do} use the eikonal approximation, in that we assume that the trajectory (velocity) of the parent quark is not modified by the emission of a soft gluon; but the information about the angle made by this trajectory w.r.t.~the longitudinal axis cannot be ignored, since it is essential for the present purposes. This is in agreement with the observation in \cite{Marchesini:2003nh} that the proper formulation of the eikonal approximation for time-like evolution is in terms of (polar) angles: the angle of the emitter is not modified by the  emission of a soft gluon.

After similarly evaluating the other Dirac factor in  \eqn{Da-eval}, performing the sum over $\lambda$, and putting together all the above results, one finds
\begin{align}\label{Bgraphfin}
G_B&=-\frac{\alpha_s \CF}{\pi^2}\int\frac{\rmd k^+}{k^+}\,\rmd^2 \bv_k\,\frac{
(\bv_k-\bv_a)\cdot (\bv_k-\bv_b)}{\left(\bv_{k}-\bv_{a}\right)^2\left(\bv_{k}-\bv_{b}\right)^2}\,.
 \end{align}
We have also used $\rmd^2 \bk_\perp=(k^+)^2\rmd^2 \bv_k/2$ to change the integration variable from $\bk_\perp$ to $\bv_k$ and thus make explicit the fact that the final integral over $k^+$ is logarithmic, as expected.


It is now straightforward to deduce the respective contributions of the other 3 graphs in Fig.~\ref{fig:1g}. The other interference graph $D$ gives the same result as shown in \eqn{Bgraphfin}. As for the direct emissions, graphs $A$ and $C$, the respective contributions are obtained from \eqn{Bgraphfin}  by changing the overall sign and replacing $(\bv_k-\bv_b)\to (\bv_k-\bv_a)$ for graph $A$ and, vice-versa,  $(\bv_k-\bv_a)\to (\bv_k-\bv_b)$ for graph $C$. Hence the result of summing over the 4 graphs amounts to replacing the kernel in \eqn{Bgraphfin} (including its sign) by
\begin{align}\label{dipolekernel}
-2\,\frac{(\bv_k-\bv_a)\cdot (\bv_k-\bv_b)}{\left(\bv_{k}-\bv_{a}\right)^2\left(\bv_{k}-\bv_{b}\right)^2}\,+
\,\frac{1}{\left(\bv_{k}-\bv_{a}\right)^2}\,+\,\frac{1}{\left(\bv_{k}-\bv_{b}\right)^2}\,=\,
\frac{\left(\bv_{a}-\bv_{b}\right)^2}{\left(\bv_{k}-\bv_{a}\right)^2\left(\bv_{k}-\bv_{b}\right)^2}\,.
 \end{align}

To establish the correspondence with the kernel of the BMS equation \eqref{BMSLO}, one needs to replace the transverse velocities of the various partons by the respective (polar and azimuthal) angles on $S^2$; e.g.~$\bv_{k}\to (\phi_k,\theta_k)$. Consider first the integration measure: one can write $\rmd^2 \bv_k =\rmd\phi_k v_k\rmd v_k\simeq\rmd\phi_k \theta_k\rmd\theta_k \simeq \rmd\Omega_k$, where we have used $v_k
\simeq \theta_k$ for small angles. Finally, the approximation
\beq\label{vtheta}
\left(\bv_{k}-\bv_{a}\right)^2 \,\simeq\, 2\big(1-\cos\theta_{ka}\big)\,\simeq\,\theta_{ka}^2
\,,\eeq
allows us to recognise \eqn{dipolekernel} as the small-angle version of the dipole kernel $w_{abc}$, cf.~\eqn{wabc}. To check \eqn{vtheta}, we write the scalar product $p_a\cdot k$ in 2 different ways (in usual coordinates and  in the LC ones) and compare the results. On one hand, $p_a\cdot k = p_a^0k^0(1- \cos\theta_{ka})$; on the other hand, 
\beq
p_a\cdot k = p^+_a k^-+p^-_ak^+- \bp_{a\perp}\cdot \bk_\perp=
\frac{p^+_ak_\perp^2}{2k^+} + \frac{k^+p_{a\perp}^2}{2p^+_a}- \bp_{a\perp}\cdot \bk_\perp=
\frac{p^+_a k^+}{4}\left(\bv_{k}-\bv_{a}\right)^2\,.
\eeq
Recalling that $p^+_a k^+\simeq 2 p_a^0k^0$ at high energies (or small angles), one immediately deduces \eqn{vtheta}. 

To summarise, the differential probability for emitting a soft gluon from the quark-antiquark antenna $(ab)$ reads:
\beq\label{P1g}
k^+\frac{\rmd P}{\rmd k^+}=\frac{2\alpha_s \CF}{\pi}\int\frac{\rmd\Omega_k}{4\pi}\,
\frac{1- \cos\theta_{ab}}{(1- \cos\theta_{ka})(1- \cos\theta_{kb})}\,,\eeq
which at large $N_c$ (where $2\CF\simeq N_c$) agrees indeed with \eqn{BMSLO}.

Let us finally comment on the physical interpretation of the energy denominators \eqref{DaDb}. Using \eqn{vtheta} for small angles, we see that e.g.~$1/D_a \simeq 4/(k^+\theta_{ka}^2)$, which is the formation time $\tau_k$ for the gluon emission by the quark $a$ (i.e.~the time it takes the gluon to lose coherence w.r.t.~its parent parton). As expected, the energy denominators encode the quantum uncertainty between energy and time. This information will be further exploited in the next subsection.

\subsection{Two-gluon emission: time-ordering from energy denominators}
\label{sec:2g}

In this subsection, we shall consider two successive gluon emissions, whose longitudinal momenta are strongly decreasing, $p_a^+\gg k_1^+\gg k_2^+\gg \mu$ with $\mu$ an infrared cutoff, but whose emission angles are strongly increasing: $\theta_{ab}\ll\theta_{1a}\ll\theta_{21}\ll 1$. This is the `anti-collinear regime' in which the LO BMS equation was previously argued to resum double-logarithmic corrections of the type  $(\abar Y\rho)^n$, where $Y=\ln(p_a^+/\mu)$ is the energy logarithm and $\rho=\ln(1/\theta_{ab}^2)$ is the collinear one. In particular,  we expect the dominant contribution of our sequence of two gluon emissions to be of order $(\abar Y\rho)^2$. Yet, as we would like to show in what follows, there is an important assumption underlying the LLA, which is not enforced in the LO BMS equation: this is the fact that the formation time $\tau_2\sim 1/(k^+_2\theta^2_{21})$ of the softer gluon is (much) larger than that, $\tau_1\sim 1/(k^+_1\theta^2_{1a})$, of the harder gluon; that is,  $\tau_{2}\gg\tau_{1}$. This condition is automatically satisfied in the usual BMS regime where the emission angles are comparable (in particular, it was always satisfied in previous applications of this equation in the literature), yet it becomes non-trivial --- and its non-enforcement spoils the convergence of the high-energy approximations --- in the anti-collinear regime of interest for us here.


Our purpose in this subsection is merely to demonstrate the emergence of this time-ordering condition from the energy denominators associated with the 2-gluon graphs. Hence, we shall not compute such graphs in full generality (that would be quite tedious even in the LLA, due to the many possible topologies), but merely exhibit the energy denominators corresponding to selected topologies, which are representative. Also, it is sufficient to consider only `real' graphs (that is, Feynman diagrams in which both gluons are produced in the final state) and to study emission {\em amplitudes} (rather than {\em probabilities}) --- indeed, the information about the formation times is separately included in the DA and in the CCA, since the respective energy denominators are simply multiplied with each other.

When computing Feynman graphs in LCPT, the time ordering of the emission vertices is important and in what follows we shall concentrate on graphs where the (harder) gluon 1 is emitted prior to the (softer) gluon 2. By itself, this ordering of the emission vertices does not guarantee that the formation times obey the expected condition $\tau_{2}>\tau_{1}$. We shall nevertheless find that the latter is respected by the contributions which matter to LLA. The discussion of the `anti-time-ordered' graphs in which the softer gluon is the first one to be emitted is quite non-trivial, but the final conclusion is that such graphs are not important to the accuracy of interest (we refer to Ref.~\cite{Iancu:2015vea} for a detailed argument in that sense, developed in the context of the space-like evolution of the dipole scattering amplitude).

\begin{figure*}
\begin{center}
\begin{minipage}[b]{0.32\textwidth}
\begin{center}
\includegraphics[width=0.8\textwidth,angle=0]{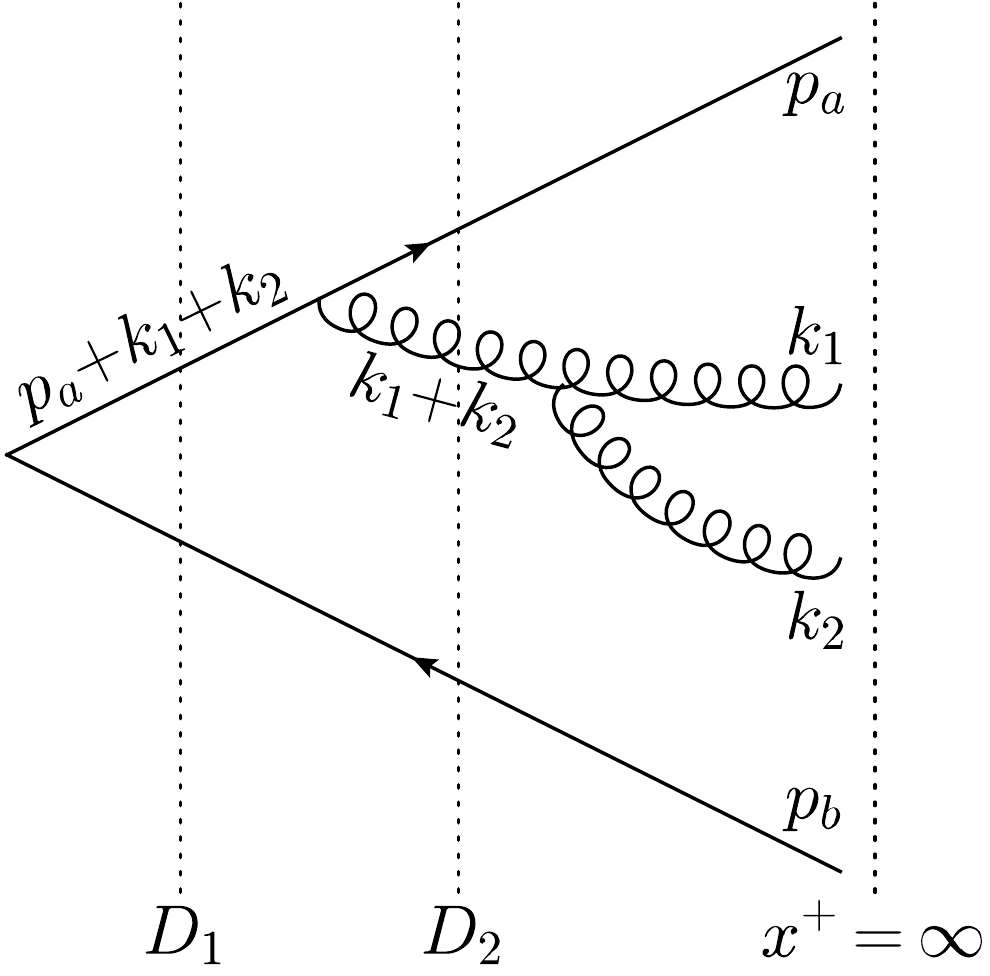}\\(a)
\end{center}
\end{minipage}
\begin{minipage}[b]{0.32\textwidth}
\begin{center}
\includegraphics[width=0.8\textwidth,angle=0]{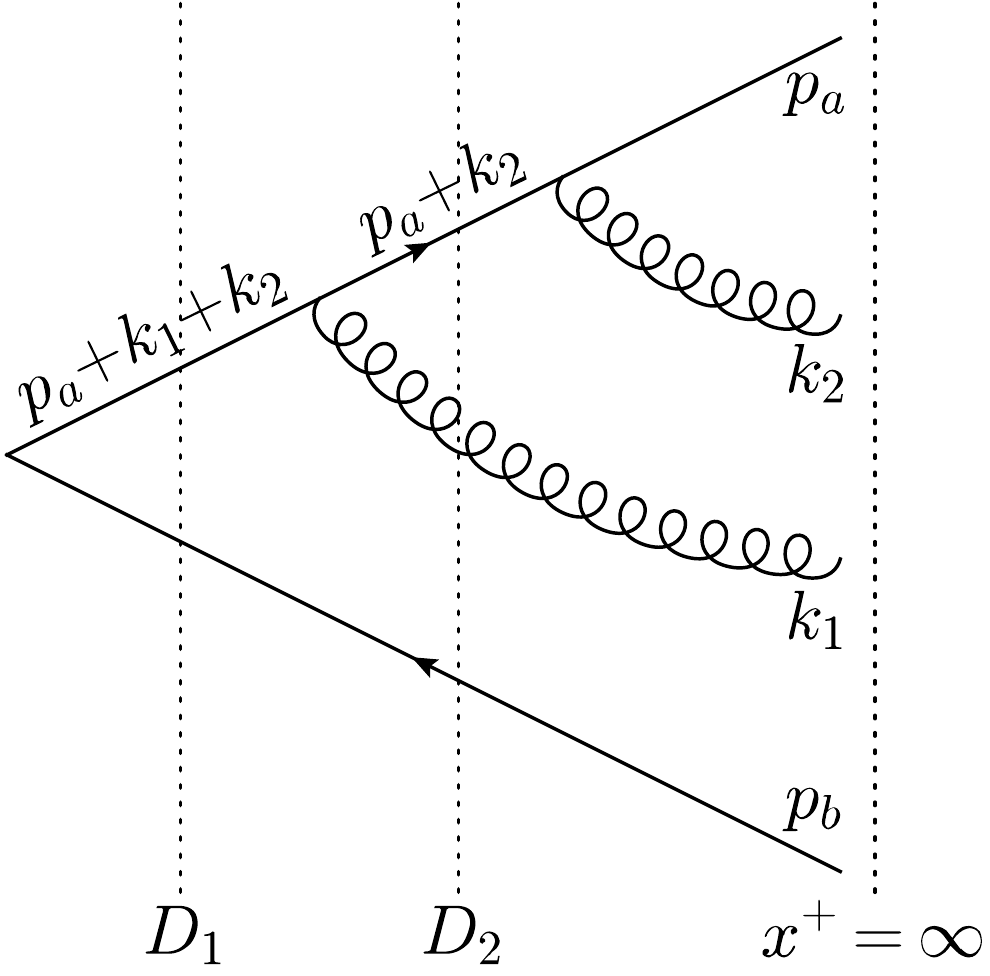}\\(b)
\end{center}
\end{minipage}
\begin{minipage}[b]{0.32\textwidth}
\begin{center}
\includegraphics[width=0.8\textwidth,angle=0]{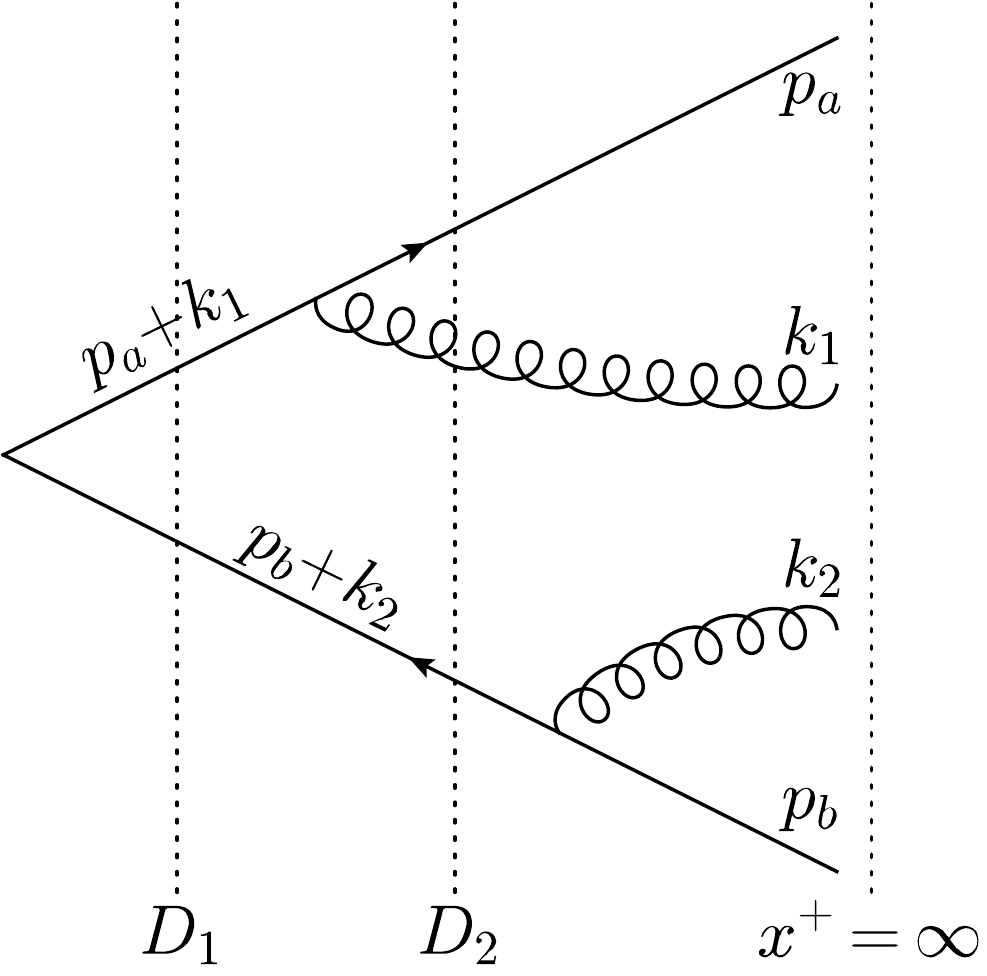}\\(c)
\end{center}
\end{minipage}
\end{center}
\caption{\label{fig:2g} \small Three over the six topologies contributing to the 2-gluon amplitude in LCPT. As discussed in the text, we consider only diagrams in which the harder gluon is emitted prior to the softer one ($k_1^+ \gg k_2^+ $). Also, albeit this is not visible in the graphical representation, we consider emission angles which are much larger than the antenna opening and strongly increasing from the first emission to the second one ($\theta_2\gg\theta_1\gg \theta_{ab}$).}
\end{figure*}

Under the present assumptions, there are 6 possible topologies contributing to the 2-gluon amplitude: the 3 graphs shown in Fig.~\ref{fig:2g}, where gluon 1 is emitted from the quark $a$, and the 3 corresponding ones where it is emitted from the antiquark $b$. As we shall see, any of these 6 topologies carries the required information about the time-ordering; for pedagogy we shall discuss those exhibited in Fig.~\ref{fig:2g}.

Consider first the graph in Fig.~\ref{fig:2g}.a. This involves the product of two energy denominators, $1/(D_1D_2)$, corresponding to the two intermediate states indicated with dashed lines. Once again, these energy denominators are generated by integrating over the emission times, $x^+_1$ and $x^+_2$, associated with the two vertices:
\beq\label{D1D2}
\frac{1}{D_1}\,\frac{1}{D_2}\,=\,-\int_0^\infty\rmd x_1^+ \int_{x_1^+ }^\infty\rmd x_2^+ \,\rme^{i\Delta E_1 x_1^+ +
i\Delta E_2 x_2^+}\,=\,\frac{1}{\Delta E_1 +\Delta E_2}\,\frac{1}{\Delta E_2}\,,\eeq
where $\Delta E_i=E_{i+1}-E_{i}$, with $i=1,2$, is the difference between the LC energies of the partonic states before and after the gluon emission at the vertex $i$; accordingly, $D_2=\Delta E_2$ and $D_1=\Delta E_1 +\Delta E_2$. As a general rule, $D_i$ is equal to the difference between the LC energy of the final state (here involving the quark-antiquark pair together with 2 gluons) and the LC energy of the intermediate state at hand. 

Let us first consider the second energy denominator $1/D_2$, since the respective discussion is simpler. By inspection of Fig.~\ref{fig:2g}.a, one finds (compare to \eqn{Da-eval})
\begin{align}
{D_2}&\,=\,k_1^- + k^-_2 -  \big(k_1+k_2\big)^-\,=\,\frac{k_{1\perp}^2}{2k^+_1} +\frac{k_{2\perp}^2}{2k^+_2} - 
\frac{(\bk_{1\perp}+\bk_{2\perp})^2}{2(k_1^++k^+_2)}\nn
&\,=\,\frac{k^+_1k^+_2}{4(k_1^++k^+_2)}\left(\bv_{1}-\bv_{2}\right)^2\,\simeq\,\frac{k^+_2}{4}\theta_{21}^2
\,,\end{align}
where the last, approximate, equality follows after using $k^+_2\ll k^+_1$ together with \eqn{vtheta}. The energy denominator $1/D_2$ is recognized as the formation time $\tau_2$ for the emission of the second gluon from the first one. This denominator contains no information about the relative time ordering of the two emissions, since the first gluon merely acts as a source for the second one. This information is rather encoded in the first  energy denominator $1/D_1$: from the uncertainty principle, we expect $\tau_1\sim 1/\Delta E_1$, but $1/D_1$ rather involves the sum $\Delta E_1 +\Delta E_2$; specifically,
\beq\label{D1}
\frac{1}{D_1}\,=\,\frac{1}{\Delta E_1 +\Delta E_2}\,=\,\frac{1}{1/\tau_1+1/\tau_2}\,\simeq\,
   \begin{cases}
        \displaystyle{\tau_1} 
        &
        \text{ if\,  $\tau_2 \gg \tau_1$,}
        \\*[0.35cm]
                 \displaystyle{\tau_2} 
        &
        \text{ if\,  $\tau_1 \gg \tau_2$.}
         \\*[0.2cm]
    \end{cases}
\eeq
In the usual formulation of the LLA, one effectively replaces $1/D_1\to \tau_1$; this ensures the factorization of the first gluon emission from the second one and also provides the logarithmic phase-space for the integration over $k_1^+$. However, as explicit in \eqn{D1} above, this  logarithmic phase-space is truly available only so long as $\tau_2 > \tau_1$; that is, for the purposes of the LLA, one should rather use $1/D_1\simeq \tau_1\Theta(\tau_2-\tau_1)$. The $\Theta$-function enforces the time-ordering condition that we were anticipating. 

The above argument may look a bit schematic, so let us rephrase it by using more explicit notations. With reference to Fig.~\ref{fig:2g}.a, one can successively write
\begin{align}\label{D1b}
D_1&\,=\, p_a^- + k_1^- + k^-_2 - \big(p_a+k_1+k_2\big)^-=\,   \big[p_a^- +
 k^- -\big(p_a+k\big)^-\big]+\big[k_1^- + k^-_2 - k^-\big] 
 \nn
&\,=\,\frac{p_a^+k^+}{4(p_a^++k^+)}\left(\bv_{k}-\bv_{a}\right)^2\,+\,\frac{k^+_1k^+_2}{4(k_1^++k^+_2)}\left(\bv_{1}-\bv_{2}\right)^2\nn
&\,\simeq\,\frac{k^+}{4}\theta_{ka}^2\,+\,\frac{k^+_2}{4}\theta_{21}^2
\,\simeq\,\frac{k^+_1 \theta_{1a}^2}{4}\,\Theta(\tau_2-\tau_1)\,+\,\frac{k^+_2 \theta_{21}^2}{4}\,\Theta(\tau_1-\tau_2)\,.
\end{align}
In writing the second equality above we have added and subtracted $k^-\equiv \bk_{\perp}^2/2k^+$ with\footnote{We recall that $(p_a+k)^-$ denotes the LC energy of the intermediate state with 3-momentum $(p_a^++k^+,\,\bp_{a\perp}+\bk_\perp)$, hence  $\big(p_a+k_1+k_2\big)^-$ is the same as $\big(p_a+k\big)^-$.} $\bk_{\perp}=
\bk_{1\perp}+\bk_{2\perp}$ and $k^+=k^+_1+k^+_2$, in order to construct the energy differences  $\Delta E_1$ and $\Delta E_2$. 
Then we have used Eqs.~\eqref{Da-eval} and  \eqref{vtheta} separately for  $\Delta E_1$ and $\Delta E_2$, together with simplifications following from the fact that $p_a^+\gg k_1^+\gg k_2^+$. The final approximation in \eqn{D1b}, where $\tau_1\equiv4/(k^+_1 \theta_{1a}^2)$ and $\tau_2\equiv4/(k^+_2 \theta_{21}^2)$, is the same as the last  estimate in \eqn{D1}.  Notice that, in writing the final result, we have approximated $\theta_{ka}\simeq \theta_{1a}$, that is, we have assumed that the direction of propagation of the parent gluon $k^+$, as described by its transverse velocity $\bv_{k}$, is not modified by the emission of the soft gluon $k_2^+$. This property is a hallmark of the eikonal approximation in the context of the time-like evolution \cite{Marchesini:2003nh}, so it is instructive to elaborate more on it. Using the definition in \eqn{veldef}, one can successively write
\beq\label{veleik}
\bv_{k}=\sqrt{2}\,\frac{\bk_{1\perp}+\bk_{2\perp}}{k^+_1+k^+_2}\,=\,
\frac{k^+_1}{k^+_1+k^+_2}\bv_1+\frac{k^+_2}{k^+_1+k^+_2}\bv_2\,\simeq\,
\bv_1+\frac{k^+_2}{k^+_1}\bv_2\,\simeq\,\bv_1
\,,\eeq
where the only non-trivial approximation is the very last one: this is not trivial since, albeit $k_1^+ \gg k_2^+ $, one also has $v_2\simeq \theta_2\gg v_1 \simeq \theta_1$. This being said, we shall shortly check that the condition $k^+_2\theta_2\ll k^+_1\theta_1$ remains satisfied.  Equivalently, the gluon transverse momenta are strongly decreasing, $k_{2\perp}\ll k_{1\perp}$, for the emissions contributing  to the LLA.

Returning to the final result in \eqn{D1b}, it is only the time-ordered regime at $\tau_2>\tau_1$ which contributes to the LLA. Indeed, in this regime $D_1\simeq ({k^+_1 \theta_{1a}^2})/{4}$ has the right dependence upon $k_1^+$ to generate, together with the other factors occurring when evaluating the emission {\em probabilities}, a logarithmic phase-space for the integral over $k_1^+$ (and similarly for the integral over $k_2^+$). Since $k_2^+\ll  k_1^+$, the condition $\tau_2 >\tau_1$ is automatically satisfied {\em except} within the anti-collinear evolution, where the emission angles are strongly increasing: $\theta_{21}\gg\theta_{1a}$.  In that context, the condition $\tau_2>\tau_1$ implies an upper limit on the angle of the softer emission: $\theta_{21}^2 < \theta_{1a}^2(k_1^+/k_2^+)$. Since $k_1^+\gg k_2^+$, this constraint still allows for angles $\theta_2$ much larger than $\theta_1$. But this constraint also implies $k^+_2\theta_{2}\ll k^+_1\theta_{1}$, as anticipated after \eqn{veleik}.

It is easy to extend the above considerations to the two 
other amplitudes in Fig.~\ref{fig:2g}. For the middle graph in Fig.~\ref{fig:2g}.b, the second energy denominator reads
\beq
D_2=p_a^-+k_2^--(p_a+k_2)^-\,\simeq\,\frac{k^+_2}{4}\theta_{2a}^2\,,
\eeq
and encodes the formation time for the emission of gluon 2 from the quark $a$. As for the first energy denominator $D_1$, this is evaluated exactly as in Eqs.~\eqref{D1}--\eqref{D1b} and carries the information about the formation time for the gluon 1 and the time-ordering of the two emissions. Finally, for the graph in Fig.~\ref{fig:2g}.c, one has $D_2=p_b^-+k_2^--(p_b+k_2)^-\simeq ({k^+_2 \theta_{2b}^2})/{4}$ (describing the emission of the softest gluon from the antiquark $b$), whereas
\begin{align}\label{D1c}
D_1&\,=\big[p_a^-+k_1^--(p_1+k_1)^-\big]+\big[p_b^-+k_2^--(p_b+k_2)^-\big]\simeq\,\frac{k^+_1}{4}\theta_{1a}^2\,+\,\frac{k^+_2}{4}\theta_{2b}^2\nn&\,\simeq\,\frac{k^+_1 \theta_{1a}^2}{4}\,\Theta(\tau_2-\tau_1)\,,
\end{align}
with the last estimate holding in the LLA. Note that the expressions of the formation time for the same gluon may involve different emission angles in different graphs; e.g.~$\tau_2 =4/(k^+_2 \theta_{21}^2)$ for the graph in Fig.~\ref{fig:2g}.a, but $\tau_2=4/(k^+_2 \theta_{2b}^2)$ for that in Fig.~\ref{fig:2g}.c. Yet this is not important to the accuracy of interest, since the difference between say $\theta_{21}$ and $\theta_{2b}$ is irrelevant for the emissions at wide angles which must be constrained by time-ordering:
$\theta_{21}\simeq \theta_{2b}\simeq \theta_2$.

To better appreciate the effects of the time-ordering within the anti-collinear evolution, let us integrate out the intermediate gluon 1 for a fixed kinematics of the gluon 2, such that $p_a^+\gg k_1^+\gg k_2^+$ and $\theta_{0}\ll\theta_{1}\ll\theta_{2}\ll 1$ (we wrote $\theta_{ab}=2\theta_0$, to match the notations in other sections of the paper). Our calculation will be schematic and restricted to the double-logarithmic approximation (DLA), where the integrals over $k_1^+$ and $\theta_1$ are both logarithmic. The DLA integral over gluon 1 reads
\begin{align}\label{DL1}
\abar\int_{\theta_0}^{\theta_{2}}\frac{\rmd \theta_1}{\theta_1}\,
\int_{k^+_2}^{p_a^+}\frac{\rmd k^+_1}{k^+_1}\,\theta(\tau_1-\tau_0)\theta(\tau_2-\tau_1)
&=\abar\int_{\theta_0}^{\theta_{2}}\frac{\rmd \theta_1}{\theta_1}\,
\int_{k^+_2({\theta_2^2}/{\theta_1^2})}^{p_a^+({\theta_0^2}/{\theta_1^2})}
\frac{\rmd k^+_1}{k^+_1}\nn
&=\frac{\abar}{2}\,
\ln\frac{p_a^+ \theta_0^2}{k^+_2\theta_2^2}\,\ln\frac{\theta_2^2}{\theta_0^2}=\frac{\abar}{2}\,\big(Y^+-\rho\big)\rho
\end{align}
and involves {\em two} time-ordering constraints: the formation time $\tau_1$ of the  intermediate gluon 1 should be smaller than the respective time, $\tau_2$, of the softer gluon 2, but also larger than the coherence time  $\tau_0=1/(p_a^+ {\theta_0^2})$ of the original dipole, as explained in Sect.~\ref{sec:boost}. These two constraints have been used to modify the integration limits for the integral over $k^+_1$: the upper limit has been reduced from $p_a^+$ to $p_a^+({\theta_0^2}/{\theta_1^2})$, whereas the lower limit has been raised from $k_2^+$ to $k^+_2({\theta_2^2}/{\theta_1^2})$. In the final result, $Y^+\equiv \ln(p_a^+/k^+_2)$ is the rapidity phase-space that would be available to gluon 1 in the absence of time-ordering, whereas $\rho\equiv \ln({\theta_2^2}/{\theta_0^2})$ is the respective collinear phase-space. As anticipated at the end of Sect.~\ref{sec:collCOM} (and further discussed from a different perspective in Sect.~\ref{sec:boost}),  the main consequence of time-ordering is to reduce the phase-space for rapidity evolution from $Y^+$ to $Y^-=Y^+-\rho$. 

In the usual organization of the high-energy resummation in perturbative QCD, the two terms in the final result in \eqn{DL1}, that is, $\abar Y^+\rho$ and $\abar \rho^2$, appear at different orders in $\abar$. To understand that, notice that the subsequent integral over the kinematics of gluon 2 will generate, in particular,  another factor $\abar Y^+\rho$. The product $(\abar Y^+\rho)^2$, which is quadratic in $Y^+$, is then interpreted as a LO piece, coming from two iterations of the LO BMS equation. On the other hand, the term linear in $Y^+$, that is, $(-\abar^2\rho^2Y^+)$,  is viewed as a NLO correction to the BMS kernel. As clear from the above, this NLO correction is negative and potentially large (in the anti-collinear kinematics), since enhanced by a double collinear logarithm. This particular correction is explicitly extracted from the general NLO result in Appendix \ref{sect:nlo}. Such large higher-order corrections, which appear when treating the time-ordering condition within a strict expansion in powers of $\alpha_s$, are expected to spoil the convergence of the perturbation series. Better methods to enforce time-ordering within the BMS evolution will be presented in the next section.

\section{The collinearly-improved BMS equation}
\label{sec:coll}

In this section, we will modify the LO BMS equation in order to incorporate the effects of time-ordering within the anti-collinear evolution of a boosted dipole. As we shall see, this amounts to a partial resummation of the perturbative expansion to all orders. We shall first perform this resummation at the level of the DLA, i.e.~in the regime where the emission angles are strongly increasing, and then generalize it to generic values for the emission angles. The subsequent construction is very similar to that presented in Ref.~\cite{Iancu:2015vea} in the context of the space-like evolution, so in what follows we shall skip some of the details.

If one was interested in the DLA {\em alone}, then the inclusion of the proper time-ordering would be straightforward: it would suffice to use the formation time $\tau=1/(k^+\theta^2)$ instead of the gluon longitudinal momentum $k^+$ as the energy variable for the evolution. More precisely, the `evolution time' should be $Y^-=\ln(\tau_{\rm max}/\tau)$, with $\tau$ comprised within the range $\tau_{0} < \tau < \tau_{\rm max}$. We recall that $\tau_0=1/(p_a^+ \theta_{0}^2)$ is the coherence time of the primary $q\bar q$ dipole, whereas $\tau_{\rm max}= 1/k^+_{\rm min}$, with $k^+_{\rm min} = \sqrt{2} \theta_0 E_0$, denotes the formation time of the softest gluon which matters to the DLA evolution. The function 
$\mathcal{A}(\rho,Y^-)$ (the relevant observable to DLA, cf.~Sect.~\ref{sec:collBOOST}) will obey the standard version of the DLA equation, that is, \eqn{DLAboost} with $Y\to Y^-$. The associated solution, as given by \eqn{Aboost} with $Y\to Y^-_{\rm max}=\ln(E/E_0)$, would precisely coincide with the corresponding approximation in the COM frame, cf.~\eqn{DLACOM}.

Our ultimate interest, however, is not in the DLA equation {\em per se}, but in the more general BMS equation. The natural energy variable for building the BMS evolution of a boosted dipole which propagates along the positive $z$ axis is $Y^+ = \ln(k^+/k^+_{\rm min})$, as shown in Sect.~\ref{sec:LCPT}.  In general, the BMS equation is {\em non-local} in the emission angles, so one cannot simply replace $Y^+\to Y^-= Y^+-\rho$ as the `evolution time'. Indeed, the very definition of the `collinear logarithm' $\rho=\ln (1/\theta^2)$ becomes ambiguous in the presence of the non-locality in angles. Besides, $\rho$ is not a monotonous variable anymore: angles can both increase or decrease during the evolution. To work out the evolution in this more general context, one must enforce the condition of time-ordering directly in terms of $Y^+$. 



For more clarity, we shall preserve the notations $Y^+$ and $\rho$ for the respective maximal values, 
$Y^+ = \ln(p_a^+/k^+_{\rm min})$ and $\rho=\ln (1/\theta_{0}^2)$, and use a subscript $c$ to indicate the kinematics of the emitted gluon:  $Y_c^+ = \ln(k_c^+/k^+_{\rm min})$,  $\rho_c = \ln(1/\theta_c^2)$, etc.
The relevant time-ordering conditions read
 \beq
 \label{tologs}
 \tau_{0} < \tau_c < \tau_{\rm max}\,
  \Longrightarrow \,
 \frac{1}{p_a^+ \theta_{0}^2} \,< \, \frac{1}{k_c^+ \theta_c^2} <\frac{1}{k^+_{\rm min}}
\, \Longrightarrow\,
\rho_c < Y_c^+ < Y^+ - \rho +\rho_c\,.
 \eeq
Clearly, the condition $Y^+> \rho$ is required in order to ensure a non-trivial phase-space for the DLA evolution. The time-ordered version of the BMS equation in the DLA limit reads  (in integral form)
 \beq
 \label{bmsdlato}
   \mathcal{A}(\rho,Y^+) = 
 \frac{\abar}{8}\,(Y^+ - \rho)\Theta(Y^+ - \rho) + 
 \abar \int\limits_0^{\rho} \dif \rho_c
 \int\limits_{\rho_c}^{Y^+ - \rho + \rho_c} \dif Y_c^+
 \mathcal{A}(\rho_c,Y_c^+).
 \eeq
Both the source  term and the evolution term in the r.h.s. of this equation have support at $Y^+ > \rho$ alone, hence the same property is valid for its solution $\mathcal{A}(\rho,Y^+)$. The integral form in \eqn{bmsdlato} is particularly convenient for constructing the exact solution via an iterative procedure; we thus obtain the following power series
 \beq
 \label{Aseries}
 \mathcal{A}(\rho,Y^+) = \frac{1}{8}\,\Theta(Y^+ -\rho)
 \sum_{k=1}^{\infty} 
 \frac{\abar^k (Y^+-\rho)^k \rho^{k-1}}{k!(k-1)!}\,,
 \eeq
which can be readily summed to give
 \beq
 \label{Abessel}
 \mathcal{A}(\rho,Y^+) =  \frac{1}{8}\,\Theta(Y^+ -\rho)
 \sqrt{\frac{\abar(Y^+ -\rho)}{\rho}}\,
 \rmI_1 \big(2 \sqrt{\abar(Y^+ -\rho)\rho} \big).
 \eeq
 As anticipated, this is the same as the DLA solution\footnote{Via the simple change of variables $Y^+\equiv Y^-+\rho$, \eqn{bmsdlato} would take the standard DLA form, cf.~\eqn{DLAboost}.} \eqref{Aboost} with $Y\to Y^+-\rho = Y^-$. In particular, the $\Theta$-function enforcing $Y^+ > \rho$ is the equivalent of the vanishing initial condition at $Y^-=0$ in the COM frame. 
 
It should be possible to promote the time-ordered DLA equation  \eqref{bmsdlato} to full BMS accuracy, by following steps similar to those detailed in \cite{Beuf:2014uia} for the case of the BK equation (i.e.~for the space-like evolution). However, still as in that case, the ensuing equation would have the drawback to be {\em non-local} in $Y^+$. Indeed, this is already the case for \eqn{bmsdlato}, due to the $\rho$-dependence of the upper limit on the integral over $Y_c^+$. Here we shall use a different strategy, which follows the treatment of the BK equation in \cite{Iancu:2015vea}. Namely, we shall first rewrite 
the DLA version of the BMS equation in a {\em local} form, that is, in a form in which the evolution kernel does not depend on $Y^+$. To that aim, we shall use an integral representation of the modified Bessel function $\rmI_1$ in the complex plane: for $Y^+>\rho$, \eqn{Abessel} is equivalent to
 \beq
 \label{Aintegral}
 \mathcal{A}(\rho,Y^+) =
 \frac{1}{8} 
 \int\limits_{\frac{1}{2}-\rmi \infty}^{\frac{1}{2}+\rmi \infty}
 \frac{\dif\xi}{2\pi \rmi}\,
 \exp \left[ 
 \frac{\abar}{1-\xi}(Y^+ - \rho) + (1-\xi)\rho
 \right]
 -\frac{\delta(\rho)}{8}.
 \eeq
The $\delta$-function in the above equation is necessary, as it ensures that the {\em boundary condition} at $Y^+=\rho$ is correct, i.e.~$\mathcal{A}(\rho,Y^+=\rho)=0$. Now the integral in the r.h.s. of \eqn{Aintegral} is well defined also for $Y^+< \rho$, where it represents the ordinary (oscillating) Bessel function $\rmJ_1$. So, we shall use \eqn{Aintegral} to define an {\em analytical continuation} of $\mcal{A}(\rho,Y^+)$ valid for all positive values of $Y^+$, including the non-physical range at $Y^+ <\rho$. For this analytic continuation, that we shall still denote as $\mathcal{A}(\rho,Y^+)$ to avoid a proliferation of symbols, we will now deduce an evolution equation which is local in $Y^+$ and is formulated as an initial-value problem (but including a source term), with the initial condition given at $Y^+=0$.
 
Namely, by making a change of the integration variable according to $\gamma = \xi +\abar/(1-\xi)$, \eqn{Aintegral} can be written in the form of a standard Mellin representation w.r.t.~to the variable $\rho$, that is
 \beq
 \label{Agamma}
 \mathcal{A}(\rho,Y^+) =
 \frac{1}{8} 
 \int\limits_{\frac{1}{2}-\rmi \infty}^{\frac{1}{2}+\rmi \infty}
 \frac{\dif\gamma}{2\pi \rmi}\,J(\gamma)
 \exp \left[ \abar \chi_{\sdla}(\gamma) Y^+ + (1-\gamma)\rho
 \right]
 -\frac{\delta(\rho)}{8}.
 \eeq   
There are two new functions appearing in the above. The first is the `characteristic function' $\chi_{\sdla}(\gamma)$, which is defined as the coefficient of $\abar Y^+$ and thus given by
 \beq
 \label{chidla}
 \abar \chi_{\sdla}(\gamma) = 
 \frac{1}{2} \left[
 -(1-\gamma) + \sqrt{(1-\gamma)^2 + 4 \abar}
 \right] = 
 \frac{\abar}{1-\gamma} - \frac{\abar^2}{(1-\gamma)^3} + 
 \frac{2 \abar^3}{(1-\gamma)^5} + \cdots,
 \eeq
and the second is the Jacobian due to the change of variables, which is related to the characteristic function and reads
 \beq
 \label{jac}
 J(\gamma) = 1 - \abar \chi'_{\sdla}(\gamma) = 
 1 - \frac{\abar}{(1-\gamma)^2}+
 \frac{3 \abar^2}{(1-\gamma)^4} + \cdots.
 \eeq
Notice that both functions are finite at $\gamma=1$; the poles  at $\gamma=1$ appearing after the second equality in each of \eqn{chidla} and \eqn{jac} are just an artifact of expanding in powers of $\abar$ and truncating the series. Also, the first term (the unity) in the r.h.s. of \eqn{jac} for $J(\gamma)$ simply cancels the contribution of the $\delta$-function explicit in the r.h.s. of \eqn{Agamma}.

Using the properties of the Mellin transform, one finds that the function $\mathcal{A}(\rho,Y^+) $ defined by \eqn{Agamma} obeys an evolution equation which is local in $Y^+$ and in integral form it reads
 \beq
 \label{bmsdlaloc}
 \mathcal{A}(\rho,Y^+) = 
 \mathcal{A}(\rho,Y^+=0)
 +\frac{\abar}{8}\,\mathcal{K}_{\sdla}(\rho) Y^+
 +\abar \int\limits_0^{Y^+} \dif Y_c^+
 \int\limits_0^{\rho}
 \dif \rho_c\,
 \mathcal{K}_{\sdla}(\rho-\rho_c)
 \mathcal{A}(\rho_c,Y_c^+). 
 \eeq 
The new evolution kernel $\mathcal{K}_{\sdla}(\rho)$ is  the inverse Mellin transform of $\chi_{\sdla}(\gamma)$ and is given by
\begin{align}
\label{kdla}
\hspace*{-0.9cm}
\mathcal{K}_{\sdla}(\rho) =\!\!\!
\int\limits_{\frac{1}{2}-\rmi \infty}^{\frac{1}{2}+\rmi \infty}\!\!\!
 \frac{\dif\gamma }{2\pi \rmi}\, 
 \chi_{\sdla}(\gamma) \exp \left[(1-\gamma)\rho
 \right]
 =\frac{\rmJ_1\big(2 \sqrt{\abar \rho^2}\big)}{\sqrt{\abar \rho^2}} =
1 - \frac{\abar \rho^2}{2} + \frac{(\abar \rho^2)^2}{12} +
\mathcal{O}\big((\abar\rho^2)^3\big),
 \end{align}
where $\rmJ_1$ is a Bessel function of the first kind. We notice not only the resummation of the source term in terms of the evolution kernel, but also the emergence of an {\em initial condition} which is formulated at the unphysical point $Y^+=0$. This initial condition is determined by \eqn{Agamma} as
 \beq
\label{ainit}
\mathcal{A}(\rho,Y^+ =0) = - \frac{\sqrt{\abar}}{8}\, \rmJ_1\big(2 \sqrt{\abar \rho^2}\big) = -
    \frac{\abar \rho}{8}
    \left[1 - \frac{\abar \rho^2}{2}+ \frac{(\abar \rho^2)^2}{12} + \mathcal{O}\big((\abar\rho^2)^3\big) \right]. 
\eeq
This result for $\mathcal{A}(\rho,Y^+ =0)$, including its perturbative expansion as shown above, can also be recovered by letting $Y^+\to 0$ in the series in \eqn{Aseries}. In particular, the leading-order term $-{\abar \rho}/{8}$ in this expansion is recognised as the piece independent of $Y^+$ from the respective  term in \eqn{Aseries}, i.e.~$\mathcal{A}(\rho,Y^+)\simeq {\abar (Y^+-\rho)}/{8}$.

Let us see in more detail how \eqn{bmsdlaloc}, and in particular the resummed source term\footnote {Such a term is absent in the respective BK problem \cite{Iancu:2015vea}.}, arise from the Mellin integral representation in \eqn{Agamma}. To this end, we first differentiate \eqn{bmsdlaloc} w.r.t.~$Y^+$  in order to get rid of the initial condition and then substitute $\mathcal{A}(\rho_c,Y^+)$ from \eqn{Agamma} and $\mathcal{K}_{\sdla}$ from its integral representation  \eqref{kdla} (but only within the last, integral, term in \eqn{bmsdlaloc}); we thus find
 \begin{align}
 	\label{dadytest}
 	\frac{\del \mathcal{A}(\rho,Y^+)}{\del Y^+}
 	=\: & \frac{\abar}{8}\,\mathcal{K}_{\sdla}(\rho)
 	+ \frac{\abar}{8}
 	\int\limits_{\frac{1}{2}-\rmi \infty}^{\frac{1}{2}+\rmi \infty}
 \frac{\dif\gamma}{2\pi \rmi}
 \int\limits_{\frac{1}{2}-\rmi \infty}^{\frac{1}{2}+\rmi \infty}
 \frac{\dif\gamma_c}{2\pi \rmi}
 \int\limits_{0}^{\rho} \dif \rho_c\,
 \chi_{\sdla}(\gamma) \exp \left[(1-\gamma)(\rho-\rho_c)
 \right]
 \nn
 & \times J(\gamma_c)
 \exp \left[ \abar \chi_{\sdla}(\gamma_c) Y^+ + (1-\gamma_c)\rho_c \right] 
 - \frac{\abar}{8} \int\limits_{0}^{\rho} \dif \rho_c\, 
 \mathcal{K}_{\sdla}(\rho-\rho_c) \delta(\rho_c). 	
 \end{align}
The resummed source term cancels the last term in the above, while for the middle term we first perform the integration over $\rho_c$ which sets $\gamma_c =\gamma$.  Thus, we arrive at
 \beq
 \label{dady}
 \frac{\del \mathcal{A}(\rho,Y^+)}{\del Y^+}
 = \frac{\abar}{8} 
 \int\limits_{\frac{1}{2}-\rmi \infty}^{\frac{1}{2}+\rmi \infty}
 \frac{\dif\gamma}{2\pi \rmi}\,J(\gamma) \chi_{\sdla}(\gamma) 
 \exp \left[ \abar \chi_{\sdla}(\gamma) Y^+ + (1-\gamma)\rho
 \right],
 \eeq
which is obviously the derivative of \eqn{Agamma} w.r.t.~$Y^+$. It is a straightforward algebraic exercise to verify order by order in $\abar$ that \eqn{bmsdlaloc} together with Eqs.~\eqref{kdla} and \eqref{ainit} lead to the series solution given in \eqn{Aseries}, thus providing a cross-check for the validity of our construction. 

\eqn{bmsdlaloc}  is valid in the regime where the angles are strongly increasing. To extend this equation valid to generic values for the emission angles (and thus covering the whole angular phase-space for the BMS evolution), we first recall the relation between the currently employed function $\mathcal{A}(\rho,Y^+)$ and the general BMS observable: namely, we wrote $R_{ab}=1-P_{ab}$ with $R_{ab}(Y)\simeq \theta_{ab}^2\calA(\theta_{ab}, Y^+)$ to DLA accuracy. We therefore propose the following, collinearly-improved, version of the BMS equation
 \begin{align}
 \label{collBMS}
 \frac{\del P_{ab}(Y^+)}{\del Y^+}=&
 - \abar\int_{\mathcal{C}_{\rm out}} 
 \frac{\dif \Omega_c}{4\pi}\, 
 w_{abc}\,
 {\cal K}_{\sdla}
 \big(\sqrt{L_{acb}L_{bca}}\big)
 P_{ab}(Y^+)
 \nn
 &+
 \abar\int_{\mathcal{C}_{\rm in}} 
 \frac{\dif \Omega_c}{4\pi}\, 
 w_{abc}\,
 {\cal K}_{\sdla}
 \big(\sqrt{L_{acb}L_{bca}}\big)
 \big[P_{ac}(Y^+) P_{bc}(Y^+)-P_{ab}(Y^+)\big],
 \end{align}
where (recall that $\theta_{ab}=2\theta_{0}$)   
 \begin{align}
 \label{Ldef}
 L_{acb}\equiv 
 \ln \frac{1-\cos \theta_{ac}}{1-\cos \theta_{ab}}, 
 \qquad 
 L_{bca}\equiv
 \ln \frac{1-\cos \theta_{bc}}{1-\cos\theta_{ab}}.
\end{align}
The BMS equation in \eqref{collBMS} is to be solved with the initial condition
 \begin{align}
 \label{IC} P_{ab}(Y^+=0) =
 1+ \frac{\sqrt{\abar}}{4}\,
 (1-\cos \theta_{ab})
 {\rm J}_1(2\sqrt{\abar\rho^2})
\end{align}
with  the `collinear logarithm' defined in the more general form $\rho \equiv - \ln[2(1-\cos \theta_{ab})]$. 

Notice that in writing  \eqn{collBMS} it has been possible to use the same argument for the `collinearly-improved' kernel $\mathcal{K}_{\sdla}$ everywhere in the angular space --- that is, in both the source term and the evolution term. Indeed, with this particular choice for the argument, one reproduces the DLA structure in \eqn{bmsdlaloc} in the anti-collinear regime, without introducing spurious logarithms in the collinear regime. 
To see this, consider first the `evolution' piece of  \eqn{collBMS}: in the anti-collinear regime at $\theta_c\gg \theta_0$, one can write $\theta_{ac}\simeq \theta_{bc} \simeq \theta_{c}
\gg \theta_{ab}$, hence $\sqrt{L_{acb}L_{bca}}\simeq \ln(\theta_{c}^2/ \theta_0^2)=\rho-\rho_c$, in agreement with \eqn{bmsdlaloc}. On the other hand, in the regime where the soft gluon is nearly collinear with say the quark $a$, that is, $\theta_{ac}\ll \theta_{ab}\simeq \theta_{bc}$, one of the two logarithms in \eqn{Ldef} becomes large, $L_{acb}\simeq 
-\ln(\theta_{ab}^2/\theta_{ac}^2)$, but the other one vanishes, $ L_{bca}\simeq 0$, so the resummation becomes trivial, $\mathcal{K}_{\sdla}\simeq 1$, as it should.

Consider similarly the source term in \eqn{collBMS}, where the integral over $\theta_c$ runs over the interval $[\pi/2,\pi]$, while $\theta_{ab}=2\theta_0\ll 1$.  One can write  $\cos \theta_{ac}\simeq \cos \theta_{bc}\simeq \cos \theta_c$, where $1-\cos \theta_{c}$ is of order one whereas $1-\cos \theta_{ab}\simeq \theta_{ab}^2/2$ is small. Accordingly, to logarithmic accuracy one can neglect the dependence on $\theta_c$ inside the logarithms and thus deduce
\beq
L_{acb}L_{bca}\,\simeq\, \ln^2\frac{2(1-\cos\theta_c)}{\theta_{ab}^2}\,\simeq\, \ln^2\frac{1}{\theta_{0}^2}\,=\,\rho^2\,,\eeq
in agreement with the source term in \eqn{bmsdlaloc}.

The resummed initial condition \eqref{IC} is clearly non-physical (e.g.~the initial `probability' $P_{ab}(Y^+=0)$ can be larger than one), but from the above discussion it should be clear that this is just an artefact of reshuffling the higher-order perturbative corrections --- more precisely, of redistributing the effects of time-ordering between a local (in $Y^+$) evolution kernel and an effective initial condition at $Y^+=0$. The physical probability is eventually obtained by restricting the solution to the collinearly-improved equation to the domain at $Y^+\ge \rho$. In that domain, the solution $P_{ab}(Y^+)$ is expected to be well-defined, that is, positive semidefinite and smaller than 1.

\eqn{collBMS} is our main new result in this paper. On top of the LO BMS evolution, this equation properly resums the double collinear logarithm to all orders. In Appendix \ref{sect:nlo} we will explicitly check that the first non-trivial term in this resummation, i.e.~the piece of order $\abar\rho^2$ in the perturbative expansion of $\mathcal{K}_{\sdla}$ exhibited in \eqn{kdla}, coincides indeed with the respective piece from the NLO corrections to the BMS kernel, as obtained by evaluating the latter in the anti-collinear regime (see notably Eqs.~\eqref{hong} and \eqref{hongexact} there).  


\section{Relating space-like and time-like evolutions with collinear improvement}
\label{sec:BK}

In the previous sections, we have often mentioned the treatment of the Balitsky-Kovchegov equation as a source of inspiration for the collinear improvement of the BMS equation. We have furthermore noticed that the LO versions of these two equations are related to each other by a precise mathematical transformation --- a stereographic projection  \cite{Hatta:2008st}. This correspondence was recently shown to remain valid at NLO accuracy \cite{Caron-Huot:2015bja}, but only in the `conformal' sector (i.e.~after excluding the NLO corrections associated with the running of the QCD coupling). We recall that the BK equation describes the space-like evolution of the light-cone wavefunction of an energetic color dipole (a quark-antiquark pair in a color singlet state), as probed by the multiple scattering between the dipole and a dense target (a `nucleus'). The existence of such a mathematical correspondence between the space-like evolution of a hadronic wavefunction and the time-like evolution of a jet is highly non-trivial and suggests the existence of a deeper equivalence at a physical level.

In this section, we would like to show that, within the context of the LLA, this stereographic projection --- more precisely, its extension to a conformal mapping as introduced in  \cite{Cornalba:2007fs,Hofman:2008ar} --- correctly predicts the need for time-ordering, or, equivalently, for collinear improvement.  That is, the condition of energy ordering on one side of the correspondence is mapped onto the condition of time-ordering on the other side of the correspondence, in all the situations where an explicit time-ordering is indeed necessary. On the other hand, the {\em details} of the collinear improvement are {\em not} predicted by this correspondence, that is, one cannot use this conformal mapping to obtain the collinearly-improved BMS equation from the corresponding version of the BK equation. This is so since the double logarithms which express the effect of time-ordering order-by-order in perturbation theory explicitly break the conformal symmetry.

For what follows, it is useful to exhibit the LO BK equation. The  natural variables are the transverse coordinates, since they are not modified by the scattering at high energy. The equation describes the $Y$-evolution of the $S$-matrix $S_{\bx\by}(Y)$ for the elastic scattering between a quark-antiquark pair with transverse coordinates $\bx=(x^1,x^2)$ and $\by=(y^1,y^2)$ and a dense target. $Y$ is the rapidity separation between the right-moving projectile (the dipole) and the dense target, as measured by the rapidity difference $Y=\ln(P^+/k^+)$ between the valence quark-antiquark pair (with longitudinal momentum $P^+$) and the softest gluon from the dipole wavefunction which participates in the scattering (with comparatively small longitudinal momentum $k^+\ll P^+$). The BK equation reads   \cite{Balitsky:1995ub,Kovchegov:1999yj}
\begin{align}\label{BK}
 \frac{\del }{\del Y} \,S_{\bx\by}(Y)=\abar
 \int  \frac{\rmd^2\bz}{2\pi}\, 
\frac{(\bx-\by)^2}{(\bx-\bz)^2(\bz-\by)^2}\,\Big[
 S_{\bx\bz}(Y) S_{\bz \by}(Y)
 -S_{\bx\by}(Y)\Big]\,,
 \end{align}
where $\bz$ denotes the transverse coordinate of the soft gluon emitted in one step of the evolution and $\abar=\alpha_s N_c/\pi$ as before. This equation is strictly valid in the large-$N_c$ limit where the soft gluon emission can be described as the splitting of the original dipole $(\bx,\by)$ into two new dipoles $(\bx,\bz)$ and $(\bz,\by)$, which then can independently scatter off the target. The constant function $S_{\bx\by}=1$ is a fixed point of \eqn{BK}, but this special value is only achieved in the absence of any scattering (or target). Conversely, in the presence of a non-trivial scattering, this equation must be solved with an initial condition $S_{\bx\by}(Y_0) < 1$ at the rapidity scale $Y_0$ where one starts the evolution. Then the solution will obey $0< S_{\bx\by}(Y) <1$ at any $Y>Y_0$.

The formal similarity between the evolution term in the BMS equation \eqref{BMSLO} and the BK equation \eqref{BK} is manifest by inspection. In particular, in the limit where all the angles in the BMS problem are small, $\theta_{ab}\ll 1$, $\theta_{ac}\ll 1$, $\theta_c\ll 1$, etc., the antenna kernel takes the simpler form shown in \eqn{wabcboost}; this becomes identical with the dipole kernel in the BK equation provided one identifies relative angles with transverse separations: $\theta_{ab}^2\to (\bx-\by)^2$, $\theta_{ac}^2\to (\bx-\bz)^2$, etc. Similarly, for small polar angles, the solid-angle measure
$\rmd \Omega_c\simeq \rmd \phi_c\,\theta_c\rmd\theta_c$ becomes formally similar to the flat measure in the transverse plane. We thus see emerging a precise mathematical correspondence between the two problems, which will be discussed in more details and for arbitrary angles  in the next section.

\comment{

An explicit mathematical correspondence between the two equations can be easily written down in the limit where all the angles on the BMS side are small, and similarly for the transverse separations on the BK side.

as is indeed the case for the typical evolution in the COM frame (cf.~Fig.~\ref{fig:Cout} right). In that case, the product of the measure and the antenna kernel in  \eqn{BMSLO} can be approximated as
\beq\label{dOwsmall}
\int \frac{\rmd \Omega_c}{4\pi}\,w_{abc}\,\simeq\,\int \frac{\rmd \phi_c\,\theta_c\rmd\theta_c}{2\pi}\,
\frac{\theta_{ab}^2}{\theta_{bc}^2\theta_{ac}^2}\,.
\eeq
If one formally extends the integration over $\theta_c$ to $0\le\theta_c<\infty$, then the r.h.s. of \eqn {dOwsmall} becomes formally identical with the corresponding product in the BK equation \eqref{BK} if one identifies transverse coordinates with angles: $\bz\to {\bm \theta}_c$, $\bx\to {\bm \theta}_a$,
etc (which of course implies $(\bx-\by)^2\to \theta_{ab}^2$ etc). By ${\bm \theta}_c$, we understand the vector in the 2-dimensional transverse plane with modulus $\theta_c$ and azimuthal angle $\phi_c$.
In the next section, we shall present the generalization of this transformation to arbitrary values for the angles
Strictly speaking, the correct relation between transverse coordinates and angles is different from the one discussed after \eqn {dOwsmall} even in the small angle limit
}

\subsection{The conformal mapping}
\label{sec:stereo}

\begin{figure}[t] \centerline{
\includegraphics[width=.8\textwidth]{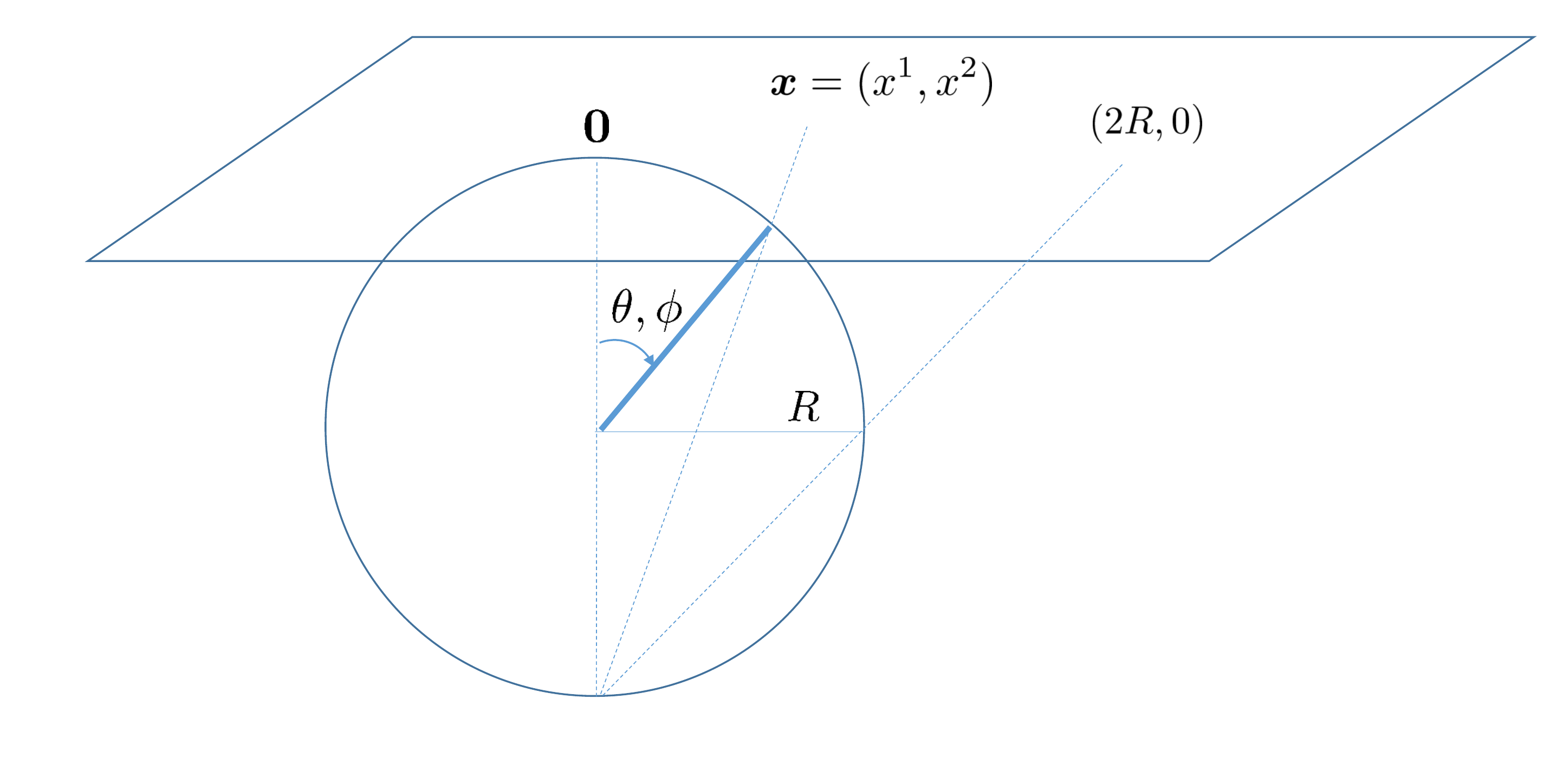}}
 \caption{\small The geometry of the stereographic projection. The `transverse' plane is tangent
 to the sphere at the `North Pole', while the projection point is the `South Pole'. The Equator on the sphere is projected on a circle in the plane with radius equal to $D=2R$.}
 \label{fig:stereo}
\end{figure}

Amply used in cartography, the stereographic projection is a mapping that projects a sphere onto a plane tangent to it. The projection point is the pole on the sphere opposite to the tangent plane (see Fig.~\ref{fig:stereo}). This projection is conformal, that is, it preserves angles at which the curves meet.

Consider a point on the sphere with angular coordinates $\Omega=(\theta,\phi)$. The point on the tangent plane which is associated to it has coordinates $\bx=(x^1,x^2)$, with [we recall that $\tan(\theta/2)=\sin\theta/
(1+\cos\theta)$]
\beq
x^1=D\,\frac{\sin \theta \cos \phi}{1+\cos \theta}\,,
\quad x^2 =D\,\frac{\sin \theta \sin \phi}{1+\cos \theta}\,, \label{ste}
\eeq
where $D=2R$ is the diameter of the sphere. The actual value of $D$ is irrelevant for what follows, since it would anyway cancel out in the scale-invariant mapping between the BMS and BK equations. With this is mind, we shall henceforth choose $D=1$, to simplify writing. The transformation \eqref{ste} can be easily inverted to yield (with $D=1$, as mentioned)
 \beq\label{steinv}
 \cos \theta=\frac{1-\bx^2}{1+\bx^2}, \quad \sin \theta=\frac{2|\bx|}{1+\bx^2},
\quad \cos \phi=\frac{x^1}{|\bx|}, \quad
 \sin \phi=\frac{x^2}{|\bx|}\, .
\eeq
  The squared length transforms as
\beq
 (\rmd\bx)^2=\frac{1}{(1+\cos \theta)^2}(\rmd\theta^2 + \sin^2 \theta d\phi^2)
\equiv \frac{1}{(1+\cos \theta)^2}\rmd\Omega^2\,, \label{ste2}
\eeq
and the area element as
 \begin{align}
 \rmd^2\Omega=(1+\cos \theta)^2 \rmd^2\bx =\frac{4}{(1+\bx^2)^2}\rmd^2\bx \, . \label{area}
 \end{align}
 
 Two limiting cases will be interesting in what follows. When the polar angle $\theta$ is small, $\theta\ll 1$, the transverse coordinate of the projection point is small as well, $|\bx|\ll 1$, and the above equations imply $\theta\simeq 2|\bx|$. On the other hand, when $\theta$ is close to $\pi$, \eqn{steinv} implies $|\bx|\gg 1$ and $\pi-\theta\simeq 2/|\bx|$.
 
Let us now check that this stereographic projection indeed relates the BMS and BK equations, as anticipated. We shall identify a point  $\Omega=(\theta,\phi)$ on the sphere with the direction of a parton in the final state of the jet and the corresponding projection point $\bx$, cf.~\eqn{ste}, with the transverse coordinate of a parton from the dipole wavefunction. Writing $\cos\theta_{ab}=\vec n_a\cdot\vec n_b$, with $\vec n_a=(\sin\theta_a\cos\phi_a,\, \sin\theta_a\sin\phi_a,\,\cos\theta_a)$ etc., and then using \eqn{steinv}, one easily finds
 \beq\label{conf}
 1-\cos\theta_{ab}\,=\,\frac{2(\bx_a-\bx_b)^2}{(1+\bx_a^2)(1+\bx_b^2)}\,,\eeq
 which together with \eqn{area} for the transformation of the integration measure immediately implies
 \beq\label{dOw}
\int \frac{\rmd \Omega_c}{4\pi}\,\frac{1- \cos\theta_{ab}}{(1- \cos\theta_{ac})(1- \cos\theta_{cb})}\,
=\int  \frac{\rmd^2\bx_c}{2\pi}\, 
\frac{(\bx_a-\bx_b)^2}{(\bx_a-\bx_b)^2(\bx_b-\bx_c)^2}\,.
\eeq
Notice that all the would-be scale dependent factors like $(1+\bx_c^2)$ have compensated in the global transformation, as previously mentioned.

\eqn{dOw} establishes the sought-for correspondence between the angular distribution and the transverse coordinate distribution in the LO BMS and respectively BK equations. Here however we will need an additional piece of information, which refers to the respective energy (or rapidity) phase-spaces, and which follows from a more general, 4-dimensional, conformal mapping that encompasses the stereographic projection \cite{Cornalba:2007fs,Hofman:2008ar}. We shall not need all the details of this mapping (see \cite{Cornalba:2007fs,Hofman:2008ar,Avsar:2009yb} for more general discusisons), but only the correspondence implied by it between the {\em energy density in the jet} --- in the sense of energy ${\mathcal E}_{J}(\Omega)$ per unit solid angle --- and the  {\em energy density in the `hadron'} (the dipole together with its descendants) --- in the sense of longitudinal momentum ${\mathcal E}_H(\bx)$ per unit transverse area.

For what follows, it is useful to keep in mind the following difference between the BMS and the BK equations. As already discussed, the BMS equation  \eqref{BMSLO} holds as it stands in {\em any frame} --- in particular, in the COM frame of the original quarks, or in a boosted frame where both quarks have large longitudinal momenta. In any such a frame, the rapidity variable $Y$ can be computed as $Y=\ln(1/x)$, with $x$ the {\em energy} fraction carried by a gluon from the evolution\footnote{This energy fraction is however not boost invariant, which explains why the {\em range} for $Y$ is generally different in different frames, as explained at the end of Sect.~\ref{sec:collCOM}.}. By contrast, the BK equation \eqref{BK} is necessarily written in a `boosted' frame in which the dipole has a large longitudinal momentum $P^+$; the respective rapidity $Y=\ln(P^+/k^+)$ is computed from the {\em longitudinal momentum} fraction $x=k^+/P^+$. 

The total momentum $P^+$ of the hadron can be computed as
\beq
P^+_H=\int \rmd^2\bx\int_{-\infty}^{\infty} \rmd x^-  T^{++}(x^+=0,x^-,\bx)\equiv
\int \rmd^2\bx\, {\mathcal E}_H(\bx).
\eeq
The conformal mapping allows one to compute the total 4-momentum of the jet in terms of the hadronic energy density ${\mathcal E}_H(\bx)$ \cite{Hofman:2008ar} :
 \beq\label{PJ}
 P^+_J=\int \rmd^2\bx\, {\mathcal E}_H(\bx),\qquad
 P^-_J=\int \rmd^2\bx \,\bx^2 {\mathcal E}_H(\bx),\qquad
 {\bm P}_J =\sqrt{2}\int \rmd^2\bx\, \bx \,{\mathcal E}_H(\bx)\, . 
 \eeq
From these relations, the jet energy is obtained as
 \beq\label{EJ}
 E_J \equiv \int \rmd^2\Omega \,  {\mathcal E}_J(\Omega)= \frac{1}{\sqrt{2}}(P^+_J+P^-_J) =
 \frac{1}{\sqrt{2}}\int \rmd^2\bx\, (1+\bx^2)\, {\mathcal E}_H(\bx)\, .
 \eeq
 By also using the transformations \eqref{steinv}--\eqref{area}, one can deduce the following relation between the respective energy densities:
  \beq
{\mathcal E}_J(\Omega) = \frac{\sqrt{2}}{(1+\cos \theta)^3}\,{\mathcal E}_H(\bx)\,. \label{kon}
\eeq
However, this (rigorous) relation is not the most useful one for our present purposes. Rather, we shall use the {\em global} relations \eqref{PJ}--\eqref{EJ} to heuristically infer a correspondence between the 4-momenta of the {\em individual} gluons in the (time-like and space-like) cascades. This reads
\beq\label{JH}
p^+ \leftrightarrow k^+,\qquad p^- \leftrightarrow \bx^2 k^+,\qquad \bp \leftrightarrow \sqrt{2} \bx\, k^+\,,
\eeq
where we employ the convention that the 4-momentum of a gluon in the time-like cascade is denoted by $p^\mu$, while that in the space-like cascade is denoted by $k^\mu$. Notice that these relations are indeed consistent with the mass-shell condition $p^-=\bp^2/2p^+$. For what follows, it is perhaps more suggestive to rewrite these relations in terms of the respective energies, $p^0=(p^++p^-)/\sqrt{2}$ and respectively $k^0\simeq k^+/\sqrt{2}$. One finds
\beq
\bp&\leftrightarrow& 2\bx \, k^0\,, \label{11}\\
p^0 &\leftrightarrow& (1+\bx^2)k^0\,, \label{22}
\eeq
As a further check of these equations, we note that they are consistent with the trigonometrical relation 
\begin{align}
\frac{|\bp|}{p^0}=\sin \theta =
\frac{2|\bx| }{1+\bx^2}\,.
\end{align}

The relation \eqref{22} between the parton energies in the two types of cascades will play an important role in what follows. To better appreciate its consequences, let us consider 2 limiting cases:

\texttt{(i)} The parton in the time-like evolution of the jet propagates at a small polar angle $\theta\ll 1$. This is the typical situation when the jets are boosted along the positive $z$ axis, like the hadron in the space-like problem. In this case, the stereographic projection predicts that the  transverse position of the corresponding `space-like' parton is small as well, $|\bx|\simeq\theta/2 \ll 1$, hence \eqn{22}  implies that the respective energies can be simply identified with each other: $p^0  \leftrightarrow k^0$.  Incidentally, the fact that  $|\bx|\sim \theta \sim 1/\gamma$, with $\gamma$ the parton boost factor, suggests that the factor $\bx^2$ in the relation \eqref{JH} between $p^-$ and $k^+$ can be viewed as the expression of Lorentz contraction.

\texttt{(ii)} The parton in the time-like evolution of the jet propagates along the negative $z$ direction, that is $\theta\simeq \pi$.  In that case, $|\bx|\simeq 2/(\pi-\theta)\gg 1$ and  \eqn{22} reduces to $p^0 \leftrightarrow \bx^2 k^0$. By the uncertainty principle, one has $\bx^2\sim 1/\bk^2$, with $\bk$ the transverse momentum of the parton from the space-like wavefunction. Then the quantity $\bx^2 k^0
\sim k^+/\bk^2$ is recognised as the {\em lifetime} of that parton. We thus see that, quite remarkably,  the energy distribution of the partons in the time-like cascade provides information about the time-ordering in the corresponding space-like cascade. This information will be discussed in more detail in the next section.

\subsection{Connecting space-like and time-like evolutions}
\label{sec:map}

In what follows, we shall use the conformal mapping to study the correspondence between space-like and time-like evolutions in various set-ups. We shall consider the two situations for jet evolution that we already discussed --- di-jets in the COM frame and in a boosted frame, respectively --- together with two related configurations that will bring some new features.

\begin{figure}[t] \centerline{
\includegraphics[width=.7\textwidth]{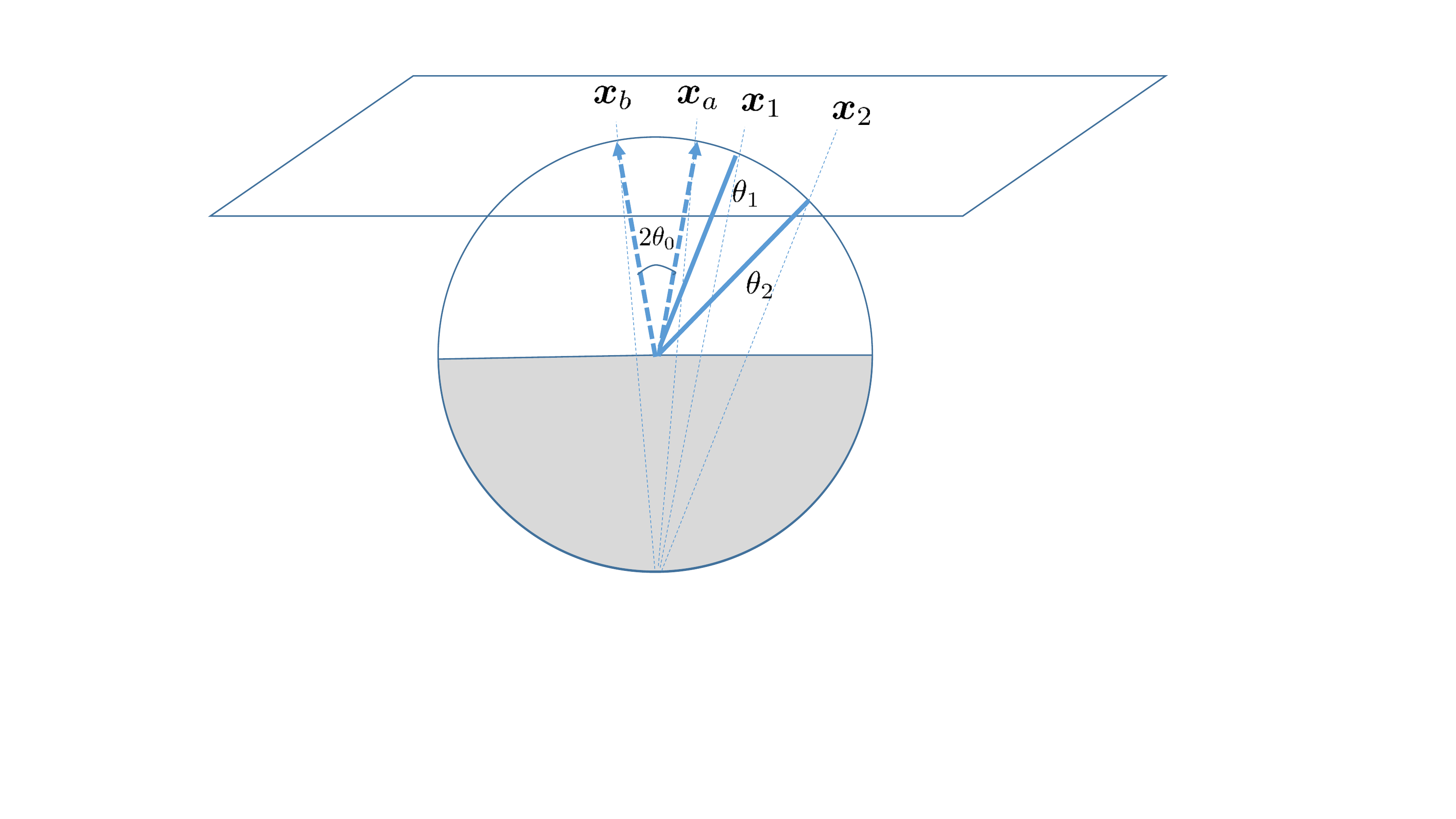}}
 \caption{\small The stereographic projection for the case where the di-jets are boosted along the positive $z$ axis. The original quark and antiquark, as represented by dashed lines, make a small angle $2\theta_0$ and are projected onto the points with coordinates $\bx_a$ and $\bx_b$ in the transverse plane. Two successive soft gluon emissions, with increasing angles $\theta_0\ll \theta_1\ll \theta_2$, are projected onto points with increasing transverse coordinates, $|\bx_a-\bx_b|\ll |\bx_1|\ll |\bx_2|$. The projection relates the anti-collinear time-like evolution to the anti-collinear space-like evolution.}
 \label{figI}
\end{figure}

I. {\em Di-jets boosted along the positive $z$ axis \& excluded region at $\pi/2 < \theta <\pi$.} In this case, the two original quarks make a small angle $\theta_0=1/\gamma\ll 1$ w.r.t.~the $z$ axis: $\theta_a=\theta_b=\theta_0$. (The corresponding azimuthal angles can be chosen as $\phi_a=0$ and $\phi_b=\pi$; see Fig.~\ref{figI}.)  From the previous sections, we know that the dominant emissions --- those which matter in the double-logarithmic approximation (DLA) --- are such that the successive angles are strongly increasing, yet they remain small: $\theta_0\ll \theta_1\ll \theta_2\ll \dots\ll 1$. The 
stereographic projection implies a similar ordering for the corresponding dipole sizes: $|\bx_0|\ll
|\bx_1|\ll |\bx_2|\ll \dots\ll 1$, where $|\bx_0|=|\bx_a-\bx_b|/2$ (half of the size of the original dipole). We recall that a condition like $|\bx|\ll 1$ truly means $|\bx|\ll D$, with $D=2R$ the diameter of the projection sphere. But as we shall shortly see, the actual value of $R$ plays no role, it is only the ordering of the dipole sizes that matters. Such an evolution from small to large dipole sizes (``hard-to-soft'' or ``anti-collinear'') is indeed the typical dipole evolution in the case of asymmetric, dilute-dense, collisions (like in the applications of the BK equation to deep inelastic scattering or proton-nucleus collisions).
 
Still from the previous discussion in this paper, we know that the LO BMS evolution of the boosted jets might violate the proper time ordering --- the fact that the formation times $\tau_i \simeq {1}/({p^0_i \theta_i^2})$ must increase from one emission to the next one ---, hence this condition must be enforced by hand. A similar discussion applies to the BK evolution, but in that context it refers to the {\em lifetime} of the space-like fluctuations $\Delta\tau_i\sim  \bx_i^2 k^+_i$, which must {\em decrease} along the cascade: $\Delta\tau_{i+1} < \Delta\tau_i$. This condition might be violated by the LO BK evolution, which proceeds with decreasing $k^+$ and increasing transverse sizes. As we now explain, if the need  for time-ordering has been properly understood on one side of the correspondence, then via the conformal mapping it is also predicted for the other side.

\begin{figure}[t] \centerline{
\includegraphics[width=.95\textwidth]{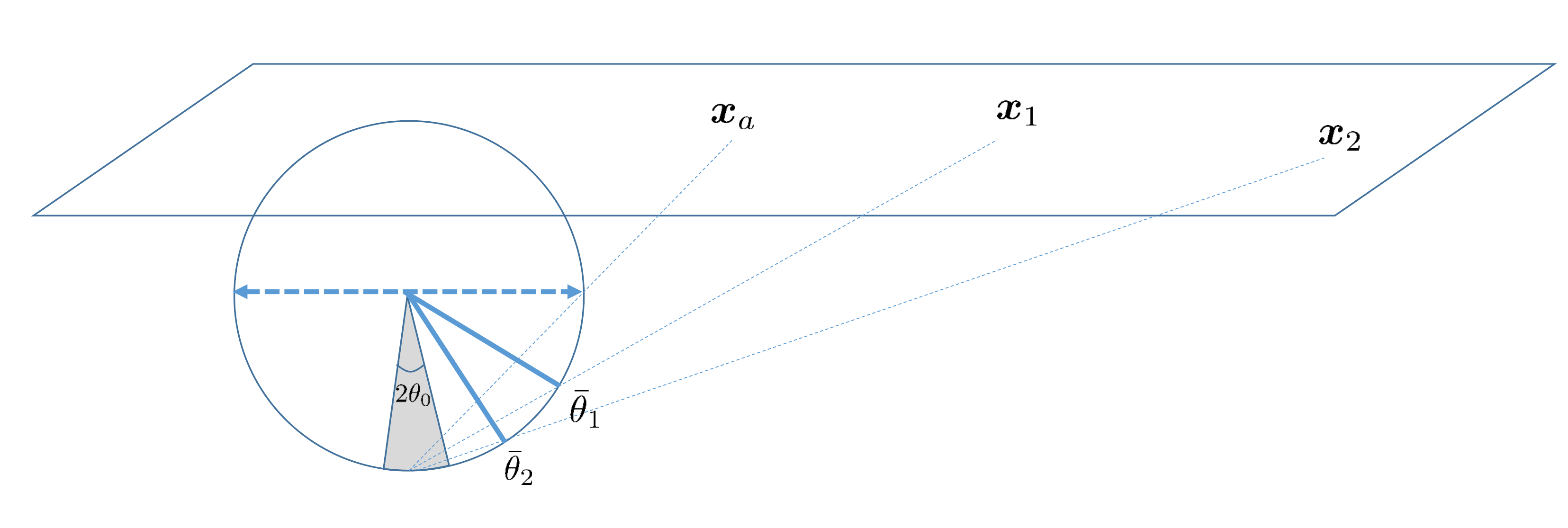}}
 \caption{\small The stereographic projection in the COM frame of the di-jets and for an exclusion region making a small angle $\theta_0$ around the negative $z$ axis. The successive, soft, gluon emissions make smaller and smaller angles w.r.t.~the negative $z$ axis, $\theta_0\ll \bar\theta_2\ll \bar\theta_1$. They are projected onto points with increasing transverse coordinates in the tangent plane: $|\bx_a-\bx_b|\ll |\bx_1|\ll |\bx_2|$. The projection relates the collinear time-like evolution to the anti-collinear space-like evolution.}
 \label{figII}
\end{figure}

Notice first that, since $|\bx|\ll 1$, the energy correspondence in \eqn{22} instructs us to simply identify the gluon energies in the two problems, $p^0\leftrightarrow k^0$. This is of course consistent with the first correspondence in \eqn{JH} together with the fact that, in this boosted frame, we have $p^0\simeq p^+/\sqrt{2}$ and $k^0\simeq k^+/\sqrt{2}$. Assume now that, in view of our previous experience with the BK equation, we know that successive gluon emissions in the dipole wavefunction in dilute-dense scattering must be ordered with decreasing lifetime $\Delta\tau \sim  \bx^2 k^+$. The 4-dimensional conformal mapping identifies $\bx^2 k^+ \leftrightarrow \theta^2 p^+/4 $ and thus instructs us that the corresponding jet evolution should be ordered with {\em increasing} formation time $\tau \sim {1}/{p^+ \theta^2}$. We thus see that, via this conformal mapping, we could have anticipated the need for collinear improvement in the BMS evolution on the basis of the corresponding improvement of the BK equation  \cite{Beuf:2014uia,Iancu:2015vea,Iancu:2015joa}.

II. {\em Di-jets in the COM frame \& small exclusion region around $\theta=\pi$.} Now $\theta_a=\theta_b=\pi/2$ and the exclusion region itself makes a small angle $\theta_0$ w.r.t.~the negative $z$ axis, see Fig.~\ref{figII}. The emissions which matter at DLA are those which accumulate towards $\theta_0$: writing $\bar\theta_i\equiv \pi-\theta_i$, one has $\theta_0\ll\cdots\ll \bar\theta_2
\ll \bar\theta_1 \ll 1$. The transverse coordinates $|\bx_i|\simeq 2/\bar\theta_i$ in the corresponding space-like evolution are again strongly increasing from one emission to the next one, but now they are all large: $1 \ll |\bx_1|\ll |\bx_2|\ll \dots\ll 2/\theta_0$. (Notice that, in these dimensionless variables, the original quark and antiquark legs of the dipole have $|\bx_a|=|\bx_b|=1$ and $|\bx_a-\bx_b|=2$; see Fig.~\ref{figII}.) The energy correspondence \eqref{22} therefore implies $p^0_i  \leftrightarrow 
\bx_i^2 k^0_i$. As already discussed, the jet evolution with decreasing $p^0$ and decreasing $\bar\theta$ automatically obeys the correct time-ordering. But the above mapping of the energy variables implies that, on the space-like side, the evolution of the dipole should be ordered with {\em decreasing lifetimes} $\Delta\tau_i\sim  \bx_i^2 k^0_i$.

\begin{figure}[t] \centerline{
\includegraphics[width=.9\textwidth]{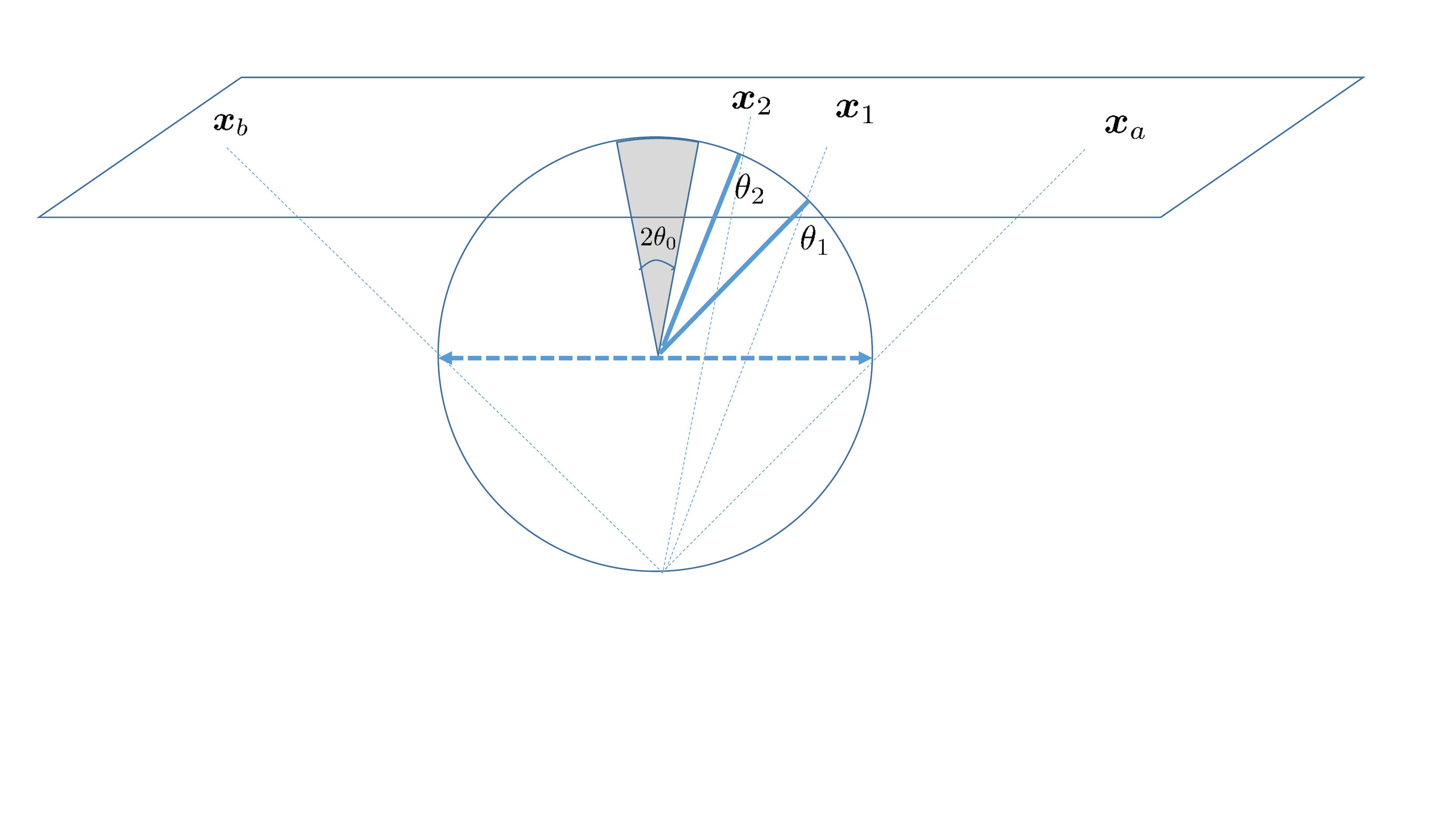}}
 \caption{\small The stereographic projection in the COM frame of the di-jets and for an exclusion region making a small angle $\theta_0$ around the positive $z$ axis. The successive, soft, gluon emissions make smaller and smaller angles w.r.t.~the positive $z$ axis, $\theta_0\ll \theta_2\ll \theta_1$. They are projected onto points with decreasing transverse coordinates in the tangent plane: $|\bx_2|\ll |\bx_1|\ll |\bx_a-\bx_b|$. The evolution is collinear on both sides of the correspondence.}
 \label{figIII}
\end{figure}

This discussion is in agreement with the fact that the problem of the dipole evolution is actually {\em the same} in the two above configurations I and II, albeit the respective jet problems are indeed different (one requires time-ordering, the other one does not) and in spite of the fact that all the dipole sizes are ``small'' ($|\bx_i|\ll 1$) in case I, but ``large'' ($|\bx_i|\gg 1$) in case II. This confirms the fact that what matters at a physical level is the direction of the evolution --- from small to large dipoles in both problems above --- and not the absolute sizes of the dipoles predicted by the stereographic projection (which depend upon the unphysical parameter $D$). Note also that the Lorentz boost which relates the two jet problems, I and II, and which has indeed physical consequences for the time-like evolution, corresponds via the conformal mapping to an overall dilation of the dipole sizes, which is totally irrelevant as it can be undone via a rescaling of the parameter $D$.

III. {\em Di-jets in the COM frame \& small exclusion region around $\theta=0$.} From the viewpoint of the jet dynamics, this situation, which is illustrated in Fig.~\ref{figIII}, is physically equivalent to that discussed at  point II above: it is obtained from the latter via a reflection w.r.t.~the $(x,y)$ plane
(the plane $\theta=\pi/2$ crossing the projection sphere along its diameter). Hence, clearly, the dominant evolution consists in small angle emissions which accumulate towards the positive $z$ axis: $\theta_0\ll\cdots\ll \theta_2 \ll \theta_1 \ll 1$. Yet, the dipole evolution which is associated to it by the conformal mapping is opposite to that described at point II: the transverse sizes $|\bx_i|\simeq\theta_i/2$ are small and {\em strongly decreasing} from one emission to the next one: $ 2/\theta_0 \ll \dots\ll |\bx_2|\ll |\bx_1|\ll  1$. Physically, this ``collinear'' (or  ``soft-to-hard'') evolution is realized in situations where the target size $r_0$ is much smaller than the size $r=|\bx_a-\bx_b|$ of the original $q\bar q$ pair from the projectile: $r_0\ll r$. More precisely, the situation presented here corresponds to the case where the small target with size $r_0$ is located (in the transverse plane) at the middle of the `big' projectile dipole --- indeed, the transverse coordinates $\bx_i$ accumulate towards $\bx=0$. This is the counterpart of the fact that, in the corresponding jet problem, the excluded region makes a small angle around an axis which is perpendicular on the jet axis.

For such collinear evolutions, the proper time-ordering conditions are automatically satisfied --- both for the jet problem and for the dipole one ---, so one can use the respective energies as the right ordering variables. This is indeed consistent with the conformal mapping, which simply identifies the gluon energies in the two problems: $p^0\leftrightarrow k^0$.

\begin{figure}[t] \centerline{
\includegraphics[width=1.\textwidth]{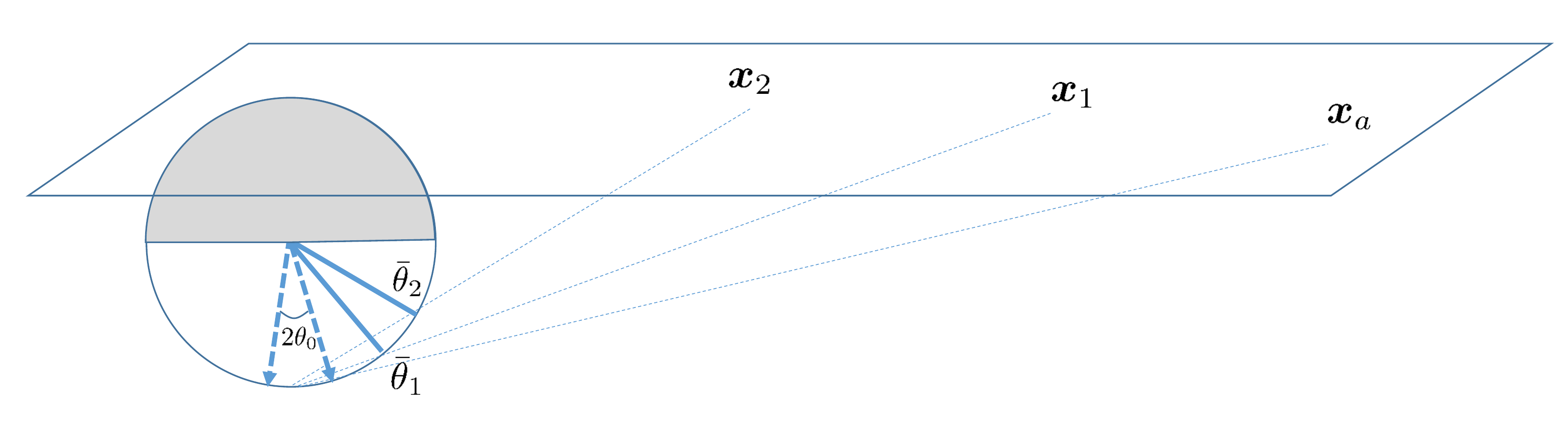}}
 \caption{\small The stereographic projection for the case where the di-jets are boosted along the negative $z$ axis. Two successive soft gluon emissions, which make increasing angles $\theta_0\ll \bar\theta_1\ll \bar\theta_2$ w.r.t.~the negative $z$ axis, are projected onto points with decreasing transverse coordinates, $|\bx_2|\ll |\bx_1|\ll |\bx_a-\bx_b|$. The projection relates the anti-collinear time-like evolution to the collinear space-like evolution.}
 \label{figIV}
\end{figure}

IV. {\em Di-jets boosted along the negative $z$ axis \& excluded region at $0 < \theta <\pi/2$.} This situation (see  Fig.~\ref{figIV}) is similar to that discussed at the previous point, in that the jet evolution is not new --- it is equivalent to that discussed at point I (so in particular it requires time-ordering) ---, whereas the dipole evolution proceeds towards smaller and smaller dipoles sizes, hence it automatically fulfils the proper time-ordering for a space-like cascade. We have indeed $|\bx_i|\simeq 2/\bar\theta_i\gg 1$ where $\bar\theta_i\equiv \pi-\theta_i$ is small, but strongly increasing from one emission to the next one. Once again, the need for time ordering on the jet side is correctly predicted by the conformal mapping: using $p^0  \leftrightarrow 
\bx^2 k^0$ and $\bx^2 \leftrightarrow 4/\bar\theta^2$, we conclude that the ordering with decreasing $k_0$ in the space-like evolution corresponds to an ordering with increasing lifetime $\tau\sim 1/(p^0\bar\theta^2)$ in the time-like evolution.

\comment{

\begin{table}
\begin{tabular}{|l|c|r|r|} \hline
& BK & BMS \\ \hline
I & Yes & Yes \\
II & Yes & No \\
III & No & No \\
IV & No & Yes \\ \hline
\end{tabular}
\end{table}
}

Since the need for time-ordering is correctly predicted by the conformal mapping, one may wonder whether the collinearly-improved versions of the BMS and BK equations can be directly related to each other via this mathematical operation. For instance, could one obtain the evolution term in the resummed BMS equation \eqref{collBMS} by directly applying the conformal mapping to the corresponding term in the resummed BK equation, that is, Eq.~(32) in Ref.~\cite{Iancu:2015vea} ? The answer is no, as we argue now: the double collinear logarithms which are concerned by these resummations are not invariant under conformal transformations (they break the invariance under inversion). To see this, consider the argument $\rho\equiv \sqrt{L_{acb}L_{bca}}$ of the corrective factor ${\cal K}_{\sdla}$ in the BMS kernel in \eqn{collBMS}. This is rewritten here for convenience:
\beq
\label{rho2}
\rho^2 = \ln \frac{1- \cos\theta_{ac}}{1-\cos\theta_{ab}}\,
\ln \frac{1- \cos\theta_{bc}}{1-\cos\theta_{ab}}.
\eeq
Via the conformal transformation \eqref{conf}, this double angular logarithm gets mapped onto
\beq
\label{rho2x}
\rho^2  \longleftrightarrow 
 \ln \frac{(\bx_a-\bx_c)^2(D^2+ \bx_b^2)}{(\bx_a-\bx_b)^2(D^2 + \bx_c^2)}\,
\ln \frac{(\bx_b-\bx_c)^2(D^2 +\bx_a^2)}{(\bx_a-\bx_b)^2(D^2 + \bx_c^2)}\,,
\eeq
where we have reintroduced a generic value $D$ for the diameter of the projection sphere, for more clarity. Clearly, the r.h.s. of \eqn{rho2x} is not satisfactory in that it depends upon the unphysical scale $D$. This reflects the lack of conformal symmetry of \eqn{rho2}, as anticipated. The would-be argument of  ${\cal K}_{\sdla}$ in the resummed BK equation is the same as the limit of  \eqn{rho2x} when $D\to \infty$, namely
\beq
\label{rhoBK}
\rho^2\Big |_{\rm BK}=\, \ln \frac{(\bx_a-\bx_c)^2}{(\bx_a-\bx_b)^2}\,
\ln \frac{(\bx_b-\bx_c)^2}{(\bx_a-\bx_b)^2}\,.
\eeq
This is merely the statement that the collinearly-improved BK and BMS equations can indeed be related to each other in the limit where all the angles on the BMS side are {\em small}. In that case, the conformal transformation \eqref{conf} reduces to
\beq\label{simpleconf}
\theta_{ab}^2\,\leftrightarrow \,4(\bx_a-\bx_b)^2\,,\eeq
which correctly relates the respective limits of the variables $\rho^2$ in both problems. This covers our cases I and III above, in which {\em both} evolutions obey indeed the same pattern: the variables $\rho^2$ are both large in case I (and then they require all-order resummations), but they are both small, i.e.~of $\order{1}$, in case III, when no improvement is needed.
On the other hand, in the more interesting situations at point II and IV, where one of the two evolutions requires collinear-improvement but the other one does not, it is not possible to relate the respective versions of the evolution equations via a conformal mapping. For instance, in case IV, the double logarithm angular  for the first emission can be estimated as (recall that $\bar\theta_c\equiv \pi-\theta_c$ is much larger than the original angle  $\theta_{ab}=2\theta$)
\beq\label{rho2IV}
\rho^2 \,\simeq\, \ln^2 \frac{\bar\theta_{c}^2}{\theta_{0}^2}\,.
\eeq
This is large, $\rho^2\gg 1$, which signals the need for collinear resummation. The prediction of the conformal mapping for the corresponding double transverse log, that is (cf.~\eqn{rho2x} where $|\bx_a|=|\bx_b|=2/\theta_0 \gg |\bx_c| \gg D$ and $|\bx_a-\bx_b|=2|\bx_a|$)
\beq
\rho^2  \,\simeq\,
 \ln^2 \frac{ \bx_a^2}{\bx_c^2}\,.\eeq
This is large too (it is numerically the same as \eqn{rho2IV}, by construction), which would suggest the need for resummation. But this conclusion would be wrong, as we know by now. And indeed the correct result for the double transverse logarithm, cf.~\eqn{rhoBK}, is actually of $\order{1}$.

\section*{Acknowledgments}
\vspace*{-0.3cm}
We would like to thank Gr\'egory Soyez for discussions and comments on the manuscript.
Y.H., A.H.M.~and D.N.T.~would like to acknowledge l'Institut de Physique Th\'eorique de Saclay for 
hospitality during the early stages of this work.
 The work of E.I.~is supported in part by the Agence Nationale de la Recherche project 
 ANR-16-CE31-0019-01.   The work of A.H.M.~is supported in part by the U.S. Department of Energy Grant \# DE-FG02-92ER40699. Figures were made with Jaxodraw \cite{Binosi:2003yf}.

\appendix

\section{NLO kernels and double collinear logarithms}
\label{sect:nlo}

As we have seen in the main text, time ordering for the emissions in the boosted frame leads to a resummation of large double collinear logarithms to all orders in $\abar$. Here we shall see how the lowest order double logarithm of such a series emerges from the complete NLO calculation of \cite{Caron-Huot:2015bja}. The result there has been given in terms of an evolution Hamiltonian which can act on desired observables and determine the evolution of jets. This Hamiltonian is expanded as
\begin{equation}
\label{hexpand}
	H = -\frac{\alpha_s}{4\pi} K^{(1)} 
	- \frac{\alpha_s^2}{16\pi^2} K^{(2)}, 
\end{equation}
where the NLO piece in the non-conformal scheme is given as a sum of three terms to be found in Eq.~(3.20) of \cite{Caron-Huot:2015bja}. Only the part of the first term which contains one or no adjoint unitary matrices is relevant for our purposes and we rewrite it here as
\begin{align}
	\label{k2unou}
	K^{(2)} \supset \sum_{a,b,c}
	\int \frac{\dif^2 \Omega_1}{4\pi}
	\frac{\dif^2 \Omega_2}{4\pi}
	K^{(2)}_{abc;12}
	\rmi f^{\ssA\ssB\ssC}
	\Big[
	&-L_a^{\ssA'} U_1^{\ssA'\ssA} R_b^{\ssB} R_c^{\ssC}
	-R_a^{\ssA} L_b^{\ssB'} U_2^{\ssB'\ssB} R_c^{\ssC}
	+L_a^{\ssA} U_2^{\ssB\ssB'} R_b^{\ssB'} L_c^{\ssC}
	\nn
	&+U_1^{\ssA\ssA'} R_a^{\ssA'} L_b^{\ssB} L_c^{\ssC}
	+R_a^{\ssA} R_b^{\ssB} R_c^{\ssC}
	-L_a^{\ssA} L_b^{\ssB} L_c^{\ssC}
	\Big],
\end{align}   
where the kernel $K^{(2)}_{abc;12}$ reads (cf.~Eq.~(3.12a) of \cite{Caron-Huot:2015bja})  
\begin{equation}
	\label{k2nc}
	K^{(2)}_{abc;12} = 
	\frac{1}{\alpha_{1a}\alpha_{2b}}
	\left( 
	\frac{\alpha_{ab}}{\alpha_{12}}
	+\frac{\alpha_{ac}\alpha_{bc}}{\alpha_{1c}\alpha_{2c}}
	-\frac{\alpha_{1b}\alpha_{ac}}{\alpha_{1c}\alpha_{12}}
	-\frac{\alpha_{bc}\alpha_{2a}}{\alpha_{12}\alpha_{2c}}
	\right)
	\ln \frac{\alpha_{1c}^2}{\alpha_{2c}^2},
\end{equation}
with the notation $\alpha_{ab} = (1- \cos\theta_{ab})/2$. In general, the left and right Lie derivatives in \eqn{k2unou} obey 
\begin{equation}
 	[R_a^{\ssA},R_b^{\ssB}] = 
 	\rmi \delta_{ab} f^{\ssA\ssB\ssC} R_{a}^{\ssC}, 
 	\quad
 	[L_a^{\ssA},L_b^{\ssB}] = 
 	-\rmi \delta_{ab} f^{\ssA\ssB\ssC} L_{a}^{\ssC},
 	\quad
    [R_a^{\ssA},L_b^{\ssB}] =0, 
\end{equation}
and act on unitary matrices\footnote{They don't act on those $U$'s appearing in the Hamiltonian.} according to 
\begin{equation}
	L_a^{\ssA} U_d = \delta_{ad} T^{\ssA} U_a, \quad
	R_a^{\ssA} U_d = \delta_{ad} U_a T^{\ssA}, \quad
	L_a^{\ssA} U_d^{\dagger} = - \delta_{ad} U_a^{\dagger} T^{\ssA}, \quad
	R_a^{\ssA} U_d^{\dagger} = - \delta_{ad} T^{\ssA} U_a^{\dagger},
\end{equation} 
where $U$ in the above can belong to an arbitrary  representation of SU$(N_c)$ and $T^{\ssA}$ are the respective generators.

Let us focus on the terms of the Hamiltonian which contain two right derivatives, one left and one $U$-matrix, i.e.~on the first two terms in \eqn{k2unou}. Given the property $K^{(2)}_{bac;21} = - K^{(2)}_{abc;12}$ of the kernel in \eqn{k2nc}, the two terms contribute equally, thus
\begin{equation}
	\label{k2ulrr}
	K^{(2)}_{\rm ULRR} = -2 \sum_{a,b,c}
	\int \frac{\dif^2 \Omega_1}{4\pi}
	\frac{\dif^2 \Omega_2}{4\pi}
	K^{(2)}_{abc;12}
	\rmi f^{\ssA\ssB\ssC}
	R_a^{\ssA} L_b^{\ssB'} U_2^{\ssB'\ssB} R_c^{\ssC}.
\end{equation}
It is clear that one can integrate over $\Omega_1$ and we naturally define a new kernel\footnote{We use the notation $K^{(2)}$ for many different quantities, but there should be nowhere any source of confusion since we shall also always write the associated indices.}
\begin{equation}
	\label{k2int}
	K^{(2)}_{abc;2} = \int \frac{\dif^2 \Omega_1}{4\pi}
	K^{(2)}_{abc;12}.
\end{equation}
We shall do the integration first in the small angle approximation, in which one has $\alpha_{ab} \simeq \theta_{ab}^2/4$ and similarly for the other angles, and in the strongly ordered regime
\begin{equation}
	\label{strord}
	\theta_{ab}, \theta_{ac}, \theta_{bc} \ll \theta_1 \ll \theta_2.
\end{equation} 
In the above $\theta_1$ stands collectively for the angles between gluon 1 and any of the partons $a,b$ and $c$, while $\theta_2$ stands for those between gluon 2 and any of the partons $a,b,c$ and 1. In this region the kernel in \eqn{k2nc} becomes
\begin{equation}
\label{ksord}
	K^{(2)}_{abc;12} \simeq
	-\frac{32}{\theta_2^4}\,
	\frac{\theta_{ab}^2-\theta_{ac}^2-\theta_{bc}^2}{\theta_1^2}
	\ln\frac{\theta_2^2}{\theta_1^2},
\end{equation}
and now it is straightforward to integrate over $\dif^2\Omega_1 \simeq \pi \dif \theta_1^2$. We easily see that the logarithmically dominated integration gives
\begin{equation}
	\label{k2intapp}
	K^{(2)}_{abc;2} \simeq
	\int_{\theta^2_{\rm min}}^{\theta^2_2}
	\frac{\pi \dif \theta_1^2}{4\pi}\,K^{(2)}_{abc;12}
	\simeq - \frac{4(\theta_{ab}^2-\theta_{ac}^2-\theta_{bc}^2)}{\theta_2^4}\,
	\ln^2\frac{\theta_2^2}{\theta_{\rm min}^2},
\end{equation}
where $\theta_{\rm min}$ is any of $\theta_{ab},\theta_{ac},$ or $\theta_{bc}$ to the order of accuracy. Therefore, we already see the large double logarithm in the Hamiltonian kernel, which is further accompanied by the multiplicative factor $\sim \theta_{ab}^2/\theta_{2}^4$ representing the ``standard'' collinear behavior. Notice also the symmetry $K^{(2)}_{bac;2} = K^{(2)}_{abc;2}$.

In order to calculate the respective NLO contribution to the BMS equation, we shall act with the Hamiltonian on the fundamental dipole
 \begin{equation}
 	\label{gde}
 	P_{de} = \frac{1}{N_c} {\rm tr} \big(V_d V_e^{\dagger} \big),
 \end{equation}
which in the current framework is the probability appearing in the BMS equation. Let us note here that the two right derivatives in \eqn{k2ulrr} are implicitly assumed to be symmetrized \cite{Caron-Huot:2015bja}, i.e.~$R_a^{\ssA} R_c^{\ssC} \to \{R_a^{\ssA} R_c^{\ssC},R_c^{\ssC}R_a^{\ssA} \}/2$. This of course becomes relevant only when $a=c$, and due to the antisymmetry of the structure constants $f^{\ssA\ssB\ssC}$, the two right derivatives must act on different $V$-matrices in order to obtain a non-vanishing result. Thus we have
\begin{align}
	\rmi f^{\ssA\ssB\ssC} R_a^{\ssA} U_2^{\ssB'\ssB} L_b^{\ssB'} R_c^{\ssC} P_{de}
	& = \frac{\rmi f^{\ssA\ssB\ssC}}{N_c} 
	(\delta_{bd}-\delta_{be})
	(\delta_{ad}\delta_{ce} - \delta_{ae}\delta_{cd})
	U_2^{\ssB'\ssB}
	{\rm tr} \big(t^{\ssB'} V_d t^{\ssC} t^{\ssA}
	V_e^{\dagger} \big)
	\nn
	& = -\frac{1}{2}(\delta_{bd}-\delta_{be})
	(\delta_{ad}\delta_{ce} - \delta_{ae}\delta_{cd})
	U_2^{\ssB'\ssB}
	{\rm tr} \big(t^{\ssB'} V_d t^{\ssB} V_e^{\dagger} \big),
\end{align} 
and using first $U_2^{\ssB'\ssB} t^{\ssB'} = V_2 t^{\ssB} V_2^{\dagger}$ and subsequently standard Fierz rearrangement, we get  
\begin{equation}
	\label{actong}
	\rmi f^{\ssA\ssB\ssC} R_a^{\ssA} U_2^{\ssB'\ssB} L_b^{\ssB'} R_c^{\ssC} P_{de}
	= - \frac{N_c^2}{4}
	(\delta_{bd}-\delta_{be})
	(\delta_{ad}\delta_{ce} - \delta_{ae}\delta_{cd})
	\left(P_{d2} P_{2e} - \frac{1}{N_c^2} P_{de}\right).
\end{equation} 
It remains to ``contract'' the indices in the structures appearing in Eqs.~\eqref{k2intapp} and \eqref{actong} and we readily obtain
\begin{equation}
	\label{contract}
	\sum_{a,b,c} (\delta_{bd}-\delta_{be})
	(\delta_{ad}\delta_{ce} - \delta_{ae}\delta_{cd})
	(\theta_{ab}^2-\theta_{ac}^2-\theta_{bc}^2)
	= -4 \theta_{de}^2.
\end{equation} 
Putting everything together we arrive at
\begin{equation}
	\label{k2ulrrong}
	K^{(2)}_{\rm ULRR} P_{de} = 
	2 N_c^2 \int\frac{\dif \theta_2^2\, \theta_{de}^2}{\theta_2^4}\,\ln^2\frac{\theta_2^2}{\theta_{de}^2}
	\left(P_{d2} P_{2e} - \frac{1}{N_c^2} P_{de}\right).
\end{equation}
It is not hard to understand that the two terms with two left derivatives, one right and one $U$-matrix in \eqn{k2unou} will give a contribution equal to the above. Furthermore, the terms with three (same) derivatives and no $U$-matrix will add a contribution $-(2\CF/N_c) P_{de}$, so that we simply need to replace $(1/N_c^2) P_{de} \to P_{de}$ in \eqn{k2ulrrong}. Taking into account both the LO and the NLO terms in the expansion given in \eqn{hexpand} we can finally write (with the relabeling $d,e \to a,b$)  
\begin{equation}
	\label{hong}
	H P_{ab} = 
	\frac{\abar}{2}
	\int \frac{\dif \theta_2^2\, \theta_{ab}^2}{\theta_2^4}
	\left(1 - \frac{\abar}{2}
	\ln^2\frac{\theta_2^2}{\theta_{ab}^2}
	\right)
	\left(P_{a2} P_{2b} - P_{ab}\right),
\end{equation}
valid in the regime $\theta_{ab} \ll \theta_2$.

It becomes natural to ask whether similar double logarithms occur when gluons are radiated at smaller and smaller angles. We shall show that this is not the case, but given the individual strong UV singularities of each term in \eqn{k2nc}, we must perform the exact integration of the kernel in \eqn{k2int}. In order to simplify our task, we shall directly work with the BMS equation (and not at the Hamiltonian level). Eqs.~\eqref{gde}-\eqref{actong} are still valid and therefore the kernel to be integrated simplifies, more precisely we have
\begin{equation}
	\label{sumk2}
	\sum_{a,b,c} (\delta_{bd}-\delta_{be})
	(\delta_{ad}\delta_{ce} - \delta_{ae}\delta_{cd})
	K^{(2)}_{abc;12}
	= K^{(2)}_{dde;12} - K^{(2)}_{edd;12} 
	-K^{(2)}_{dee;12} + K^{(2)}_{eed;12}.
\end{equation}
It is easy to see that $K^{(2)}_{edd;12}=K^{(2)}_{dee;12}=0$, while
\begin{equation}
\label{k2dde}
	K^{(2)}_{dde;12} = 
	\frac{\alpha_{de}}{\alpha_{d2}\alpha_{2e}}
	\left( 
	\frac{\alpha_{de}}{\alpha_{d1}\alpha_{1e}}
	-\frac{\alpha_{2e}}{\alpha_{21}\alpha_{1e}}
	-\frac{\alpha_{d2}}{\alpha_{d1}\alpha_{12}}
	\right) \ln \frac{\alpha_{1e}^2}{\alpha_{2e}^2}
	\equiv \frac{\alpha_{de}}{\alpha_{d2}\alpha_{2e}}\, 
	\tilde{K}^{(2)}_{dde;12},
\end{equation}
and of course similarly for $K^{(2)}_{eed;12}$. We immediately observe that the prefactor and all three terms in the parenthesis in \eqn{k2dde} correspond to elementary antenna patterns. Since we shall integrate over $\dif^2 \Omega_1$, we have temporarily defined the kernel $\tilde{K}^{(2)}_{dde;12}$ by leaving aside the antenna pattern in the prefactor which does not involve the gluon 1.

In order to perform the integration over $\dif^2 \Omega_1$, we shall find it very convenient to make a change of variables from angles to transverse coordinates according to the stereographic projection. That is, we define (as before, $D$ denotes the diameter of the sphere used for the projection)
\begin{equation}
	\label{ster}
	\alpha_{ab} = 
	\frac{D^2x_{ab}^2}{(D^2+x_a^2)(D^2+ x_b^2)},
\end{equation}
where the dependence upon $D^2$ cancels when considering a conformal invariant quantity, like the antenna pattern accompanied by the relevant integration measure:
\begin{equation}
\label{sterint}	
	\int
	\frac{\dif^2 \Omega_1}{4\pi}\,
	\frac{\alpha_{ab}}{\alpha_{a1}\alpha_{1b}} = 
	\int \frac{\dif^2 x_1}{\pi}\,
	\frac{x_{ab}^2}{x_{a1}^2 x_{1b}^2}.
\end{equation}
Then we have
\begin{equation}
	\label{k2tildeint}
	\tilde{K}^{(2)}_{dde;2} = 
	\int \frac{\dif^2 \Omega_1}{4\pi}\,
	\tilde{K}^{(2)}_{dde;12} =
	2 \int \frac{\dif^2 x_1}{\pi}\,
	\left( 
	\frac{x_{de}^2}{x_{d1}^2x_{1e}^2}
	-\frac{x_{2e}^2}{x_{21}^2x_{1e}^2}
	-\frac{x_{d2}^2}{x_{d1}^2x_{12}^2}
	\right)
	\left(
	\ln \frac{x_{1e}^2}{x_{2e}^2}
	+\ln\frac{D^2 +  x_2^2}{D^2+x_1^2}
	\right).
\end{equation}  
For the moment we shall neglect the last logarithmic term, but we shall comment towards the end of the calculation on its importance and the changes it brings to our results. 

Now, the three terms will be integrated separately in $d=2+2\epsilon$ dimensions and we shall need three basic integrals. The first one is
\begin{equation}
	\label{j1}
	J_1 = \frac{1}{\pi}\int \frac{\dif^d z\, x^2}{z^2(x-z)^2} = 
	(\pi x^2)^{\epsilon}\, 
	\frac{\Gamma^2(\epsilon) \Gamma(1-\epsilon)}{\Gamma(2\epsilon)}
	= \frac{2}{\epsilon} + 
	2 \left( 
	\ln x^2 + \ln \pi +\gamma_{\ssE}
	\right), 
\end{equation}
where of course after the last equality we have dropped terms that vanish when $\epsilon \to 0$, and with $\gamma_{\ssE}=0.577 \dots$ the Euler-Mascheroni constant. The second one is
\begin{align}
	\label{j2}
	J_2 & = \frac{1}{\pi}\int \frac{\dif^d z\, x^2}{z^2(x-z)^2} \ln z^2 = 
	\lim_{\gamma\to 0}
	\frac{\dif}{\dif \gamma}
	\frac{1}{\pi}\int \frac{\dif^d z\, x^2 z^{2\gamma}}{z^2(x-z)^2}
	\nn
	 & = 
	\lim_{\gamma\to 0}
	\frac{\dif}{\dif \gamma}
	\pi^{\epsilon} \big(x^2\big)^{\gamma+\epsilon}
	\frac{(1-\gamma)\Gamma(1-\gamma-\epsilon)\Gamma(\gamma+\epsilon) \Gamma(\epsilon)}{\Gamma(2-\gamma) \Gamma(\gamma+2\epsilon)}
	\nn 
	& = -\frac{1}{\epsilon^2} +
	\frac{1}{\epsilon}
	\left(\ln x^2 -\ln\pi -\gamma_{\ssE} \right)
	+ 
	\left[ 
	\frac{3}{2}\ln^2 x^2 + (\ln \pi +\gamma_{\ssE}) \ln x^2
	+ \frac{\pi^2}{12} - \frac{1}{2} (\ln \pi + \gamma_{\ssE})^2
	\right]. 
\end{align}
The above two integrals have been calculated in the ``standard way''; the denominators (including the $z^{2\gamma}$ in $J_2$) have been combined by introducing a Feynman parameter via the general formula
\begin{equation}
	\label{feyn}
	\frac{1}{A^a B^b} = \frac{\Gamma(a+b)}{\Gamma(a)\Gamma(b)}
	\int_0^1 \dif u\,\frac{u^{a-1} (1-u)^{b-1}}{[A u +B (1-u)]^{a+b}},
\end{equation}
then, after a shift in $z$, the $\dif^2 z$ integration has been performed by making use of 
\begin{equation}
	\label{solid}
	\int \frac{\dif^d z}{\big(z^2 + \Delta \big)^n} = \pi^{d/2}
	\frac{\Gamma(n-d/2)}{\Gamma(n)}\Delta^{d/2-n},
\end{equation}
with $\Delta = u(1-u)x^2$, and finally the integration over $u$ has been done by relying once again on \eqn{feyn} (this time from ``right to left'' and with $A=B=1$, but with different exponents). The third integral we need is a finite one and reads (cf.~Eq.~(120) in \cite{Balitsky:2009xg})
\begin{equation}
	\label{j3}
	J_3 = \frac{1}{\pi}\int 
	\frac{\dif^2 z\, x^2}{z^2(x-z)^2}
	\ln\frac{(z-y)^4}{y^2(x-y)^2}
	= \ln^2 \frac{(x-y)^2}{y^2}.    
\end{equation}
The integrand in \eqn{j3} becomes singular when $z\to 0$ or $z\to x$, however the two singularities cancel each other due to the presence of the logarithm. 

Employing the expressions for $J_1$ and $J_2$ we find for the first two contribution in \eqn{k2tildeint}
\begin{align}
\label{firstint}
	2 \int \frac{\dif^d x_1}{\pi}\,
	\frac{x_{de}^2}{x_{d1}^2x_{1e}^2}
	\ln \frac{x_{1e}^2}{x_{2e}^2} = 
	&-\frac{2}{\epsilon^2} 
	+ \frac{2}{\epsilon}
	\left(\ln x_{de}^2 -\ln\pi -\gamma_{\ssE} \right)
	+ \Big[ 3\ln^2 x_{de}^2 + 2(\ln \pi +\gamma_{\ssE}) 
	\ln x_{de}^2
	\nn
	&+ \frac{\pi^2}{6} -(\ln \pi + \gamma_{\ssE})^2
	\Big]
	-\frac{4}{\epsilon}\ln x_{2e}^2
	-4 \left( 
	\ln x_{de}^2 + \ln \pi +\gamma_{\ssE}
	\right)\ln x_{2e}^2,
\end{align}
and
\begin{align}
\label{secondint}
\hspace*{-0.1cm}
	-2 \int \frac{\dif^d x_1}{\pi}\,
	\frac{x_{2e}^2}{x_{21}^2 x_{1e}^2}
	\ln \frac{x_{1e}^2}{x_{2e}^2} = 
	&+\frac{2}{\epsilon^2} 
	- \frac{2}{\epsilon}
	\left(\ln x_{2e}^2 -\ln\pi -\gamma_{\ssE} \right)
	- \Big[ 3\ln^2 x_{2e}^2 + 2(\ln \pi +\gamma_{\ssE}) 
	\ln x_{2e}^2
	\nn
	&+ \frac{\pi^2}{6} -(\ln \pi + \gamma_{\ssE})^2
	\Big]
	+\frac{4}{\epsilon}\ln x_{2e}^2
	+4 \left( 
	\ln x_{2e}^2 + \ln \pi +\gamma_{\ssE}
	\right)\ln x_{2e}^2.
\end{align}
Regarding the contribution from the third term in \eqn{k2tildeint} we first ``decompose'' it as
\begin{align}
\label{thirddec}
	-2 \int \frac{\dif^d x_1}{\pi}\,
	\frac{x_{d2}^2}{x_{d1}^2x_{12}^2}
	\ln \frac{x_{1e}^2}{x_{2e}^2} = 
	-\int \frac{\dif^d x_1}{\pi}\,
	\frac{x_{d2}^2}{x_{d1}^2x_{12}^2}
	\ln \frac{(x_{12}-x_{e2})^4}{x_{e2}^2 x_{de}^2}
	- \int \frac{\dif^d x_1}{\pi}\,
	\frac{x_{d2}^2}{x_{d1}^2x_{12}^2}
	\ln \frac{x_{de}^2}{x_{2e}^2}.
\end{align} 
Using the expressions for $J_3$ (after setting $d=2$ in the first term in \eqn{thirddec}) and $J_1$ we get\begin{align}
\label{thirdint}
	-2 \int \frac{\dif^d x_1}{\pi}\,
	\frac{x_{d2}^2}{x_{d1}^2x_{12}^2}
	\ln \frac{x_{1e}^2}{x_{2e}^2} =
	-\ln^2\frac{x_{2e}^2}{x_{de}^2}
	- \frac{2}{\epsilon}\ln \frac{x_{de}^2}{x_{2e}^2}
	- 2 \big(\ln x_{d2}^2 + \ln \pi +\gamma_{\ssE}\big)
	\ln \frac{x_{de}^2}{x_{2e}^2}.
\end{align} 
Putting together Eqs.~\eqref{firstint}, \eqref{secondint} and \eqref{thirdint} we see that all the divergencies cancel (notice that for the cancellation of the single pole all three terms contribute) and the remaining finite piece after some rearrangement reads\footnote{All the finite terms involving $\pi^2$, $\ln\pi$ and $\gamma_{\ssE}$ cancel.} 
\begin{equation}
\label{k2tilde}
	\tilde{K}^{(2)}_{dde;2} =
	2 \ln \frac{x_{d2}^2}{x_{de}^2}
	\ln \frac{x_{2e}^2}{x_{de}^2}.
\end{equation}
Since this is symmetric in $d$ and $e$, we eventually obtain an identical double logarithm from $\tilde{K}^{(2)}_{eed;2}$.
 
At this point of the calculation let us go back to the angle coordinates. The inverse to the transformation in \eqn{ster} is given by
\begin{equation}
	\label{sterinv}
	 \frac{x_{ab}^2}{D^2} = \frac{\alpha_{ab}}{(1-\alpha_a)(1-\alpha_b)} = 
	 \frac{2(1-\cos\theta_{ab})}{(1+\cos\theta_a)
	 (1+\cos\theta_b)},
\end{equation}
and returning to \eqn{k2tilde} we have
\begin{equation}
\label{k2tildeangle}
	\tilde{K}^{(2)}_{dde;2} =
	2 \ln 
	\frac{\alpha_{d2}(1-\alpha_e)}{\alpha_{de}(1-\alpha_2)}
	\ln 
	\frac{\alpha_{2e}(1-\alpha_d)}{\alpha_{de}(1-\alpha_2)}.
\end{equation}
We note that there is no large logarithmic contribution when the gluon 2 is very close to one of the legs $d$ or $e$: although one of the logarithms gets large, the other vanishes. It looks like issues will arise when the gluon 2 is radiated close to $\pi$, however such singularities cannot be present as one can see immediately by a direct inspection of the unintegrated kernel in \eqn{k2dde}. Eventually they should cancel with the logarithmic term neglected in \eqn{k2tildeint}. Instead of performing the exact calculation of this contribution, we shall equivalently drop all the factors $(1-\alpha_i)$ in \eqn{k2tildeangle}, and the error in such a procedure is at most of order $\mathcal{O}(\abar^2)$, but not enhanced by a large logarithm. Thus, for what follows we shall simply take
\begin{equation}
\label{k2tildeanglenew}
	\tilde{K}^{(2)}_{dde;2} =
	2 \ln 
	\frac{\alpha_{d2}}{\alpha_{de}}
	\ln 
	\frac{\alpha_{2e}}{\alpha_{de}}.
\end{equation} 

It only remains to assemble the various parts of the calculation, and the steps are identical to those we took in order to derive \eqn{k2ulrrong} and subsequently \eqn{hong}. We arrive at (again with the relabeling $d,e \to a,b$)
\begin{equation}
	\label{hongexact}
	H P_{ab} = 
	\frac{\abar}{2}
	\int \frac{\dif \Omega_2}{4\pi}
	\frac{\alpha_{ab}}{\alpha_{a2}\alpha_{2b}}
	\left(1 - \frac{\abar}{2}
	\ln\frac{\alpha_{a2}}{\alpha_{ab}}
	\ln\frac{\alpha_{2b}}{\alpha_{ab}}
	\right)
	\left(P_{a2} P_{2b} - P_{ab}\right).
\end{equation}
Needless to say, when angles are small and strongly ordered according to $\theta_{ab} \ll \theta_{a2} \simeq \theta_{2b} \ll 1$, \eqn{hongexact} reduces to \eqn{hong}. Still, we would like to emphasize once again that the double logarithm in the above equation doesn't get large in any other special ``corner'' of the phase space. 

To go from the non-conformal to the conformal scheme one changes the observables on which the Hamiltonian is acting \cite{Balitsky:2009xg,Caron-Huot:2015bja}. This induces a change in the kernels of the Hamiltonian, and the particular logarithmic structure of \eqn{k2nc}, which led to the large double logarithm in \eqn{hong}, dissappears. However, a new logarithmic term arises in a different kernel of the Hamiltonian, as can be readily seen in \cite{Caron-Huot:2015bja} by comparing the $\mathcal{N}=4$ SYM part of Eq.~(3.12b) with Eq.~(3.33) there, and is the one which will give large double logarithmic contributions. To be more precise, the change in the kernel of interest is
\begin{equation}
	\label{deltak2}
	\delta K^{(2) \rm cs}_{ab;12} =
	\frac{2 \alpha_{ab}}{\alpha_{a1}\alpha_{12}\alpha_{2b}}
	\ln \frac{\alpha_{ab}\alpha_{12}}{\alpha_{a2}\alpha_{1b}},
\end{equation}
and the corresponding term in the NLO BMS equation in the conformal scheme reads
\begin{equation}
	\label{hcongc}
	\delta H^{\rm cs} P_{ab}^{\rm cs} = \frac{\abar^2}{8}
	\int \frac{\dif^2 \Omega_1}{4 \pi}
	\frac{\dif^2 \Omega_2}{4 \pi}\,
	\delta K^{(2) \rm cs}_{ab;12}\,
	( P_{a1}^{\rm cs} P_{12}^{\rm cs} P_{2b}^{\rm cs} - P_{a1}^{\rm cs} P_{1b}^{\rm cs}).
\end{equation}
 The kernel in \eqn{deltak2} in the small angle approximation and with strongly increasing angles, i.e.~when $\theta_{ab} \ll \theta_1 \ll \theta_2$ (with the notation already explained after \eqn{strord}), becomes
\begin{equation}
	\label{deltak2approx}
	\delta K^{(2) \rm cs}_{ab;12} \simeq
	-\frac{32 \theta_{ab}^2}{\theta_2^4\theta_1^2}
	\ln \frac{\theta_1^2}{\theta_{ab}^2},
\end{equation}
thus it exhibits a logarithmic enhancement. We shall work in the region where all probabilities are close to unity, so let us define $P=1-R$ as in the main text. Then one has $P_{a1}^{\rm cs} P_{12}^{\rm cs} P_{2b}^{\rm cs} - P_{a1}^{\rm cs} P_{1b}^{\rm cs} \simeq - R_{12}^{\rm cs} - R_{2b}^{\rm cs} + R_{1b}^{\rm cs} \simeq -2 R_2^{\rm cs}$, where the last approximate equality is valid in the aforementioned strongly ordered regime, and \eqn{hcongc} reads
\begin{equation}
	\label{hcongcord}
	\delta H^{\rm cs} R_{ab}^{\rm cs} = -\frac{\abar^2}{2}
	\int \frac{\dif \theta_2^2\, \theta_{ab}^2}{\theta_2^2}
	\,R_2^{\rm cs}
	\int_{\theta_{ab}^2}^{\theta_2^2}
	\frac{\dif \theta_1^2}{\theta_1^2}
	\ln \frac{\theta_1^2}{\theta_{ab}^2}.
\end{equation}
Now it is straightforward to integrate over $\dif \theta_1^2$ and by furthermore including the LO term of the BMS equation in the regime of interest we finally arrive at 
\begin{equation}
	\label{hconqcfull}
	H^{\rm cs} R_{ab}^{\rm cs} = \abar
	\int \frac{\dif \theta_2^2\, \theta_{ab}^2}{\theta_2^4}
	\left(1 - \frac{\abar}{4}
	\ln^2 \frac{\theta_2^2}{\theta_{ab}^2}
	\right) R_2^{\rm cs}.
\end{equation}  
Notice that the coefficient of the double logarithmic contribution in the conformal scheme above is half of the respective coefficient in the non-conformal scheme (cf.~\eqn{hong}).     
Furthermore, contrary to what happens in the non-conformal scheme, large logarithms emerge also in the case that angles get smaller and smaller. One can see that there are two strongly ordered regimes in which the logarithm of the kernel in \eqn{deltak2} can become large: (i) when $\theta_{ab} \simeq \theta_{a1} \simeq \theta_{a2} \gg \theta_{1b} \simeq \theta_{2b} \gg \theta_{12}$ and (ii) when $\theta_{ab} \simeq \theta_{1b} \simeq \theta_{2b} \gg \theta_{a1} \simeq \theta_{a2} \gg \theta_{12}$. They contribute the same and a straightforward calculation, analogous to the one that led us to \eqn{hconqcfull} above, gives 
\begin{equation}
	\label{hconqcfull2}
	H^{\rm cs} R_{ab}^{\rm cs} = \abar
	\int \frac{\dif \theta_{12}^2}{\theta_{12}^2}
	\left(1 - \frac{\abar}{4}
	\ln^2 \frac{\theta_{ab}^2}{\theta_{12}^2}
	\right) R_{12}^{\rm cs}.
\end{equation}  

\bigskip
\bibliographystyle{utcaps}

\providecommand{\href}[2]{#2}\begingroup\raggedright\endgroup

\end{document}